\documentclass{emulateapj}
\lefthead{Siverd et al.} \righthead{KELT-1b}
\newcommand{\be}{\begin{equation}}
\newcommand{\ee}{\end{equation}}
\newcommand{\bea}{\begin{eqnarray}}
\newcommand{\eea}{\end{eqnarray}}
\newcommand{\mjup}{M_{\rm Jup}}
\newcommand{\rjup}{R_{\rm Jup}}
\newcommand{\teff}{\ensuremath{T_{\rm eff}}}
\newcommand{\feh}{{\rm [Fe/H]}}
\newcommand{\mum}{\mu{\rm m}}
\newcommand{\logg}{\ensuremath{\log{g}}}
\newcommand{\pchisq}{\ensuremath{P\left(>\chisq \right)}}
\newcommand{\chisq}{\ensuremath{\chi^{\,2}}}
\newcommand{\vsini}{\ensuremath{\,{v\sin{I_*}}}}
\newcommand{\bjdtdb}{\ensuremath{\rm {BJD_{TDB}}}}
\newcommand{\ecosw}{\ensuremath{e\cos{\omega_*}}}
\newcommand{\esinw}{\ensuremath{e\sin{\omega_*}}}
\newcommand{\msun}{\ensuremath{\,{\rm M}_\Sun}}
\newcommand{\rsun}{\ensuremath{\,{\rm R}_\Sun}}
\newcommand{\lsun}{\ensuremath{\,{\rm L}_\Sun}}
\newcommand{\mj}{\ensuremath{\,{\rm M}_{\rm J}}}
\newcommand{\rj}{\ensuremath{\,{\rm R}_{\rm J}}}
\newcommand{\fave}{\langle F \rangle}
\newcommand{\fluxcgs}{10$^9$ erg s$^{-1}$ cm$^{-2}$}
\newcommand{\kms}{\ensuremath{\,{\rm km~s^{-1}}}}
\newcommand{\ms}{\ensuremath{\,{\rm m~s^{-1}}}}
\usepackage{apjfonts}

\begin{document}
\title{KELT-1\lowercase{b}: A Strongly Irradiated, Highly Inflated, Short Period, $27$ Jupiter-mass Companion Transiting a mid-F Star}
\author{
  Robert J.\ Siverd\altaffilmark{1},
  Thomas G.\ Beatty\altaffilmark{2},
  Joshua Pepper\altaffilmark{1},
  Jason D.\ Eastman\altaffilmark{3,4},
  Karen Collins\altaffilmark{5}, 
  Allyson Bieryla\altaffilmark{6},
  David W.\ Latham\altaffilmark{6},
  Lars A.\ Buchhave\altaffilmark{7,8},
  Eric L.\ N.\ Jensen\altaffilmark{9},
  Justin R.\ Crepp\altaffilmark{10},
  Rachel Street\altaffilmark{3},
  Keivan G.\ Stassun\altaffilmark{1,11},
  B.\ Scott Gaudi\altaffilmark{2},
  Perry Berlind\altaffilmark{6}
  Michael L.\ Calkins\altaffilmark{6},
  D.\ L.\ DePoy\altaffilmark{12,13},
  Gilbert A.\ Esquerdo\altaffilmark{6},
  Benjamin J.Fulton\altaffilmark{3},
  G\'{a}bor F\H{u}r\'{e}sz\altaffilmark{6},
  John C.\ Geary\altaffilmark{6},
  Andrew Gould\altaffilmark{2},
  Leslie Hebb\altaffilmark{1},
  John F. Kielkopf\altaffilmark{5},
  Jennifer L.\ Marshall\altaffilmark{12}, 
  Richard Pogge\altaffilmark{2},
  K.Z.Stanek\altaffilmark{2},
  Robert P.\ Stefanik\altaffilmark{6},
  Andrew H.\ Szentgyorgyi\altaffilmark{6},
  Mark Trueblood\altaffilmark{14},
  Patricia Trueblood\altaffilmark{14},
  Amelia M.\ Stutz\altaffilmark{15,16},
  and
  Jennifer L.\ van Saders\altaffilmark{2}
}
\altaffiltext{1}{Department of Physics and Astronomy, Vanderbilt University, Nashville, TN 37235, USA}
\altaffiltext{2}{Department of Astronomy, The Ohio State University, 140 W.\ 18th Ave., Columbus, OH 43210, USA}
\altaffiltext{3}{Las Cumbres Observatory Global Telescope Network, 6740 Cortona Drive, Suite 102, Santa Barbara, CA 93117, USA}
\altaffiltext{4}{Department of Physics Broida Hall, University of California, Santa Barbara, CA 93106, USA}
\altaffiltext{5}{Department of Physics \& Astronomy, University of Louisville, Louisville, KY 40292, USA}
\altaffiltext{6}{Harvard-Smithsonian Center for Astrophysics, 60 Garden Street, Cambridge, MA 02138, USA}
\altaffiltext{7}{Niels Bohr Institute, University of Copenhagen,Juliane Maries vej 30, 21S00 Copenhagen, Denmark}
\altaffiltext{8}{Centre for Star and Planet Formation, Geological Museum, {\O}ster Voldgade 5, 1350 Copenhagen, Denmark}
\altaffiltext{9}{Department of Physics and Astronomy, Swarthmore College, Swarthmore, PA 19081, USA}
\altaffiltext{10}{Department of Astrophysics, California Institute of Technology, Pasadena, CA 91125, USA}
\altaffiltext{11}{Department of Physics, Fisk University, Nashville, TN 37208, USA}
\altaffiltext{12}{Department of Physics \& Astronomy, Texas A\&M University, College Station, TX 77843, USA}
\altaffiltext{13}{George P. and Cynthia Woods Mitchell Institute for Fundamental Physics and Astronomy}
\altaffiltext{14}{Winer Observatory, Sonoita, AZ 85637, USA}
\altaffiltext{15}{Max Planck Institute for Astronomy, Heidelberg, Germany}
\altaffiltext{16}{Department of Astronomy \& Steward Observatory, University of Arizona, Tucson, AZ 85721, USA}

\begin{abstract}
We present the discovery of KELT-1b, the first transiting low-mass
companion from the wide-field Kilodegree Extremely Little
Telescope-North (KELT-North) transit survey, which surveys $\sim 40\%$
of the northern sky to search for transiting planets around bright
stars.  The initial transit signal was robustly identified in the
KELT-North survey data, and the low-mass nature of the occultor was
confirmed via a combination of follow-up photometry, high-resolution
spectroscopy, and radial velocity measurements.  False positives are
disfavored by the achromaticity of the primary transits in several
bands, a lack of evidence for a secondary eclipse, and insignificant
bisector variations.  A joint analysis of the spectroscopic, radial
velocity, and photometric data indicates that the $V=10.7$ primary is
a mildly evolved mid-F star with $\teff=6518\pm 50~{\rm K}$,
$\log{g_*}=4.229_{-0.019}^{+0.012}$ and $\feh=0.008\pm0.073$, with an
inferred mass $M_*=1.324\pm0.026~M_\odot$ and radius
$R_*=1.462_{-0.024}^{+0.037}~R_\odot$. The companion is a low-mass
brown dwarf or a super-massive planet with mass
$M_P=27.23_{-0.48}^{+0.50}~\mjup$, radius
$R_P=1.110_{-0.022}^{+0.032}~\rjup$, surface gravity
$\log{g_{P}}=4.738_{-0.023}^{+0.017}$, and a density
$\rho_P=24.7_{-1.9}^{+1.4}~{\rm g~cm^{-3}}$.  The companion is on a
very short ($\sim 29$ hour) period circular orbit, with an ephemeris
$T_c({\bjdtdb})=2455909.292797\pm 0.00024$ and $P=1.2175007\pm
0.000018~{\rm d}$, and a semimajor axis of $a=0.02466\pm
0.00016$AU. KELT-1b receives a large amount of stellar insolation,
with $\langle F \rangle = 7.81_{-0.33}^{+0.42} \times 10^9~{\rm
erg~s^{-1}~cm^{-2}}$, implying an equilibrium temperature assuming
zero albedo and perfect redistribution of
$T_{eq}=2422_{-26}^{+32}~{\rm K}$.  Upper limits on the secondary
eclipse depth in $i$ and $z$ bands indicate that either the
companion must have a non-zero albedo, or it must experience some
energy redistribution.  Comparison with standard evolutionary models
for brown dwarfs suggests that the radius of KELT-1b is likely to be
significantly inflated. Adaptive optics imaging reveals a candidate
stellar companion to KELT-1 with a separation of $588\pm 1$mas, which
is consistent with an M dwarf if it is at the same distance as the primary.
Rossiter-McLaughlin measurements during transit imply a projected
spin-orbit alignment angle $\lambda = 2\pm16$ degrees, consistent with
the orbit pole of KELT-1b being aligned with the spin axis of the primary.
Finally, the $\vsini=55.4\pm 2.0~{\rm km~s^{-1}}$ of the primary is
consistent at $\sim 2~\sigma$ with tidal
synchronization.  Given the extreme parameters of the
KELT-1 system, we expect it to provide an important testbed for
theories of the emplacement and evolution of short-period companions,
as well as theories of tidal dissipation and irradiated brown dwarf
atmospheres.
\end{abstract}
\section{Introduction}

The most information-rich exoplanetary systems are those in which the
companion happens to transit in front of its parent star.  Transiting
systems are enormously useful for enabling detailed measurements of a
seemingly endless array of physical properties of extrasolar planets
and their host stars (see reviews by \citealt{winn2009,winn2010}).
The most basic properties that can be measured using transiting
planets are the planet mass and radius, and so average density.  These
parameters alone allow for interesting constraints on the internal
composition and structure of planets
\citep{guillot05,fortney2007,rogers2010,miller2011}.  In addition to
these basic parameters, transiting planets enable the study of their
atmospheres \citep{seager2000,charbonneau02,vidal03,seagerd2010} and 
thermal emission \citep{deming05,charbonneau2005,knutson2008}.
They also allow measurement of planetary and stellar oblateness,
rotation rate, and spin-orbit alignment 
\citep{seager02,spiegel2007,carter2010,rossiter1924,mclaughlin1924,winn2005,gaudi06}.
Transiting planets may also be searched 
for associated rings and moons \citep{brown01,barnes04,tusnski2011}.
Further, variations in transit timing may indicate the presence of other
bodies in the system \citep{holman05,agol05,steffen05,ford06,ford07,kipping2009}. 
With sufficiently precise observations, one may constrain
the presence of planets with masses smaller than that of the Earth \citep{agol06,carter2010}.

The high scientific value of transiting planet systems motivated the
first dedicated wide-field transit surveys, which by now have identified
over 100 transiting systems (TrES, \citealt{alonso2004}; XO,
\citealt{mccullough2006}; HATNet, \citealt{bakos2007}; SuperWASP,
\citealt{cameron2007a}, QES, \citealt{alsubai2011}).  Although there
is substantial diversity in their design, strategy, and sensitivity,
these surveys can be grossly characterized as having relatively small cameras
with apertures of order $10$~cm, and detectors with relatively wide fields-of-view
of tens of square degrees.  These surveys are primarily sensitive to
giant, close-in planets with radii $R_P\ga 0.5\rjup$ and periods of 
$P\la 10~{\rm d}$, orbiting relatively bright FGK stars with $V\sim 10-12$.
 
The space-based missions CoRoT
\citep{baglin2003} and Kepler \citep{borucki2010} have dramatically expanded the parameter space of
transit surveys, enabling the detection of planets with sizes down to
that of the Earth and below, planets with periods of several years, and planets
orbiting a much broader range of host stars.  Furthermore, their large
target samples have allowed the detection of rare
and therefore interesting planetary systems.  These missions have
already announced over 50 confirmed planets, and the Kepler mission
has announced an additional $\sim 2300$  candidates
\citep{batalha2012}, most of which are smaller than Neptune.
Notable individual discoveries include the first detection of a
transiting Super-Earth \citep{leger2009}, the detection of a
`temperate' gas giant with a relatively long period of $ \sim 100$
days \citep{deeg2010}, the first multi-planet transiting systems
\citep{steffen2010,holman2010,latham2011,lissauer2011}, the first circumbinary planets
\citep{doyle2011,welsh2012}, and the detection of planets with radius
of $\la R_\oplus$ \citep{muirhead2012, fressin2012}.

Although Kepler and CoRoT have revolutionized our understanding of the
demographics of planets, the opportunities for follow-up of the
systems detected by these missions are limited.  By design, 
both missions primarily monitor relatively faint stars with
$V\ga 12$.  Consequently, many of the follow-up observations discussed
above that are generically enabled by transiting systems are not
feasible for the systems detected by Kepler and CoRoT.  Detailed
characterization of the majority of these systems will therefore be
difficult or impossible.  There is thus an ongoing need to discover
transiting planets orbiting the bright stars, as well as to increase
the diversity of such systems.

All else being equal, the brightest stars hosting transiting planets
are the most valuable.  Larger photon flux permits more instruments
and/or facilities to be employed for follow-up, allows subtler
effects to be probed, reduces statistical uncertainties, and
generally allows for improved or more extensive calibration procedures
that help to control systematic errors.  Furthermore, brighter
stars are also easier to characterize, and are more likely to have
pre-existing information, such as proper motions, parallaxes,
metallicities, effective temperatures, angular diameters, and
broadband colors.  

The majority of the brightest ($V\la 8$) FGK dwarfs in the sky have been
monitored using precision radial velocity surveys for many years, and
as a result most of the giant planets with periods of less than a few
years orbiting these stars have already been discovered (e.g., \citealt{wright2012}).  A
smaller subset of these stars have been monitored over a shorter time
baseline with the sensitivity needed to detect Neptune- and SuperEarth-mass
planets.  Because of the low a priori transit probability for all but short period
planets, the transiting systems constitute a very small
fraction of this sample.  To date, seven planets first discovered via
radial velocity have subsequently been discovered to also transit;
all of the host stars for these planets are brighter than $V=9$.  
Although there are projects that aim to increase this sample \citep{kane2009},
the overall yield is expected to be small.  

Because RV surveys generically require spectroscopic observations that
are observationally expensive and must be obtained in series, it is
more efficient to discover transiting planets around the much more
abundant fainter stars by first searching for the photometric transit
signal, and then following these up with targeted RV observations to
eliminate false positives and measure the planet mass.  However, in
order to compensate for the rarity and low duty cycle, many stars 
must be monitored over
a long time baseline.  Photometric transit surveys that target
brighter stars therefore require larger fields of view.  Most of the
original transit surveys had fields of view and exposure times that were optimized to
detect planets orbiting stars with $V\ga 10$.  Indeed, only $\sim 20$
transiting planets orbiting stars with $V\le 10$ are currently known ($\sim 40$
with $V\la 11$). Of those with $V\le 10$, $\sim 40\%$ were originally detected by RV surveys.

The Kilodegree Extremely Little Telescope-North (KELT-North) transit survey
\citep{KELT_SYNOPTIC}
was designed to detect giant, short-period transiting planets orbiting
the brightest stars that are not readily accessible to RV
surveys. \citet{KELT_THEORY} determined the optimal hardware setup
specifically to detect transiting planets orbiting stars with $V\sim
8-10$, and based on the specified design requirements in that paper,
the KELT-North survey telescope system was constructed using
off-the-shelf, high-end consumer equipment.  In fact, 
as the current detection demonstrates, KELT has exceeded its
design goals, and is sensitive to transiting systems in some
favorable cases down to $V\sim 12$.

In addition to the goal of
filling in the magnitude gap between radial velocity and other transit
surveys, the KELT-North survey also has the potential to detect fainter
systems with $V\ga 10$ that are in the magnitude range
of previous surveys, but were missed or overlooked for various reasons.  
The detection discussed in this paper is an example of this opportunity. 
Here the
fact that the KELT-North survey is only now starting to vet candidates, more than
eight years after the first candidates were announced by other transit
surveys, can be seen an advantage.  In particular, previous surveys
have established the existence of massive brown dwarf companions
\citep{deleuil2008,irwin2010,bouchy2011a,johnson2011,bouchy2011b}, and
have demonstrated the feasibility of detecting low-mass companions to
hot, rapidly rotating stars \citep{cameron2010}.  Partially in
response to these results, the KELT-North survey deliberately broadened our
search targets to include hot and/or rapidly-rotating stars, which
were previously neglected by many transit surveys.  The evolving perception
of what kinds of stars constitute viable transit search
targets played an interesting role in the discovery of KELT-1b, as discussed
in \S\ref{sec:hat}.

The KELT-North survey has been collecting data since September 2006, and has
acquired a sufficient number of high-quality images to detect transit
candidates.  We have been systematically identifying and vetting
transit candidates since February 2011, and in this paper we report our first
confirmed low-mass transiting companion, which we designate KELT-1b.
KELT-1b has a mass of $\sim 27~\mjup$, and we will therefore 
follow convention
and refer to it as a `brown dwarf' throughout the majority of this paper.  
However, as we discuss
in \S\ref{sec:bdd}, we are, in fact, agnostic about its true
nature and therefore how it should be categorized. 

The outline of this paper is as follows. 
In order to introduce the survey and provide the appropriate context for
our discovery, in \S\ref{sec:survey} we summarize the properties
of the KELT-North survey and our procedure for candidate selection.
In \S\ref{sec:observations} we review the observations of KELT-1,
starting with the properties of the candidate in the KELT-North data, and then summarize
the follow-up photometry, spectroscopy, and high-contrast imaging.  
\S\ref{sec:char} describes our analysis and characterization of the host star and its
substellar companion. In \S\ref{sec:discussion}
we provide a speculative discussion of the possible implications of this unique
system for theories
of the emplacement and tidal evolution of short-period substellar companions, models
of the structure and atmosphere of brown dwarfs, and the demographics of substellar companions
to stars.  We briefly summarize in \S\ref{sec:summary}.

\section{The KELT-North Survey}\label{sec:survey}

Because this is the first paper from the KELT-North survey, we describe
the survey, selection criteria, and follow-up observations and reduction
methodology in some detail.  Readers who are  not interested in these
details, but are rather primarily interested in the properties and implications
of the KELT-1b system, can skip to \S\ref{sec:char}.  

\subsection{KELT-North Instrumentation and Survey Strategy}\label{sec:instrument}

The KELT-North survey instrument
consists of a collection of
commercially-available equipment, chosen to meet the requirements of
\cite{KELT_THEORY} and tuned to find the few brightest stars with transiting
planets in the Northern sky. The optical system consists of an Apogee
AP16E (4K x 4K 9$\mu$m pixels) thermo-electrically cooled CCD camera
attached using a custom mounting plate  to a Mamiya 
camera lens with a 80mm focal length and 42mm aperture (f/1.9). 
The resultant field of view of the detector is $26^\circ
\times 26^\circ$ at roughly $23$\arcsec per pixel, allowing simultaneous
observation of nearly 40,000 stars in typical high Galactic latitude fields. 
The medium-format image size is markedly larger than the
CCD detector (which measures $37 \times 37$mm) which greatly reduces
the severity of vignetting across the large field of view. At the same
time, the small aperture permits longer exposures, which improve observing
efficiency (assuming fixed camera read-out time). 
A Kodak Wratten \#8
red-pass filter is mounted in front of the lens to further reduce the
impact of atmospheric reddening (which primarily affects blue
wavelengths) on our photometry.  The resultant bandpass resembles a
widened Johnson-Cousins $R$-band.  This optical system is
mounted atop a Paramount ME robotic mount from Software Bisque on a
fixed pier at Winer Observatory in Sonoita, AZ (Latitude 31$^\circ$
39\arcmin 56.08\arcsec N, Longitude 110$^\circ$ 36\arcmin 06.42\arcsec W, elevation 1515.7
meters).  See \citet{KELT_SYNOPTIC} for additional details about the
system hardware.

The primary KELT-North transit survey consists of 13 fields centered at 31.7
declination, spanning all 24 hours of right
ascension.  Including the slight overlap between fields, the total
survey area is $\approx 40\%$ of the Northern sky.  Survey observations consist of 150-second exposures with a typical
per-field cadence of 15-30 minutes. The KELT-North telescope has
been collecting survey data in this manner since September 2006, and
to date has acquired between 5000 and 9300 images per field.  Given
this quantity of data and the typical achieved photometric precision
of $\sim 1\%$ for $V\la 11$, the KELT-North survey is able to detect
short-period giant transiting planets orbiting most FGK stars with
magnitudes from saturation near $V\sim 8$ down to $V\sim 12$.

\subsection{KELT-North Pipeline}\label{sec:pipeline}

Relative photometry is generated from
flat-fielded images using the ISIS image subtraction package
(\citealt{ISIS,alard2000}, see also \citealt{hartman2004}), in combination with point-spread function fitting 
using the stand-alone DAOPHOT II \citep{stetson1987,stetson1990}.  Although highly effective, the image
subtraction procedures are highly computer-intensive. To improve
reduction performance, the default ISIS scripts were modified to
facilitate distributed image reduction across many computers in
parallel.  ISIS operation in this fashion permits thorough exploration
of various reduction parameters, which would be intractable if
executed serially.  Other elements of the ISIS reduction package have
also been modified or replaced with more robust alternatives.  For
example, the standard ISIS source-identification routines and
utilities are ill-equipped to deal with the nature and ubiquity of
aberrations in KELT-North images.  In response, we have replaced the ISIS
`extract' utility with the popular SExtractor program
\citep{SExtractor}.  A more complete explanation of these modifications and driver
scripts
that implement them are available online\footnote{http://astro.phy.vanderbilt.edu/$\sim$siverdrj/soft/is3/index.html}.

\subsection{KELT-North Candidate Selection}

Once we have the light curves created by {\sc ISIS} for all of the {\sc
DAOPHOT}-identified point sources in the reference image, we begin a
series of post-processing steps before doing the initial candidate
selection. To begin, we convert the {\sc ISIS} light curves from
differential flux to instrumental magnitude using the results of the
{\sc DAOPHOT} photometry on the reference image. We also apply
5$\sigma$ iterative clipping to all of the light curves at this stage;
this typically removes $\sim 0.6\%$ of the data points.  All of the
uncertainties for the converted and clipped light curves in a
given field are then scaled as an
ensemble.  The scaling is chosen such that the $\chi^2/{\rm dof}=1$
for the main locus of the light curves on a magnitude versus
$\chi^2/{\rm dof}$ plot. Typically this scaling is around a factor of
1.2, implying that the uncertainties are somewhat
underestimated.

We next attempt to match all of the {\sc DAOPHOT}-identified point
sources in the reference image to stars in the Tycho-2 catalog.
We obtain a full-frame WCS with sub-pixel accuracy on our reference
frame using Astrometry.net \citep{lang2010}. Using this solution,
we match stars by
taking the closest Tycho-2 entry within $45$\arcsec. This typically generates
matches for 98\% of the Tycho-2 stars within each field. A successful
Tycho-2 match also will provide a 2MASS ID.  We use
the proper motions and $JHK$ apparent magnitudes from these two catalogs. 

With this catalog information, we next identify and exclude giant
stars by means of a reduced proper motion ($H_J$) diagram \citep{gould2003}. Following
the specific prescription of \citet{cameron2007b}, we place each of our matched stars on a $J$
vs. $H_J$ plot. We compute the reduced proper motion of a
star as
\begin{equation}
H_{J} = J + 5\log(\mu/{\mathrm{mas~yr^{-1}}})
\label{eqn:rpm}
\end{equation}
and determine the star to be a giant if it fails to satisfy
\begin{eqnarray}\nonumber
H_J &>& -141.25(J-H)^4 + 473.18(J-H)^3\\
&& - 583.6(J-H)^2 + 313.42(J-H) - 43.0
\label{eqn:rpmcut}
\end{eqnarray}
This process leaves us with anywhere from 10,000 to 30,000 catalog-matched putative dwarf stars
and subgiants (hereafter dwarfs) per field, depending primarily on the location of the field relative to the Galactic plane.

The dwarfs are then run through the Trend Filtering Algorithm
\citep[TFA,][]{kovacs2005}\footnote{We used the versions of TFA and BLS (described
later) found in the {\sc vartools} package
\citep{hartman2008}.} to reduce systematic noise. We select a new set
of detrending stars for each light curve by taking the 150 closest
stars -- outside of a 20 pixel exclusion zone centered on the star
being detrended -- that are within two instrumental magnitudes of the
star being detrended.

KELT's Paramount ME is a German Equatorial mount, which requires a "flip"
as it tracks stars past the meridian. Therefore, the optics and detector 
are rotated 180 degrees with respect to the stars between observations in
the Eastern and Western hemispheres, and detector defects, optical
distortions, PSF shape, flat fielding errors, etc., for a given star can be
completely different. This requires us to treat observations in the East   
and West essentially as two separate instruments.  Thus the preceding steps (magnitude
conversion, error scaling, dwarf identification, TFA) are each
performed separately on the East and West images of each field. After
the dwarf stars in the East and West have been run through TFA, we
then combine the two light curves of each target into one East+West
light curve.  We first match stars from the East and the West pointings
by their Tycho IDs, and then determine the error-weighted scaling
factor of the Western light curve needed to match the error-weighted
mean instrumental magnitude of the East light curve.  

All of the light curves from the matched Tycho dwarf stars in a field
are given an internal ID. 
We next search the combined East+West light curves of the dwarfs for
transit-like signals using the box-fitting least squares algorithm
(BLS; \citealt{kovacs2002}). We use a custom version of BLS modified to
skip over integer and half integer multiples of the sidereal day to
reduce the effect of spurious signals due to diurnal systematics and
their aliases on the BLS statistics.  We perform selection cuts along
six of the statistics that are output by the {\sc vartools}
implementation of the BLS algorithm: signal detection efficiency SDE,
signal to pink noise SPN\footnote{See \citet{kovacs2002}
and \citet{hartman2009}, respectively, for the definitions
of SDE and SPN}, the fraction of transit points from one
night $f_{1n}$, depth $\delta$, the ratio of $\Delta\chi^2$ for the
best transit model to best inverse transit model
$\Delta\chi^2/\Delta\chi^2_{-}$ \citep{burke2006}, and the fraction of
the orbit spent in transit or duty cycle $q$. 
In order to determine the appropriate threshold values
for these statistics, we injected realistic transit signals with a
range of properties into a large sample of light curves, and then
attempted to recover these using the BLS algorithm.  We then
determined the values of these statistics that roughly maximize the
overall detection efficiency while minimizing the number of spurious
detections.  The final adopted values are given in Table
\ref{tab:criteria}.

In addition to the cuts we make on the BLS statistics, we also impose
restrictions on the effective temperature and inferred density of the candidate
host stars. For the temperature, we require that
$\teff<7500$K. We calculate the stellar effective
temperature of each candidate from its 2MASS $J-K$ colors. We used the
Yonsei-Yale isochrones \citep{demarque2004} at 5 Gyr with solar
metallicity and no alpha enhancement to create a simple polynomial fit
for $\teff$ as a function of $J-K$:
\begin{eqnarray}\nonumber
\log\teff&=&3.94808-0.7353(J-K)\\
&&+1.0116(J-K)^2-0.8334(J-K)^3.
\label{eqn:teffrho}
\end{eqnarray}
As we have conducted our follow-up spectroscopy, we have found that this relation generally predicts $\teff$ to within $\sim 100$K
for $\teff\la 7000$K and to within $\sim 300$K for stars with $\teff=7000-7500$K.

\begin{deluxetable}{ll|ll}
\tabletypesize{\small}
\tablecaption{\sc KELT-North BLS Candidate Selection Criteria}
\tablewidth{0pt}
\startdata
\hline
\\
Signal Detection Efficiency & SDE$>$7.0    & Depth            & $\delta <0.05$ \\ 
Signal to Pink-noise        & SPN$>$7.0    & $\chi^2$ ratio   & $\frac{\Delta\chi^2}{\Delta\chi_{-}^2} > 1.5$ \\
Fraction from one night     & $f_{1n}<0.8$ & Duty cycle       & $q< 0.1$\\
\enddata
%\tablecomments{Table comment here.}
\label{tab:criteria}
\end{deluxetable}  

We also require that the stellar density, $\rho_*$, as inferred from the BLS transit
fit to the KELT-North light curve, to be within 1.0 dex of the stellar density
calculated for each star using its $J-K$ colors, assuming the star is on
the main sequence. A large disparity in the observed versus the
calculated density is indicative of a blend or of a giant that made it
through the reduced proper motion cuts \citep{seager2003}. Again using the Yonsei-Yale
isochrones at 5 Gyr with solar metallicity and no alpha enhancement,
we made a fit for density as a function of $J-K$:
\begin{eqnarray}\nonumber
\log (\rho_{*,\rm{calc}}/\rho_\odot)&=&-1.00972+2.82824(J-K)\\
&&-1.19772(J-K)^2.
\label{eqn:rhocalc}
\end{eqnarray}
We require that this value be within 1.0 dex of the stellar density we calculate from the KELT-North lightcurve
\begin{equation}
\log \rho_{*,\rm obs}=\log\left[\frac{3}{G\pi^2q^3P^2}\right],
\label{eqn:rhoobs}
\end{equation}
where $P$ and $q$ are the orbital period and duty cycle (transit
duration relative to the period) as returned by BLS.  This equation
assumes circular orbits and that the companion mass is much smaller
than the host star mass, $M_P \ll M_*$.  Also, because BLS does not
attempt to fit for the ingress/egress duration, and furthermore KELT-North
data typically do not resolve the ingress or egress well, we are not
able to determine the transit impact parameter and thus the true
stellar radius crossing time.  Equation \ref{eqn:rhoobs} therefore
implicitly assumes an equatorial transit, and so formally
provides only an upper limit to the true stellar density.  For a
transit with an impact parameter of $b=0.7$, the true density is 
$\sim 0.5$ dex smaller than that inferred from Equation \ref{eqn:rhoobs}.

All of the light curves that pass these selection criteria are
designated as candidates, and a number of additional diagnostic tests
are then performed on them, including Lomb-Scargle (LS, \citealt{lomb76,scargle82})
and AoV \citep{AoV,devor2005} periodograms.  The results of these tests, the
values of the BLS statistics, the light curves themselves, as well as
a host of additional information, are all collected into a webpage for
each candidate.  Members of the team can then use this information to
vote on the true nature of the candidate (probable planet, eclipsing
binary, sinusoidal variable, spurious detection, blend or other).  All
candidates with at least one vote for being a probable planet are then
discussed, and the most promising are then passed along for follow-up
photometry, reconnaissance spectroscopy, or both.

\section{Observations}\label{sec:observations}

\subsection{KELT-North Photometry, Candidate Identification, and Vetting Overview}

\begin{figure}[t]
\epsscale{1.1}
\plotone{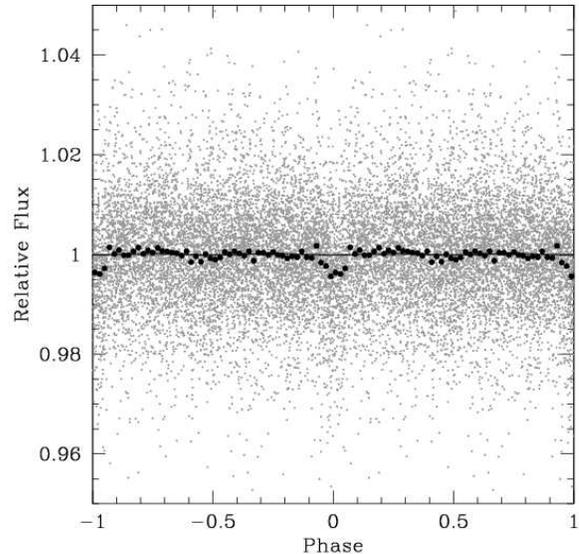}
\caption{
\label{fig:keltlc}
The KELT-North light curve of KELT-1 phased to the BLS determined period of $P=1.2175$ days
is shown in the grey points.  The black points show the data binned 0.02 in phase.  
}
\end{figure}

KC20C05168 emerged as a strong candidate from the analysis of
the combined light curves from stars
in the overlap area between fields 1 and 13.  The KC20C05168 light
curve contains 8185 epochs distributed over $\sim 4.2$ years, between
UT October 25, 2006  and UT December 28, 2010, with a weighted RMS of $9.8$ millimagnitudes
(mmag).  This RMS is typical for KELT-North light curves of stars with this
magnitude ($V \sim 10.7$).  A strong BLS signal was found at a period
of $P \simeq 1.2175$ days, with a depth of $\delta \simeq 3.8$ mmag,
and detection statistics SPN=8.53, SDE=12.41, $q=0.09$,
$\Delta\chi^2/\Delta\chi^2_-=2.06$, and
$\log(\rho_{*,obs}/\rho_{*,cal})=-0.06$.  The phased
KELT-North light curve is shown in Figure \ref{fig:keltlc}.  A significant signal also
appeared in SuperWASP data \citep{butters2010} of this star at the
same period.  The KELT-North data exhibit some evidence for out-of-transit
variability at the mmag level and exhibit some relatively weak peaks
in the LS and AoV periodograms, but we did not consider these signals
to be strong enough to warrant rejection of the candidate. 
In addition, the depth of the photometric transit
signal in the original KELT-North light curve is substantially smaller than we find in
the high-precision follow-up data (see \S\ref{sec:phot}).   Further analysis 
indicates that the out-of-transit variability and smaller
depth were likely due to a minor problem with the original data

Based on the strength of the K20C05168 signal, the estimated effective
temperature of the host star of $\teff \sim 6500$K, and the fact that
the star was sufficiently isolated in a DSS image, we submitted the
candidate for reconnaissance spectroscopy with the Tillinghast
Reflector Echelle Spectrograph (TRES; \citealt{furesz2008}) on the
1.5m Tillinghast Reflector at the Fred Lawrence Whipple Observatory (FLWO)
on Mount Hopkins in Arizona.  The first observation on UT November 9,
2011 at the predicted quadrature confirmed the $\teff$ estimate of the
star, and also demonstrated that it was a slightly evolved dwarf with
$\logg \sim 4$, and that it was rapidly rotating with $\vsini \sim
55~\kms$.  A second observation was obtained on UT November 11, 2011
separated by $\sim 1.9$ days, or $\sim 1.54$ in phase, from the first
observation and thus sampled the opposite quadrature.  The two
observations exhibited a large and significant radial velocity shift
of $\sim 8\kms$, consistent with a brown dwarf companion.  

Efforts to
obtain photometric follow-up during the primary transit and secondary
eclipse were then initiated.
Concurrently, additional spectra with TRES were taken to characterize
the spectroscopic orbit.  In addition, we obtained adaptive optics
imaging of the target to search for close companions.  Finally, once
we were fairly confident that the signals were due to a low-mass
transiting companion, we obtained continuous spectroscopic time series
with TRES during the primary transits on UT December 21, 2011 and UT
January 7, 2012 for the purposes of measuring the Rossiter-McLaughlin
(RM) effect.  All of these observations are described in greater detail
in the subsequent sections and summarized in Table \ref{tab:obs}.

\subsection{Previous Identification of the Photometric Candidate by HATNet}\label{sec:hat}.

KELT-1b was also recognized as a photometric transiting planet
candidate by the HATNet project, based on observations obtained in
2006.  In September 2009 the candidate HTR162-002 was forwarded to
D.\ Latham's team for spectroscopic follow up.  An initial observation
with TRES confirmed that the target was a late F main-sequence star,
as expected from 2MASS color $J-K_s$=0.245. The synthetic template with
$\teff = 6250$K and $\logg=4.0$ and assumed solar metallicity, gave the best
match to the observed spectrum.  However, that first TRES spectrum
also revealed that the star was rotating rapidly, with $\vsini=55~\kms$.
At that time, D.\ Latham's team routinely put aside candidates rotating
more rapidly than about $\vsini=30~\kms$, arguing that it would not be
possible to determine velocities with a precision sufficient for an
orbital solution for a planetary companion.

HTR162-002 remained on the HATNet "don't observe with TRES" list until it was
independently rediscovered by the KELT-North team and was forwarded as
candidate KC20C05168 to D.\ Latham's team in November 2011 for
spectroscopic follow up with TRES.  During the intervening 26 months,
there were two relevant developments in the procedures and tools used
by Latham's team, both resulting from contributions by L.\ Buchhave.  
The first
development, enabled by convenient tools in the observing website, was
the practice of observing new candidates only near opposite
quadratures, according to the discovery ephemeris and assuming
circular orbits.  The second development was a much improved procedure
for deriving radial velocities for rapidly-rotating stars, initially
motivated by the Kepler discovery of hot white dwarfs transiting
rapidly-rotating A stars \citep{rowe2010}.  As it turned out, the
second observation of KC20C05168 with TRES described above was taken
before the first observation was reviewed, so the candidate was not
relegated to the rejected list due to its rapid rotation before the
opposite quadrature was observed.  When the results were reviewed
after the second observation, the evidence for a significant radial
velocity shift between the two observations was obvious, despite the
rapid rotation, therefore suggesting that the unseen companion was
probably a brown dwarf, if not a giant planet.

It should also be recognized that over the 26 months since the first
observation of HTR162-002, the attitude against pursuing rapidly
rotating stars as possible hosts for transiting planets had gradually
softened among the exoplanet community.  An important
development was the demonstration that slowly-rotating subgiants that
have evolved from rapidly-rotating main-sequence A stars do
occasionally show the radial-velocity signatures of orbiting planetary
companion (e.g., \citealt{johnson2007}).  A second insight came from
the demonstration that the companion that transits the
rapidly-rotating A star WASP-33 must be a planet, using Doppler
imaging \citep{cameron2010}.  Finally, the discovery of transiting
brown dwarf companions suggested the possibility of detecting their
large amplitude RV signals even when they orbit stars with
large $\vsini$ and thus poor RV precision.

In the early days of wide-angle photometric surveys for transiting
planets, Latham's team had established procedures for handling
candidates forwarded for spectroscopic follow up by more than one
team.  Such duplications were fairly common, and the goal was to
assign credit to the initial discovery team, which was especially
important in an era when few transiting planets had been confirmed.
By the time it was noticed in mid December 2011 that KC20C05168 was
the same as HTR162-002, the KELT-North team already had in hand a convincing
orbital solution from TRES and high-quality light curves from
several sources, confirming that KELT-1b was indeed a
substellar companion.  

\subsection{Spectroscopy from FLWO/TRES}\label{sec:flwos}

A total of 81 spectra of KELT-1 were taken using the TRES spectrograph on the
1.5m Tillinghast Reflector at FLWO.  These were
used to determine
the Keplerian parameters of the spectroscopic orbit, measure bisector variations in order to
exclude false positive scenarios, measure the spectroscopic
parameters of the primary, and measure anomalous RV shift of the
stellar spectral lines as the companion transits in front of the
rapidly-rotating host star, i.e., the RM effect
\citep{rossiter1924,mclaughlin1924}.  The TRES
spectrograph provides high resolution, fiber-fed echelle spectroscopy
over a bandpass of $3900-8900\AA$ \citep{furesz2008}. The observations
obtained here employed the medium fiber for a resolution of $R\sim
44,000$.  The data were reduced and analyzed using the methods
described in \citet{quinn2012} and \citet{buchave2010}.  

A subset of six spectra were combined in order to determine the
spectroscopic parameters of the host star using the Spectral Parameter Classification (SPC)
fitting program (Buchhave et al., in preparation).  SPC cross-correlates
the observed spectrum against a grid of synthetic Kurucz \citep{kurucz1979} spectra.  This analysis yielded
$\teff=6512 \pm 50$K, $\logg = 4.20 \pm 0.10$, [Fe/H]$=0.06 \pm 0.08$,
and $\vsini=55.2 \pm 2\kms$.  These parameters were used as priors
for the joint global fit to the RV, RM, and photometric data as
described in \S\ref{sec:analysis}.  

Spectra were taken at a range of phases in order to characterize the
spectroscopic orbit and search for bisector span variations indicative of a
blend.  One of these spectra happened to be taken during a primary
transit on UT 2011-11-18, and so was not used in the
analysis because it is likely to be affected by the RM effect.  
The RV and bisector data for the remaining 23 spectra are
listed in Table \ref{tab:rvorbit}.  These observations span $\sim 88$
days from UT 2011-11-09 through UT 2012-02-05.  The uncertainties on
the listed radial velocities have been scaled by a factor of 1.214 based on an
independent fit to these data, as described in \S\ref{sec:analysis}.
The scaled median RV uncertainty is $\sim 230~\ms$.  The uncertainties
in the bisector measurements have not been scaled.  The median bisector
uncertainty is $\sim 110\ms$.

\begin{figure}[t]
\epsscale{1.2}
\hskip-0.5in
\plotone{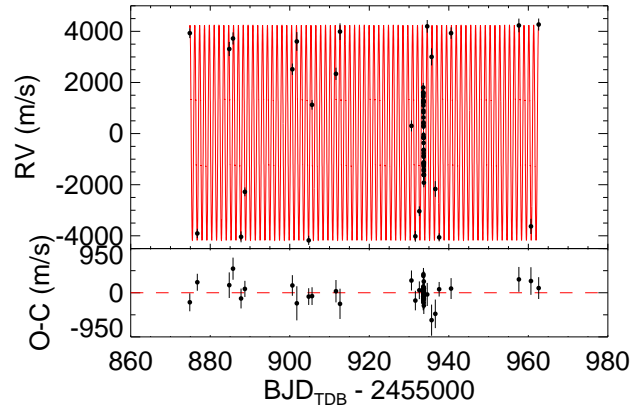}
\caption{
\label{fig:unphased}
(Top panel) The black points with uncertainties show the measured RVs for
KELT-1 as a function of time in $\bjdtdb$.  The barycentric velocity of the 
system, as determined from the model fit shown in red (see \S\ref{sec:analysis}), has been been subtracted from the data.
(Bottom panel) The residuals from the model fit.  
}
\end{figure}

Time series spectroscopy was obtained with TRES on two different
nights of primary transits in order to measure the spin-orbit alignment of
the companion via the RM effect.  Fifteen observations were
obtained on UT 2011-12-21 and forty-two observations on UT 2012-01-07.  
Conditions were relatively 
poor for the first run, resulting in a factor $\sim 2$ larger uncertainties and 
incomplete coverage of the transit.  We therefore decided not to include
these data in our final analysis, although we confirmed that this has no
effect on our final inferred parameters.  The RV and bisector data for the RM run
on UT 2012-01-07 are listed in Table \ref{tab:rvrm}.  The RV uncertainties
have been scaled by a factor of 0.731, also based on the global model fit 
described in \S\ref{sec:analysis}.  We note that the majority of the
$\chi^2$ for these data are due to a few outliers. The median scaled RV uncertainty
is $\sim 160~\ms$.  The bisector uncertainties were not scaled.

\begin{figure}[t]
\epsscale{1.2}
\hskip-0.5in
\plotone{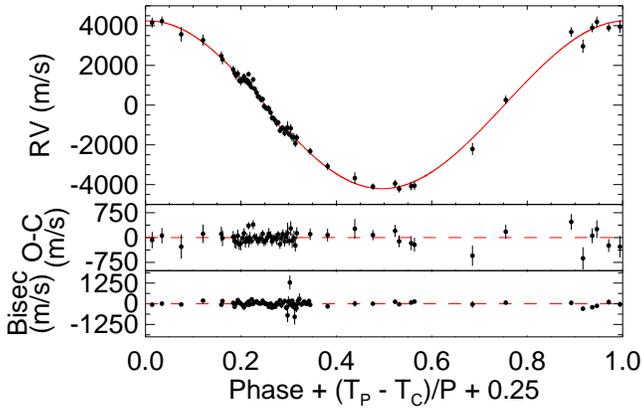}
\caption{
\label{fig:phased}
The black points with uncertainties show the measured RVs for KELT-1 relative to
the barycentric velocity of the system, phased to the
best-fit period as determined from the model fit shown in red (see \S\ref{sec:analysis}).  
The phases are normally
referenced to the time of periastron ($T_P$), but have been shifted
such that a phase of 0.25 corresponds to the time of inferior
conjunction $T_C$ or primary transit.  RV data near this phase show
deviations from the Keplerian expectation due to the RM effect, which
was included in the model.  (Middle panel) The residuals of the RV data from the
model fit.  (Bottom panel) Bisector spans as a function of phase.
}
\end{figure}

\begin{figure}[t]
\epsscale{1.2}
\hskip-0.5in
\plotone{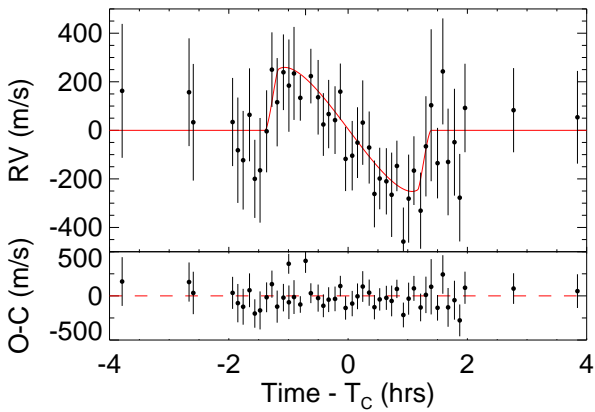}
\caption{
(Top panel) The black points with uncertainties show the measured RVs relative to
the barycentric velocity of the system for KELT-1, as a function
of the time since primary transit $T_C$, for data taken
near $T_C$.  The Keplerian RV variation as determined from the best-fit model has
been removed from both the data and model.
Data taken within $\sim 1.4$ hours of $T_C$
occur during primary transit, and are thus strongly affected by the RM
effect.  The shape of the RM signal indicates that the projected obliquity 
of the host star with respect to the orbit is small.
(Bottom panel) The residuals of the data to the RM fit.
\label{fig:RM}
}
\end{figure}

All of the RV and bisector measurements used in the subsequent
analysis are shown as a function of epoch of observation in \bjdtdb\ in
Figure \ref{fig:phased}.  The measurements phased to the best-fit
companion period from the joint fit to photometric and RV data are
shown in Figure \ref{fig:phased}, demonstrating
the very high signal-to-noise ratio with which the RV signal of the companion was detected, and
the good phase coverage of the orbit.  A detail of the RV data near the
time primary transit with the orbital
(Doppler) RV signal removed is shown in Figure \ref{fig:RM}, showing the clear
detection of the RM effect and a suggestion that the orbit normal is well
aligned with the projected stellar spin axis.

Finally, we determined the absolute radial velocity of the system
barycenter using a simple circular orbit fit to radial velocities
determined from the full set of spectra, which were determined using a
single order near the Mg b line.  (Note that the relative RVs 
used for determining the orbit were
determined using the full, multi-order analysis of the spectra.)  The zero point correction to these
velocities were determined using contemporaneous monitoring of five RV
standard stars.  The final value we obtain is $\gamma_{\rm obs}= -14.2
\pm 0.2~\kms$, where the uncertainty is dominated by the systematic
uncertainties in the absolute velocities of the standard stars.  This
zero point, along with the global fit to the data in
\S\ref{sec:analysis}, were used to place the instrumental relative
radial velocities on an absolute scale.  Therefore, the RVs 
listed in Tables \ref{tab:rvorbit} and \ref{tab:rvrm} are
on an absolute scale.

\begin{figure}[t]
\epsscale{1.1}
%\vskip4.3in
%\hskip-0.5in
\plotone{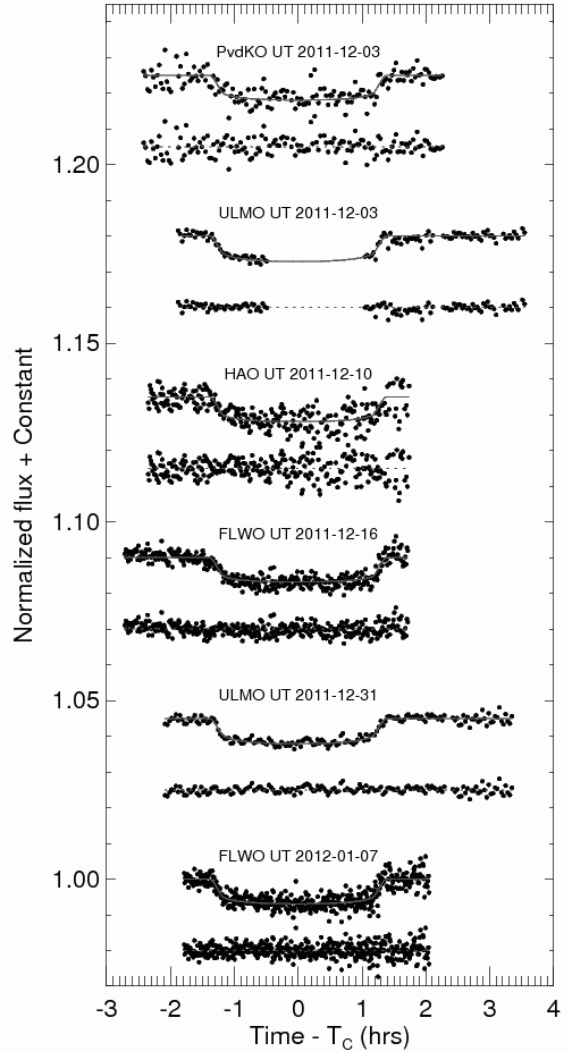}
\caption{
The black points show the relative flux as a function of time from
primary transit ($T_C$) for the six sets of follow-up observations of
primary transits we analyze here. The data sets are labeled and
summarized in Table \ref{tab:obs}.  The data are normalized by the fitted
out-of-transit flux, and a linear trend with airmass has been removed
(see \S\ref{sec:analysis}).  In addition, an arbitrary offset has been
applied to each light curve for clarity. For each observation, we plot
the data above and the residuals below.  In all cases, the red lines
show the model fit from the analysis in \S\ref{sec:analysis}.
\label{fig:transits}
}
\end{figure}

\subsection{Follow-Up Photometry}\label{sec:phot}

\begin{figure}[t]
\epsscale{1.0}
\plotone{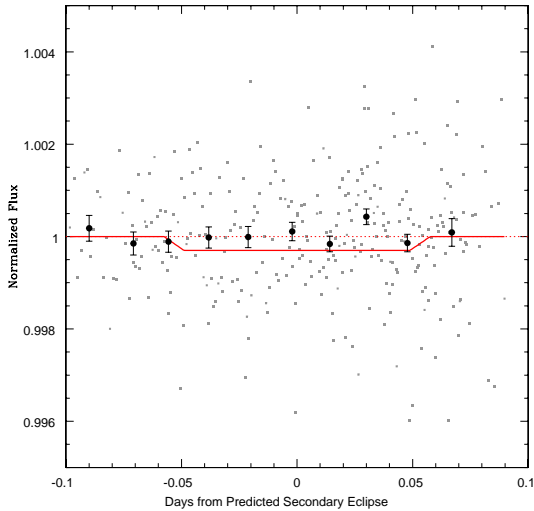}
\caption{The grey points show the combined $i$ and Pan-STARRS-$Z$ relative
photometry of  KELT-1 as a function
of the predicted time of secondary eclipse of KELT-1b ($T_S$), obtained from
the observatories listed in Table \ref{tab:obs}.  The data have been
corrected for a linear trend with airmass and normalized by the zero
point of the linear fit (see \S\ref{sec:second}).  The larger circles
with error bars show the binned observations. 
Note we do not fit to the binned data; these are shown for the
purposes of illustration only.  The over plotted example light curve
is the secondary eclipse depth we would expect if KELT-1b had a
geometric albedo of $A_g= 0.1$ and instantaneously reradiated its
incident stellar flux ($f'=2/3$). We would have detected this event with a confidence
level of $\ga 95\%$.}
\label{fig:secondary}
\end{figure}

We obtained high-precision follow-up photometry of KELT-1 in order to
confirm the K20C05168 transit signal, search for evidence of a
strongly wavelength-dependent transit depth indicative of a stellar
blend, and search for evidence of a secondary eclipse.  Also, 
these data enable
precision measurements of the transit depth, ingress/egress duration,
and total duration, in order to determine the detailed parameters of the
KELT-1 system.  In all, we obtained coverage of 9 complete and 4 partial primary
transits, and two complete and one partial secondary eclipse,
using six different telescopes in all.  Many of these data were taken under
relatively poor conditions and/or suffer from strong systematics.  We
therefore chose to include only a subset for the final analysis,
including six of the primary transits and the three secondary
eclipses. In the following subsections, we detail the observatories
and data reduction methods used to obtain these data.  The dates,
observatories, and filters for these data sets are summarized in Table
\ref{tab:obs}.  The light curves for the primary transits are
displayed in Figure \ref{fig:transits} and the data are listed in
Tables \ref{tab:transit0} through \ref{tab:transit4}, whereas the
light curves for the secondary eclipse are displayed in Figure
\ref{fig:secondary} and the data are listed in Tables
\ref{tab:second0} through \ref{tab:second2}.

\begin{figure}[t]
\epsscale{1.3}
\hskip-0.5in
\plotone{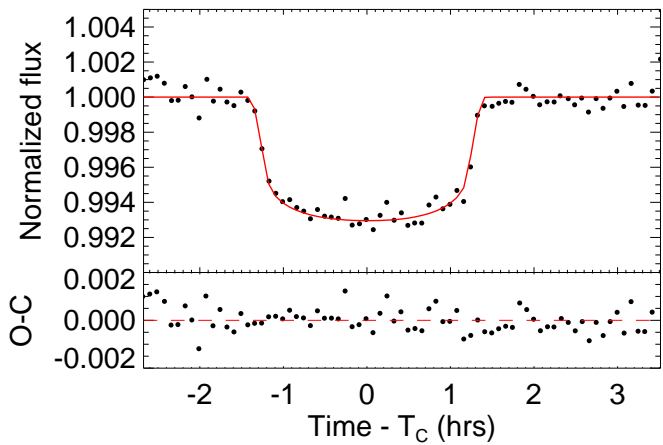}
\caption{
(Top panel) The black points show the six data sets displayed in Figure \ref{fig:transits}, combined and 
binned in 5 minute intervals.  Since these data sets were taken with different filters and have different systematics,
we do {\it not} use this combined light curve for analysis, but rather show it for the purposes of illustrating
the overall quality and statistical constraining power of the ensemble of follow-up light curves.  
The red curve shows the six transit models for each of the individual fits combined and binned in 5 minute 
intervals the same way as the data.  (Bottom panel) 
The residuals of the binned light curve from the binned model in the top panel.
\label{fig:bintrans}
}
\end{figure}

\subsubsection{Peter van de Kamp Observatory (PvdKO)}\label{sec:pvdko}

Data on the primary transit starting on UT 2011-12-03 were
acquired with the 0.6-meter, f/7.8 Ritchey-Chr\'{e}tien telescope at the
Peter van de Kamp Observatory at Swarthmore College (Swarthmore, PA,
USA).  The telescope is equipped with an Apogee U16M CCD with
4096x4096 9-micron pixels, giving a field of view 26 arcminutes on a
side.  Available filters are 50mm-square Johnson-Cousins $UBVR_cI_c$
and SDSS $ugriz$, both from Omega Optical.  The telescope is
autoguided to minimize photometric errors from imperfect
flatfielding, keeping the centroid shift of each star to within typically
3-4 pixels over the course of a night. The observations used here were
obtained with the $i$ filter and used 2x2 binning, giving a binned
pixel scale of 0.76\arcsec / pixel.

The data were reduced in IRAF using standard procedures for flat-fielding (with
twilight sky flats) and dark and bias subtraction.  Aperture photometry was
performed, and then differential magnitudes for the target star
were calculated using an ensemble of comparison stars in the same field,
chosen to minimize the scatter in the final light curve. 

\subsubsection{University of Louisville Moore Observatory (ULMO)}\label{sec:ulmo}

Data on the primary transits starting UT 2011-12-03 and 2011-12-31, and on the secondary
eclipses starting 2011-12-02 and 2012-01-04, were obtained
with the University of Louisville Moore Observatory RC24 telescope
(MORC24) located near Brownsboro, Kentucky. MORC24 is a RC Optical
Systems Ritchey-Chr\'{e}tien 0.6 m telescope on an equatorial fork
mount. The telescope is equipped with an Apogee U16M 4096$\times$4096
pixel CCD camera which has a focal plane scale of 0$\farcs$39
pixel$^{-1}$ and a field of view (FOV) of
26$\farcm$3$\times$26$\farcm$3. The UT 2011-12-03 and 2011-12-31 data were obtained
using an Astrodon Photometrics Sloan r filter, while the other two
sets of data were obtained using an Astrodon Photometrics Sloan i
filter. The MORC24 mount has excellent free-running tracking, so we did
not use a separate guide camera. Instead, minor telescope pointing
corrections are made after each exposure by comparing the CCD pixel
coordinates of the centroid of the target star to its initial position
on the CCD. KELT-1b was held to within 3-4 pixels of the starting
position on the CCD throughout each observing session. Since KELT-1b
is separated from its nearest detectable neighbor in DSS2 imagery by
$\sim18\arcsec$, we were able to defocus the telescope to allow for
longer exposures without the risk of blending from the neighbor
star. An exposure time of 100~s was used for all observations,
resulting in a 120~s cadence when combined with the 20~s CCD readout
time.

We used AstroImageJ (Collins \& Kielkopf 2012, in preparation) to
calibrate the image data. The algorithm includes bias subtraction,
CCD non-linearity correction, dark subtraction, and flat-field
division. AstroImageJ was also used to perform aperture photometry
using a circular aperture. An aperture size and an ensemble of
comparison stars in the same field were chosen to minimize the scatter
in the final light curves. AstroImageJ provides the option to use a
standard fixed radius aperture or a variable radius aperture
based on the measured FWHM of the target star in each image of the
series. When a star is well separated from other stars, the variable
aperture option tends to reduce photometric scatter under observing
conditions that result in significant changes to the PSF during the
observing session. The variable aperture produced optimal results for
all four MORC24 KELT-1b light curves.

For the observations starting on UT 2011-12-02, cirrus clouds were
present during the first half of the observations, and airmass ranged
from 1.16 at the start of observations to 3.19 at the end. For the
observations starting on UT 2011-12-04, skies were clear until clouds
moved in about 30 minutes after ingress. The clouds cleared just
prior to egress, however, sky transparency remained highly variable
until about an hour after egress. Airmass ranged from 1.05 at the
beginning of observations to 1.40 at the end. Although guiding was
maintained through the cloud cover, data during that time have been
removed. For the observations starting on UT 2011-12-31, skies were
clear with minimal variations in transparency. Airmass ranged from
1.00 at the beginning of observations to 2.17 at the end. For the
observations on UT 2012-01-04, cirrus clouds were present during the
second half of the observations, and airmass ranged from 1.03 at the
start of observations to 1.96 at the end.

\subsubsection{Hereford Arizona Observatory (HAO)}\label{sec:hao}

Data on the primary transit starting UT 2011-12-10 were obtained at
the Hereford Arizona Observatory, HAO (observatory code G95 in the IAU
Minor Planet Center). This is a private observatory in Southern
Arizona consisting of a 14-inch Meade LX-200 GPS telescope equipped
with a SBIG ST-10XME CCD, a focal reducer and a 10-position filter
wheel with SDSS filters $ugriz$. The telescope and dome are
operated via buried cables, permitting automation of observing
sessions. Calibrations usually employ a master flat frame obtained
during dusk prior to the observing session. The field-of-view (27 x 18
arcminutes) is sufficient for the use of approximately two dozen stars
as candidates for reference in a typical field. The observations
reported here were obtained with the $i$ filter.

The data were reduced and a light curve was generated as follows.  An
artificial star was inserted in each image before photometry readings
for the purpose of monitoring smooth extinction as well as extra
extinction events caused by thin clouds, dew formation, and
atmospheric seeing degradations that could swell the PSF beyond the
photometry aperture circle.  Photometry magnitude readings were made
by the program MaxIm DL and imported to a spreadsheet, where several
steps of manual reduction were completed. The first was to solve for
an extinction model (including a temporal extinction trend) based on
the sum of all candidate reference star fluxes versus air mass.
Second, subsets of reference stars were evaluated for suitability, by
toggling individual stars "on and off" in order to determine the
subset that minimize the RMS scatter in the target star light curve.

Finally, the light curve for the target was fitted using a model for
systematic effects and a transit signature.  Systematics were
represented by a temporal trend and air mass curvature (AMC). The AMC
is caused by the target star having a color that differs from the
flux-weighted color of the reference stars. The transit parameters were
depth, total duration, ingress/egress duration, and a parameter related to the
stellar limb darkening.  The solution was obtained by minimizing the
$\chi^2$ of the fit.  Outliers were identified using an objective
rejection criterion based on deviations from the model solution.  Finally,
the light curve is corrected for extinction
and systematic effects and scaled to the out-of-transit model flux.

\subsubsection{FLWO/KeplerCam}\label{sec:flwop}

Data on the primary transits on UT 2011-12-16 and 2012-01-07 were
obtained with KeplerCam on the 1.2m telescope at FLWO. KeplerCam
has a single 4K $\times$ 4K Fairchild CCD with a pixel scale of 0.366
arcseconds per pixel, for a total FOV of 23.1 x 23.1 arcminutes. A
full transit was observed on UT 2011-12-16 with clear conditions.
Observations were obtained in the SDSS $z$ filter with 30-second
exposures. We also obtained a full transit on UT 2012-01-07 and
observations were obtained with the SDSS $i$ filter with 15-second
exposures. Clouds came in at the end of the transit and as a result
there is some increased scatter in the out-of-transit baseline. The data were
reduced using a light curve reduction pipeline outlined in 
\citet{carter2011} which uses standard IDL techniques.

\subsubsection{Las Cumbres Observatory Global Telescope Network (LCOGT)}\label{sec:lcogt}

Data on the secondary eclipse on UT 2011-12-30 were obtained
with the 2.0m  Faulkes Telescope North (FTN) telescope, which
is located on Haleakala on the island of Maui in Hawaii. The FTN telescope is 
part of the Las Cumbres Observatory Global Telescope Network \footnote{http://lcogt.net}.  
These observations were made using the 4K $\times$ 4K Spectral
camera (Fairchild Imaging CCD486 BI) in bin 2x2 mode for a faster
readout together with the PanSTARRS-$Z$ filter.  As scintillation noise
becomes significant ($>$1 millimag) in exposures shorter than
$\sim$30\,sec for telescopes of this aperture, the exposure time was
kept to 60 sec and the telescope defocused to avoid saturation of the
target while ensuring sufficient signal-to-noise ratio in the comparison
stars.  These data were debiased, dark-subtracted and flat fielded by
the LCOGT offline pipeline (developed by the Astrophysics Research
Institute at Liverpool John Moores) and aperture photometry was
carried out using the stand-alone DAOPHOT II \citep{stetson1987,stetson1990}.  Differential
photometry was then computed using an ensemble of 15 comparison stars.

\subsection{Keck Adaptive Optics Imaging}\label{sec:keckao}

To further assess the multiplicity of KELT-1, we acquired adaptive
optics images using NIRC2 (PI: Keith Matthews) at Keck on UT 2012-01-07. 
Our observations consist of dithered frames taken with the $K'$
($\lambda_c=2.12 \mum$) and H ($\lambda_c=1.65 \mum$) filters.  We
used the narrow camera setting to provide fine spatial sampling of the
stellar point-spread function. The total on-source integration time
was 81 seconds in each bandpass.

\begin{figure}[t]
\epsscale{1.0}
\plotone{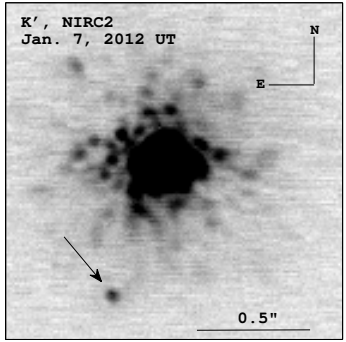}
\caption{
A Keck AO image of KELT-1 taken with NIRC2
on UT 2012-01-07 in the $K'$ filter.  North is up
and East is to the left.  A 0.5\arcsec bar is shown
for scale.  A faint companion with $\Delta K' = 5.59 \pm 0.12$ located
$\sim 558 \pm 1$ mas to the southeast is clearly visible.
\label{fig:aoimage}}
\end{figure}

Images were processed by replacing hot pixel values, flat-fielding,
and subtracting thermal background noise. No companions were
identified in individual raw frames during the observations; however,
upon stacking the images we noticed a point source ($8\sigma$) to the
south-east of KELT-1. Figure \ref{fig:aoimage} shows the final processed $K'$
image. Inspection of the companion location showed that its separation
from the star does not change with wavelength, demonstrating that it
is not a speckle.  This object is too faint and close to the primary
to be detected with seeing-limited images.

We performed aperture photometry to estimate the relative brightness
of the candidate tertiary, finding $\Delta H=5.90 \pm 0.10$ and
$\Delta K'= 5.59 \pm 0.12$. An $H-K'=0.4\pm0.2$ color is consistent
with spectral-types M1-L0 \citep{leggett2002,kraus2007}.  If the
candidate is bound to KELT-1 and thus at the same distance of $262 \pm
14$pc and suffers the same extinction of $A_V=0.18 \pm 0.10$ (see
\S\ref{sec:hostprops}), then we estimate its absolute $H$ magnitude to
be $M_H = 8.31 \pm 0.15$, corresponding to a M4-5 spectral type,
consistent with its color (see, e.g., the compilation of
\citealt{kirkpatrick2012}).   

We also measured an accurate position of the companion relative to the
star by fitting a Gaussian model to each of the point-spread function
cores.  After correcting for distortion in the NIRC2 focal plane
\footnote{http://www2.keck.hawaii.edu/inst/nirc2/forReDoc/post\_observing/dewarp/}, 
and adopting a plate scale value of $9.963 \pm 0.006$ mas
$\mbox{pix}^{-1}$ and instrument orientation relative to the sky of
$0.13^{\circ}\pm0.02^{\circ}$ \citep{ghez2008}, we find a separation
of $\rho=588 \pm 1$ mas and position angle $PA=157.4^{\circ} \pm
0.2^{\circ}$ east of north. If it
is bound to KELT-1, it has a projected physical separation of $\sim
154 \pm 8 ~{\rm AU}$, and a period of $\sim 1700$ years assuming a
circular, face-on orbit.

We used the Galactic model from \citet{dhital2010} to assess the
probability that the companion is an unrelated star (i.e., a chance
alignment). The model uses empirical number density distributions to
simulate the surface density of stars along a given line-of-sight and
thus determine probability of finding a star within a given angular
separation from KELT-1.  We estimate an a priori probability of $\sim
0.05\%$ of finding a star 
%with $K'<XX$ 
separated by $\la 0.59$\arcsec \
from KELT-1b.  We therefore conclude that the companion is likely to
be a bona fide, physically associated binary system.  With a total
proper motion of $\sim 20$~mas/year, it will be possible to
definitively determine whether the candidate tertiary is physically
associated with KELT-1 within one year.

We note that the 
companion is unresolved in
our follow-up primary transit photometry, and thus in principle leads
to a dilution of the transit signal and a bias in the parameters
we infer from a fit to the photometry described in \S\ref{sec:analysis}.
However, the effect is negligible.  As we discuss in the next section, 
we are confident that the primary is being eclipsed.  Thus the fractional effect on the transit depth is of the
same order as the fractional contribution of the companion flux to the total flux, 
which is $<1\%$.

\section{Evidence Against a Blend Scenario}\label{sec:blend}

One of the many challenges of photometric surveys for transiting
planets is the relatively high rate of astrophysical false positives,
blended eclipsing stellar binary or triple systems 
that can mimic some of the observable signatures of transiting
low-mass companions to single stars.  In the case of the KELT-North survey,
one generically expects a higher rate of false positives as compared
to other wide-field transit surveys such as HATNet or SuperWASP, because
of the poorer image quality arising from the comparatively smaller
aperture, larger pixel scale, and wider FOV. For KELT-1 in particular, the extreme properties
of the companion, relatively high $\vsini$ of the primary, and the
fact that the primary is somewhat evolved, are all reminiscent of 
false positives that have been reported in previous surveys, e.g.,
\citet{mandushev2005}.  

In the case of KELT-1b, however, we have a number of lines of evidence that
strongly disfavor a blend scenario. 

\begin{figure}[t]
\epsscale{1.0}
\plotone{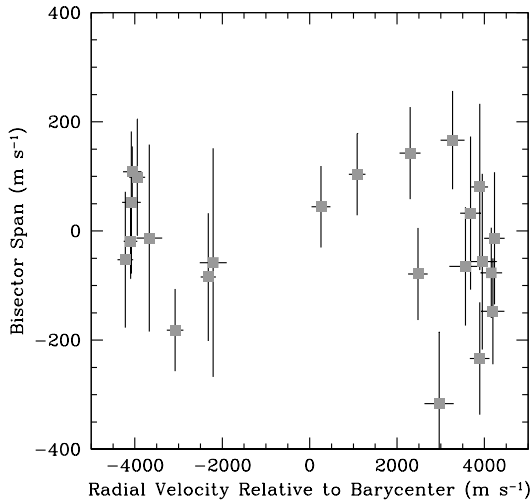}
\caption{
\label{fig:RVvsBS}
Bisector spans versus the RV relative to the system
barycenter, excluding observations taken on the night of the
primary transit on UT 2012-01-07.  There is no evidence of a significant
correlation between the bisector and RV variations, and the RMS of the 
bisector span variations is $\sim 30$ times smaller than the
RMS of the RV measurements.  
}
\end{figure}

\begin{figure}[t]
\epsscale{1.0}
\plotone{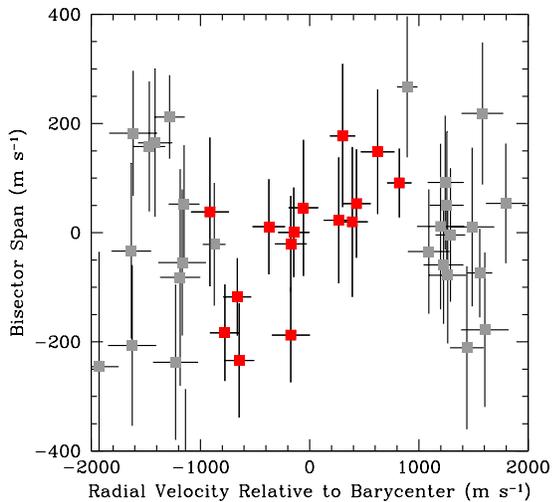}
\caption{
\label{fig:RVvsBSrm}
bisector spans versus the RV relative to the system
barycenter for observations taken on the night of the
primary transit on UT 2012-01-07.  The points in red
are the subset of those data that were taken within 0.03 days
of the center of the primary transit, roughly corresponding to
the middle half of the full transit duration.  
Note that these data are strongly correlated with the RV variations
due to the RM effect. 
}
\end{figure}

The most compelling arguments against blend scenarios arise
from the spectra.  First is the lack of
strong evidence for bisector span variations.  The lower panel of
Figure \ref{fig:phased} shows the bisector variations phased to the
best-fit period of the companion as determined from the joint fit to
the RV and photometry data described in \S\ref{sec:analysis}.  There
is no evidence for bisector variations correlated with the orbital
phase of the companion.  The weighted RMS of the bisector spans,
excluding the data taken on UT 2012-01-07, is $\sim 120~\ms$, only
$\sim 30\%$ larger than would be expected based on the native
uncertainties, and a factor of $\sim 30$ times smaller than the RMS of
the RV measurements themselves.  Figure \ref{fig:RVvsBS} shows the
bisector spans as a function of radial velocity relative to the system
barycenter.  There is no strong correlation; the correlation
coefficient is only -0.17.  In contrast, Figure \ref{fig:RVvsBSrm}
shows data taken on the night of UT 2012-01-07, which covered the
primary transit.  For the subset of these data taken within 0.03 days
of the transit center (approximately the middle half of the transit),
there is a clear correlation between the radial velocity and the
bisector variations, with a correlation coefficient of 0.68.  This is
expected since the anomalous radial velocity shift from the RM effect
is due to a distortion of the rotationally-broadened stellar spectral
lines as the planet progressively occults the light from different
parts of the face of the star.  Indeed, the second piece of evidence that the transit
signatures are indeed due to a small companion occulting the primary star
is the RM signal itself (Fig.\ \ref{fig:RM}), which has an amplitude consistent with the
apparent transit depth and spectroscopically-determined $\vsini$.

Third, photometric observations
in several different filters ($riz$) are all consistent with the
primary transit having nearly the same depth, and are well-modeled by
transits of a dark companion across a star with the limb
darkening consistent with its spectroscopically measured $\teff$ and $\logg$
(see Section \ref{sec:analysis}).

Fourth, photometric
observations at the predicted time of superior conjunction reveal no
evidence for a secondary eclipse at the $\la 1$ mmag level.  These
first two pieces of evidence tend to exclude or strongly disfavor blend scenarios
in which the observed transits are due to diluted eclipses of a much
fainter and redder eclipsing binary (e.g., \citealt{odonovan2006}).

Finally, our adaptive optics imaging does not reveal any sources further than
$\sim 0.25$\arcsec from the primary that could be both blended with it in
seeing-limited images {\it and} cause transits at the observed depth
of $\sim 1\%$.  The one source we do detect, the putative tertiary,
has a flux ratio relative to the primary of only $\sim 0.5\%$ in the
near-IR, and is likely considerably fainter in the optical, and
thus is too faint to explain the observed transits.

We did not perform any detailed modeling to determine the viability of
specific blend scenarios. We defer here to \citet{bakos2012}, who
argue that such analyses are generally unnecessary in situations in which
there are no significant bisector variations, the transit
ingress/egress durations are short compared to the total duration, and
the radial velocity variations phase with the predicted transit
ephemeris.

We conclude that all of the available data are best explained as due
to a Jupiter-sized, brown dwarf companion transiting a
rapidly-rotating mid-F star, with little or no evidence for
significant contamination from blended sources.  Under this
assumption, we proceed in the following section to analyze these data
in order to determine the physical properties of the KELT-1 host star
and its short-period, low-mass companion.

\section{Characterization of the Star, Companion, and Orbit}
\label{sec:char}

\subsection{Properties of the Host Star}\label{sec:hostprops}

Table \ref{tab:hostprops} lists various collected properties and
measurements of the KELT-1 host star.  Many these have been culled
from the literature, and the remainder are derived in this section.
In summary, KELT-1 is a mildly evolved, solar-metallicity, mid-F star
with an age of $\sim 1.5-2$ Gyr located at a distance of $\sim 260$~pc,
with kinematics consistent with membership in the thin disk.  

\begin{figure}
\includegraphics[scale=0.35,angle=90]{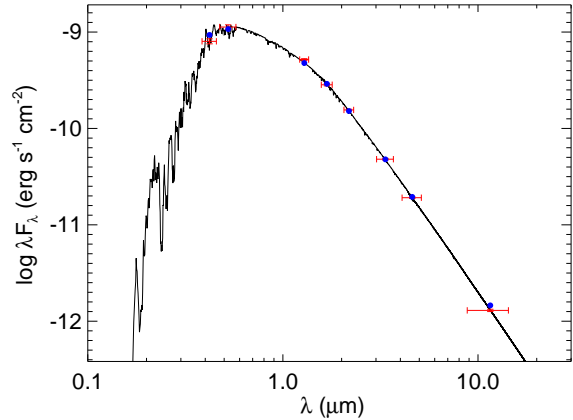}
\caption{The red errorbars indicate measurements of the flux
of KELT-1 host star in various optical and IR passbands.  The
vertical errorbar indicates the photometric uncertainty, whereas
the horizontal errorbar indicates the effective width of the passband.
The solid curve is the best-fit theoretical SED from the NextGen models of
\citet{hauschildt1999}, assuming $\teff$, $\logg$ and
[Fe/H] fixed at the values in Table
\ref{tab:physpars}, with $A_V$ and $d$ allowed to vary.  The blue
dots are the predicted passband-integrated fluxes of the 
best-fit theoretical SED.  
\label{fig:sed}}
\end{figure}

We construct an empirical spectral energy distribution (SED) of KELT-1
using optical fluxes in the $B_T$ and $V_T$ passbands from the Tycho-2
catalog \citep{hog2000}, near-infrared (IR) fluxes in the $J$, $H$ and
$K\!s$ passbands from the 2MASS Point Source Catalog
\citep{skrutskie2006,cutri2003}, and near- and mid-IR fluxes in the
four WISE passbands \citep{wright2010,cutri2012}.  This SED is
shown in \ref{fig:sed}.  We fit this SED to NextGen models from
\citet{hauschildt1999} by fixing the values of $\teff$, $\logg$ and
[Fe/H] inferred from the global fit to the light curve and RV data as
described in \S\ref{sec:analysis} and listed in Table
\ref{tab:physpars}, and then finding the values of the visual
extinction $A_V$ and distance $d$ that minimizes $\chi^2$.  We
find $A_V = 0.18 \pm 0.10$ and $d=262\pm 14$pc, with a $\chi^2 = 10.5$
for 6 degrees of freedom, indicating a reasonable fit ($P(>\chi^2)\sim
10\%$).  We also performed a fit to the SED without priors, finding
$\teff=6500 \pm 400$K,  $A_V=0.20 \pm 0.15$, $\logg=4.25 \pm 0.75$ and
[Fe/H]$=-0.5\pm 0.5$, consistent with the constrained fit.
There is no evidence for an IR excess.

We note that the quoted statistical uncertainties on $A_V$ and $d$ are
likely to be underestimated, because we have not accounted for the
uncertainties in values of $\teff$, $\logg$ and [Fe/H] used to derive
the model SED.  Furthermore, it is likely that alternate 
model atmospheres would predict somewhat different SEDs and thus values of the
extinction and distance.

\begin{figure}
\includegraphics[scale=0.35,angle=90]{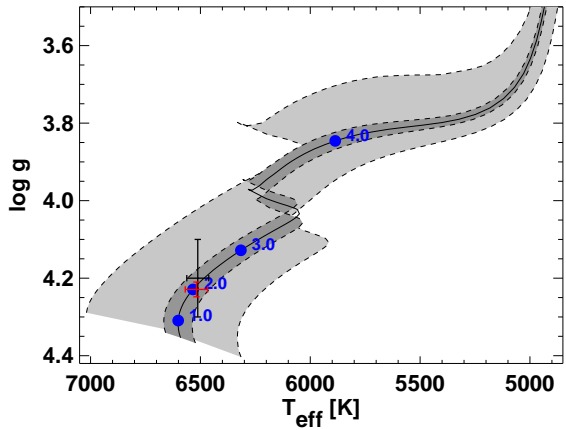}
\caption{Theoretical HR diagram based on Yonsei-Yale stellar evolution
models \citep{demarque2004}.  The solid track is the isochrone for the
best-fit values of the mass and metallicity of the host star from the
joint fit described in \S\ref{sec:analysis}, $M_*=1.32 \pm
0.03~M_\odot$ and [Fe/H]$=0.01 \pm 0.07$.   The red cross shows the best-fit
$\teff=6513\pm 50$K and $\logg=4.229_{-0.020}^{+0.012}$ from the final
analysis. The black cross shows the inferred $\teff$ and $\logg$ from the spectroscopic
analysis alone.  The blue dots represent the location of the
star for various ages in Gyr.  The host star is slightly evolved with a probable age of $\sim
2$ Gyr, although a similar analysis with a different stellar
evolutionary model prefers a slightly younger age of $\sim 1.5-1.75$
The two sets of dashed tracks
represent the tracks for the extreme range of the 1-$\sigma$
uncertainties on $M_*$ and [Fe/H] from the final analysis (dark shaded),
and from the spectroscopic constraints alone (light shaded).
\label{fig:hr}}
\end{figure}

In Figure \ref{fig:hr} we plot the predicted evolutionary track of
KELT-1 on a theoretical HR diagram ($\logg$ vs.\ $\teff$), from the
Yonsei-Yale stellar models \citep{demarque2004}. Here again we have
used the values of $M_*$ and [Fe/H] derived from the global fit
(\S\ref{sec:analysis} and Table \ref{tab:physpars}).  We also show
evolutionary tracks for masses corresponding to the $\pm 1~\sigma$
extrema in the estimated uncertainty.  In order to estimate the age of
the KELT-1 system, we compare these tracks to the values of $\teff$
and $\logg$ and associated uncertainties as determined from the global
fit.  These intersect the evolutionary track for a fairly narrow range
of ages near $\sim 2$ Gyr.  The agreement between the prediction from
the evolutionary track at this age and the inferred temperature and
surface gravity for KELT-1 is remarkably good, but perhaps not
entirely surprising.  The values of $\teff$, $\logg$, [Fe/H] and $M_*$
were all determined in the global fit to the light curve and RV data
in \S\ref{sec:analysis}, which uses the empirical relations between
($\teff$, $\logg$, [Fe/H]) and ($M_*, R_*$) inferred by
\citet{torres10} as priors on the fit, in order to break the
well-known degeneracy between $M_*$ and $R_*$ for single-lined
spectroscopic eclipsing systems.  These empirical relations are known
to reproduce the constraints between these parameters imposed by the
physics of stellar evolution quite well
(see, e.g., Section 8 in \citealt{torres10}).

Based on its $\teff$ and
$J-K$ and the empirical table of spectral type versus color and
$\teff$ for main-sequence from \citet{kenyon1995}, we infer the spectral type of KELT-1 to be F5 with an
uncertainty of roughly $\pm 1$ spectral type.

We determined the Galactic $U,V,W$ space velocities of the KELT-1
system using the proper motion of $(\mu_{\alpha},\mu_{\delta})=(-10.1
\pm 0.7,-9.4\pm 0.7)$~mas~yr$^{-1}$ from the NOMAD catalog
\citep{zacharias2004}, the distance of $d=262\pm 14$~pc from our SED
fit described above, and the barycentric radial velocity of the system
as determined from the TRES observations (\S\ref{sec:flwos}) of
$\gamma_{abs}=-14.2 \pm 0.2$ km/s.  We used a modification of the IDL
routine {\tt GAL\_UVW}, which is itself based on the method of
\citet{johnson1987}.  We adopt the correction for the Sun's motion
with respect to the Local Standard of Rest from
\citet{coskunoglu2011}, and choose a right-handed coordinate system
such that positive $U$ is toward the Galactic Center.  We find
$(U,V,W)= (19.9 \pm1.1,-9.6 \pm 0.5, -2.6 \pm 0.9)\kms$, consistent
with membership in the thin disk \citep{bensby2003}.  We note
also that the distance of KELT-1 from the Galactic plane is $\sim 80$~pc.

Finally, we use the solar evolutionary models of \citet{guenther1992},
updated with input physics from \citet{vansaders2012}, to gain some
insight into the detailed structure of the host star.  Fixing the mass
and metallicity at the values determined from the global fit
(\S\ref{sec:analysis} and Table~\ref{tab:physpars}), we evolved the
model forward until the model $\logg$ and $\teff$ approximately
matched the values inferred for KELT-1.  We found that ages of $\sim
1.5-1.75$ Gyr best matched the available constraints, and thus this
model prefers a somewhat younger age than the Yale-Yonsei model of
(Fig.\ \ref{fig:hr}, \citealt{demarque2004}).  We therefore decided to adopt an age of $1.75
\pm 0.25$ Gyr, consistent with both estimates.  For this range of
ages, the models of \citet{guenther1992} predict a radius of the
base of the convective zone of $R_{cz}=1.30\pm 0.03~R_\odot$, and a very small mass for the
convective zone of $M_{cz}=[2.8\pm 0.14]\times 10^{-5}~M_\odot$, as
expected given the effective temperature of $\teff \sim 6500$K.  In
addition, the moment of inertia for the star and convective zone are
$C_* = [1.15\pm 0.04] \times 10^{54}~{\rm g~cm^2}$ and $C_{cz}=[3.2
\pm 0.6]\times 10^{51}~{\rm g~cm^2}$, respectively.  We can also
write the moment of inertia of the star as $C_*=\alpha_*M_*R_*^2$
with $\alpha_*=0.0422$ \citep{guenther1992}.  We will use these to estimate the
angular momenta of the star, companion and orbit in \S \ref{sec:tides}.

\subsection{System Properties Derived from a Joint Fit}
\label{sec:analysis}

It is well known that a joint fit to high-quality RVs and transit
photometry of a transiting planet system allows one to determine the
mass and radius of the star and planet, as well
as the semi-major axis of the orbit, in principle to very high
precision, up to a perfect one-parameter degeneracy \citep{seager2003}.
This degeneracy arises because the duration, depth, and shape of the
primary transit, when combined with the eccentricity, longitude of
periastron, and period of the planet from RV data, allow one to
precisely estimate the density of the primary star $\rho_*$, but not
$M_*$ or $R_*$ separately.  Breaking this $M_*-R_*$ degeneracy generally
requires imposing some external constraint, such as a theoretical
mass-radius relation \citep{cody2002,seager2003}, or constraints from
theoretical isochrones (e.g., \citealt{bakos2012}).  In principle, a
measurement of $\logg$ from a high-resolution spectrum can be used to
break the degeneracy, but in practice these measurements are generally
not competitive with the constraint on $\rho_*$ and often have
systematic uncertainties that are comparable to the statistical
uncertainties.

We fitted the RV and transit data using a modified version of the IDL
fitting package EXOFAST \citep{eastman12}.  The approach of EXOFAST to
breaking the $M_*-R_*$ degeneracy is similar to the method described
in, e.g., \citet{anderson2012}, but with significant differences.  We
will review it briefly here, but point the reader to \citet{eastman12}
for more details.  We fitted the RV and transit data simultaneously with
a standard Keplerian and transit \citep{mandel2002} model using a modified MCMC method (described in more
detail below).  In addition to the standard fitting parameters, we
also included $\teff, \logg$, [Fe/H] as proposal parameters.  We then
included priors on the measured values of $\teff, \logg$, [Fe/H] as
determined from analysis of the TRES spectra and given in
\S\ref{sec:flwos}.  In addition, we included priors on $M_*$ and $R_*$,
which are based on the empirical relations between ($\teff, \logg$,
[Fe/H]) and ($M_*, R_*$) determined by \citet{torres10}.  These priors
essentially break the
$M_*-R_*$ degeneracy, as they provide similar constraints as
isochrones, i.e., they encode the mapping between the $M_*$, [Fe/H]
and age of a star to its $\teff$ and $\logg$ as dictated by stellar
physics.

We fitted the 6 primary transits, Doppler RV, stellar parameters,
and RM effect simultaneously using EXOFAST,
which employs a Differential Evolution Markov Chain Monte Carlo
(DE-MC) method \citep{braak06}. We converted all times to the \bjdtdb \ 
standard \citep{eastman10}, and then at each step in the Markov
Chain, we converted them to the target's barycentric coordinate system
(ignoring relativistic effects). Note the final times were converted
back to \bjdtdb \ for ease of use. This transformation accurately and
transparently handles the light travel time difference between the RVs
and transits.

First, we fitted the Doppler RV data independently to a simple Keplerian model, 
ignoring the RM data taken on UT 2012-01-07.  At this stage, we did not include any
priors on the stellar parameters, as they do not affect the RV-only fit.  We scaled the uncertainties such that the
probability that the \chisq \ was larger than the value we achieved,
$\pchisq$, was $0.5$, to ensure the resulting parameter
uncertainties were roughly accurate. For a uniform prior in eccentricity,
we found the orbit is consistent with circular, with a $3\sigma$ upper limit of $e < 0.04$.
Nevertheless, in order to provide conservative estimates for the fit
parameters, we allowed for a non-zero eccentricity in our final fit.
However, to test the effect of this assumption, we repeated the
final fit forcing $e=0$.  We also investigated the possibility of a
slope in the RVs, but found it to be consistent with zero, so we did
not include this in the final fit.

Next, we fitted each of the 4 transits individually, including a zero
point, $F_{0,i}$ and airmass detrending variable, $C_{0,i}$ for each of
the $i$ transits. The airmass detrending coefficient was significant
($>1 \sigma$) for all but 1 transit, so for consistency, we included it
for all.   After finding the best fit with AMOEBA
\citep{nelder65}, we scaled the errors for each transit such that
$\pchisq=0.5$.  At this stage, we included the priors on the stellar parameters
as described above.

Next, we performed a combined fit to all the data, including a prior
on the projected stellar rotation velocity ($\vsini=55.2 \pm 2~\kms$) from
the spectra\footnote{The prior on $\vsini$ improves the determination
of the spin-orbit alignment angle $\lambda$ \citep{gaudi06}.  We also performed a fit without
this prior, finding results that were roughly consistent with, although less precise than,
those with the prior.}, and a prior on the period from the KELT-North discovery light
curve ($P=1.217513 \pm 0.000015$~days). Because it is usually
systematics in the RV data that vary over long time scales (due to a
combination of instrumental drift and stellar jitter) that ultimately
set the error floor to the RVs, we expect the
uncertainties of densely packed observations to be smaller than the
RMS of all observations, but with a systematic offset relative to the
rest of the orbit. Therefore, we fitted a separate zero point during the
RM run, and also scaled the errors on the RVs during transit to force
$\pchisq = 0.5$ for those subsets of points. \pchisq depends on the
number of degrees of freedom, but it is not obvious how many degrees
of freedom there are in the RM run -- technically, the entire orbit and
the transit affect the \chisq of the RM (13 parameters), but the
freedom of the RM measurements to influence most of those parameters
is very limited, when fit simultaneously with the transits. Indeed,
even \vsini \ is constrained more by the spectroscopic prior than the
RM in this case, which means there are only two parameters (the
projected spin-orbit alignment, $\lambda$, and the zero point,
$\gamma$) that are truly free. To be conservative, we subtracted
another degree of freedom to encompass all the other parameters on
which the RM data has a slight influence, before scaling the errors.

The RM data were modeled using the \citet{ohta05} analytic
approximation with linear limb darkening. At each step in the Markov
Chain, we interpolated the linear limb darkening tables of
\citet{claret11} based on the chain's value for \logg, \teff, and \feh
\ to derive the linear limb-darkening coefficient, $u$. We assumed the
$V$ band parameters to approximate the bandpass of TRES, though we
repeated the exercise in the $B$-band with no appreciable difference
in any of the final parameters. Note that we do {\it not} fit for the
limb-darkening parameters, as the data
are not sufficient to constrain them directly..  The uncertainties in all the
limb-darkening parameters provided in Table \ref{tab:fitpars}
arise solely from the scatter in \logg, \teff, \feh. We assume no
error in the interpolation of the limb-darkening tables.

In order to search for Transit Timing Variations (TTVs), during the
combined fit, we fitted a new time of transit, $T_{C,i}$ for each of the
$i$ transits. Therefore, the constraint on $T_C$
and $P$ (quoted in Tables \ref{tab:fitpars} and \ref{tab:physpars}, respectively) come from
the prior imposed from the KELT-North light curve and the RV data, not the
follow up light curves. Using these times to constrain the period
during the fit would artificially reduce any observed TTV signal. A
separate constraint on $T_C$ and $P$ follows from fitting a linear
ephemeris to the transit times, as discussed in \S \ref{sec:ttv}. It
is the result from this fit that we quote as our final adopted
ephemeris.

The results from this global fit are summarized in Tables
\ref{tab:physpars} and \ref{tab:fitpars}.  We also show the results
for the physical parameters assuming $e=0$ in Table
\ref{tab:physpars}; the differences between the fixed and free
eccentricity fits are always smaller than their uncertainties, and
generally negligible for most of these parameters.  The values of
$\teff$, $\logg$, and [Fe/H] we infer from the global fit are in agreement
with the values measured directly from the TRES spectra to within the
uncertainties.  Since the spectroscopic values were used as priors in
the global fit, this generally indicates that there is no tension
between the value of $\rho_*$ inferred from the light curve and RV
data, and the spectroscopic values.  The median value and
uncertainty for $\teff$ is nearly unaffected by the global fit.  While
the median value for [Fe/H] has changed slightly, the uncertainty is
very similar to that from the input prior.  On the other hand, the
uncertainty in $\logg$ from the global fit is a factor of $\ga 5$
smaller than the uncertainty from the spectroscopic measurement.  This is not
surprising, since the constraint on $\rho_*$ from the RV and light curve
data provides a strong constraint on $\logg$ via the relations of \citet{torres10}.

Following papers by the HATNet collaboration (e.g.,
\citealt{bakos2011,hartman2011}), we also present in Table
\ref{tab:physpars} our estimates of the median values and
uncertainties for a number of derived physical parameters of the
system that may be of interest, including the planet equilibrium
temperature assuming zero albedo and perfect redistribution $T_{eq}$, the
average amount of stellar flux at the orbit of the companion $\left< F
\right>$, and the Safronov number $\Theta = (a/R_p)(M_P/M_*)$ (e.g.,
\citealt{handsen2007}).  In addition, in Table \ref{tab:fitpars} we 
quote our estimates of various fit parameters and intermediate derived
quantities for the Keplerian RV fit, the primary transits, and the
secondary eclipse.

We note that the final uncertainties we derive for $M_*$, $R_*$,
$\logg$ and $\rho_*$ are relatively small, $\sim 2\%-5\%$.  These
uncertainties are similar to those found for other transiting planet
systems using methods similar to the one used here (e.g.,
\citealt{anderson2012}).  Specifically, these methods derive
physical parameters from a fit to the light curve and RV data, which
simultaneously imposes an empirical constraint between the $M_*, R_*$
and $\teff, \logg$ and [Fe/H] ultimately derived from
\citet{torres10}.  This constraint helps to break the $M_*-R_*$
degeneracy pointed out by \citet{seager2003} and discussed above, and
ultimately dictates our final uncertainty on $M_*$ and $R_*$ (and thus
$M_P$ and $R_P$).  In particular, our spectroscopic measurement of
$\logg$ provides a much weaker constraint.  Given
that our results rely so heavily on the
\citet{torres10} relations, it is worthwhile to ask to what extent our
parameters and uncertainties might be affected should these relations
be systematically in error.  First, as already noted, these empirical
relations are known to agree well with stellar isochrones in general
\citep{torres10}, and for KELT-1 in particular (Fig.\ \ref{fig:hr}).
Second, analyses using stellar isochrones rather than empirical
relations produce similar uncertainties on $M_*$ and $R_*$ (e.g.,
\citealt{bakos2011}), suggesting that the small uncertainties we
derive are not a by-product of our methodology.  Finally,
\citet{southworth2009} demonstrated that the results of the analysis
of 14 transiting systems with several different sets of isochrones
generally agree to within a few percent.  We therefore conclude
that our results are likely accurate, with systematic
errors at the level of our statistical uncertainties (a few percent).

\subsection{System Ephemeris and Transit Timing Variations}
\label{sec:ttv}

\begin{figure}[h]
\epsscale{1.2}
\hskip-0.5in
\plotone{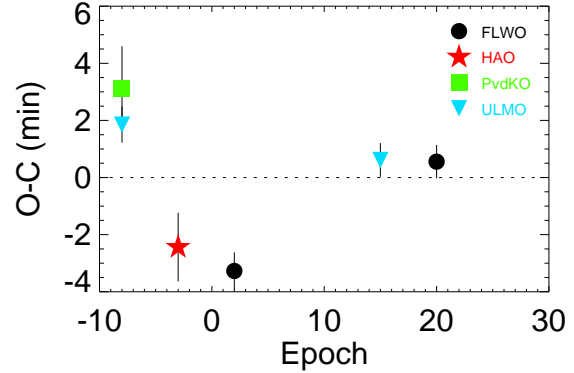} 
\caption{The residuals of the transit times from the best-fit (linear) ephemeris.
The transit times are given in Table \ref{tab:ttimes}.  
}
\label{fig:ttvs}
\end{figure}

Table \ref{tab:ttimes} lists the measured transit times for each of
the six modeled transits, and Figure \ref{fig:ttvs} shows the
residuals of these times from a fit to a linear ephemeris.  The best
fit has 
\bea
\nonumber
T_C(\bjdtdb) &=& 2455909.292797 \pm 0.00024\\
P &=& 1.2175007 \pm 0.000018,
\label{eqn:ephem}
\eea
which is consistent with the ephemeris derived from the
KELT-North light curve alone.  The $\chi^2=44.9$ for the linear fit with 4
degrees of freedom is formally quite poor.  This is mostly driven by
one nominally significant ($5 \sigma$) outlier, specifically for the
transit observed on UT 2011-12-16 from FLWO.  We note that the faint companion
to KELT-1, if indeed bound, is too distant to explain such
large TTVs.  

We have taken great care
to ensure the accuracy of our clocks and our conversion, and the fact
that the residuals from different observatories roughly follow the
same trend in time suggests that a catastrophic error in the
observatory clock cannot be blamed. Since we fit the trend with
airmass simultaneously with the transit, the potentially-large
covariance between it and the transit time should be accurately
reflected in the quoted uncertainties \citep{eastman12}. Nevertheless,
our MCMC analysis does not adequately take into account the effect of
systematic uncertainties, and in particular we do not account for
correlated uncertainties \citep{pont2006,carter2009}, which could skew
the transit time of a given event substantially.  And, given the
results from Kepler which suggest the rarity of such transit timing
variations \citep{steffen2012}, we are reluctant to over-interpret
this result.  Nevertheless, this is an interesting target for future
follow-up.

\subsection{Secondary Eclipse Limits}\label{sec:second}

\begin{figure}[h]
\epsscale{1.0}
\plotone{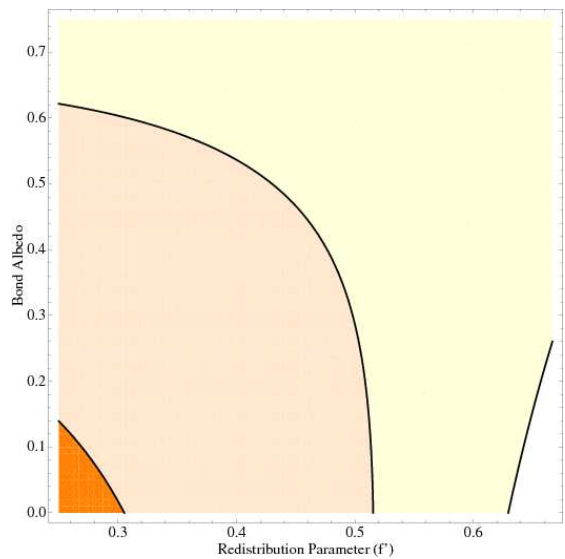}
\caption{Values of the heat redistribution parameter $f'$ 
and Bond albedo $A_B$ that are excluded at a given confidence level
based on the data taken during the secondary eclipse shown
in Figure \ref{fig:secondary}. The $f'$ parameter describes the efficiency of heat
redistribution, and is 1/4 in the case of uniform redistribution, and
2/3 in the case of no redistribution. In between these two extremes
$f'$ is not easily related to the amount of heat redistribution. The
orange section corresponds to eclipse depths detectable at less than
the $68\%$ confidence level, the light orange for 68-90\% confidence, yellow for 90-95\% confidence,
and the white for $>95\%$ confidence. }
\label{fig:secconst}
\end{figure}

We observed the predicted secondary eclipses of KELT-1b assuming
$e=0$ on UT
2011-12-02, 2011-12-30, and 2012-01-04.  In none of the three were we
able to detect a secondary eclipse. The observations on 2011-12-02 and
on 2012-01-04 were taken from the ULMO Observatory in $i$. Both
nights we were able to observe through the predicted ingress and
egress of the potential secondary. The two $i$ band light curves have a
combined 236 data points, and show an RMS scatter of 1.56 mmag. The
observations on 2011-12-30 were taken with the FTN telescope in
Pan-STARRS-$Z$. In this case we were only able to begin observations
half way through the predicted secondary eclipse. This $Z$-band light curve has
72 points, and an RMS scatter of 0.75 mmag.

We used the system parameters derived from the joint fit (\S\ref{sec:analysis}) 
to fit our three observations. Since
we did not detect a secondary eclipse, we used these fits to explore
the combination of heat redistribution efficiency and Bond albedo $A_B$ 
that would give rise to a secondary eclipse depth that is inconsistent
with our data. To do so, we calculated the
secondary eclipse depths we would expect for a range of redistribution
efficiencies and albedos, and then fit a secondary eclipse
model with the predicted depth to all three of our observations simultaneously.

In calculating the expected secondary eclipse depths, we made the
assumption that both the star and the planet were blackbodies. We also
assumed that the planet was a grey Lambert sphere, so the geometric
albedo $A_g = (2/3)A_B$, and the spherical albedo is constant as
a function of wavelength. Following \cite{seager2010}, we
parametrized the heat redistribution efficiency as $f'$, which is 1/4
in the case of uniform redistribution, and 2/3 in the case of no
redistribution. We note that in between these two extremes $f'$ is not
easily related to the amount of heat redistribution, i.e.,  $f'=0.45$
does not imply half of the incident stellar energy is redistributed
around the planet.

To test these expected secondary eclipse depths against our
observations, we fit simple trapezoidal eclipse curves with the
expected depths to all three datasets simultaneously.  Under our assumptions,
the depth, timing, shape, and duration of the secondary eclipse
are all determined by the parameters derived from the global fit 
and our specified values of $A_B$ and $f'$. We then fit this
model to our data, allowing for a normalization and a linear trend
in the flux with time. We used
the $\Delta\chi^2$ between the best fit eclipse model and the best
constant fit, which itself was allowed a free slope and offset, to
evaluate the detectability of each of the secondary eclipse depths. We
used the $\chi^2$ distribution to transform these $\Delta\chi^2$
values into detection probabilities. Figure \ref{fig:secondary} 
shows an example light curve against a median binned version of our data. This
particular curve is the secondary eclipse we would expect if KELT-1b
had $A_B=0.1$, and 
instantaneously re-radiated its incident stellar flux, i.e., $f'=2/3$. We would have
detected this event with more than 95\% confidence.

Figure \ref{fig:secconst} shows the results of our exploration of the
heat redistribution versus Bond albedo parameter space. The orange
section corresponds to eclipse depths detectable at less than the 68\%
confidence, the light orange is for depths detectable with 68-90\%
confidence, light yellow is for 90-95\%, and the white contains depths
that would have been detected at greater than 95\% confidence. The
particular shapes of the contours on this plot come from the competing
effects of reflection and blackbody emission from KELT-1b on the depth
of the secondary eclipse. Along the very bottom of the figure the
Bond albedo is zero, and thus there is only thermal emission.  We see
the strong change in eclipse depth as amount of heat redistribution
decreases, thus causing the temperature and eclipse depth for KELT-1b
to increase.  Along the top of the figure, where the Bond albedo is
0.75, the reflected starlight dominates the blackbody emission such
that changing the redistribution efficiency has little effect on the
eclipse depth.

Slightly more than half of the allowed parameter space in Figure
\ref{fig:secconst} would have caused secondary eclipses
detectable in our data at greater than 90\% confidence, while almost
all would have been detected at more than 68\%
confidence. Since we did not see a secondary eclipse in our
observations, we conclude that KELT-1b
either it has a non-zero albedo, or
it must redistribute some heat from the day side, or both.  Formally, the
scenario that is most consistent with our data is that KELT-1b has
both a low Bond albedo and is very efficient at redistributing heat
away from its day side, however we are reluctant to draw any strong
conclusions based on these data.

\begin{figure}[h]
\epsscale{2.2}
\plotone{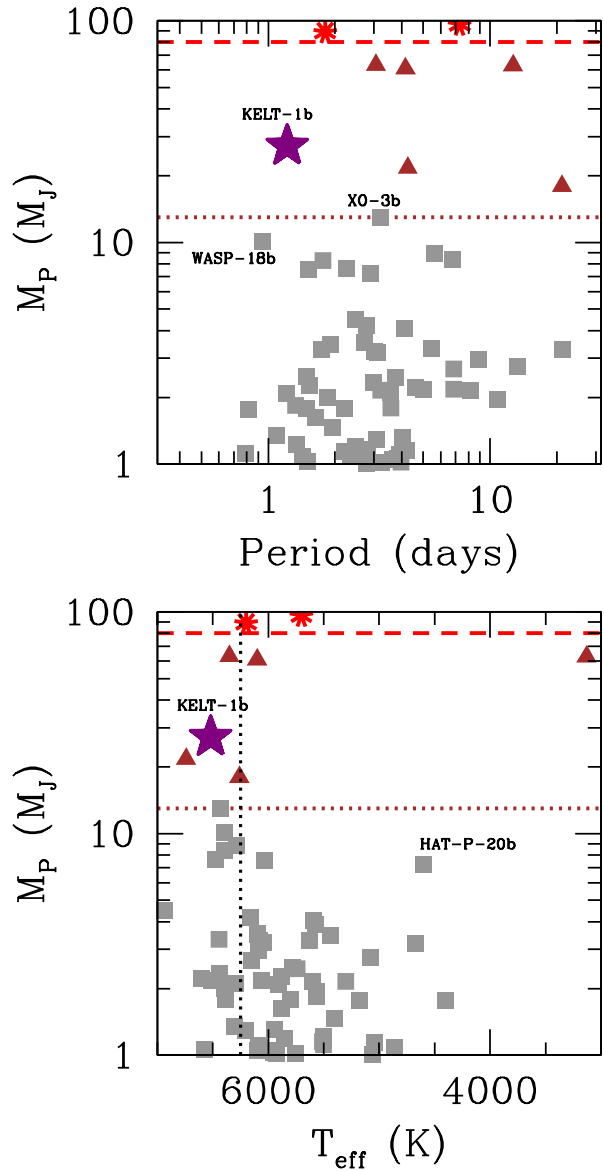}
\caption{
(Top panel) Mass versus period for the known transiting companions to
main-sequence stars with companion masses in the range $1-100~M_J$.  An
estimate of the deuterium burning limit \citep{spiegel2011} is shown
as the horizontal dotted line, while the hydrogen burning limit is
shown as the horizontal dashed line.  The vertical line
shows the division between hot and cool stars of $\teff = 6250$K
suggested by \citet{winnrm2010}.  Brown dwarfs are shown as
triangles, exoplanets as squares, and low-mass stars as asterisks.
KELT-1b is shown as the large star.  It is the shortest period
transiting brown dwarf currently known. (Bottom panel) Mass versus host
star effective temperature $\teff$ for the sample of transiting
companions shown in the top panel.  As suggested by
\citet{bouchy2011b}, there is some evidence that massive ($M_P\ga
5~\mjup$ companions are preferentially found around hot ($\teff \ga
6000$K) stars, and KELT-1b follows this possible trend.
Note that we exclude the BD companion to NLTT 41135 \citep{irwin2010},
and the double BD transit system 2M0535$-$05 \citep{stassun2007}
in this and subsequent plots.
\label{fig:mpvs}}
\end{figure}

\section{Discussion}\label{sec:discussion}

From our global fit to the light curves and RVs, we find that KELT-1b
is a low-mass companion with a measured mass
$M_P=27.23_{-0.48}^{+0.50}~\mjup$ and radius
$R_P=1.110_{-0.022}^{+0.032}~\rjup$.  It is on a circular orbit with a
semimajor axis of $a=0.02466\pm 0.00016$AU.  The host KELT-1 is a mildly evolved
mid-F star with a mass $M_*=1.324\pm0.026~M_\odot$, radius
$R_*=1.462_{-0.024}^{+0.037}~R_\odot$, effective temperature
$\teff=6518\pm 50~{\rm K}$, and a likely age of $\sim 1.5-2$Gyr.  
Because of its small semimajor axis and
hot host, KELT-1b receives a large stellar insolation flux of
$\langle F \rangle = 7.81_{-0.33}^{+0.42} \times 10^9~{\rm
erg~s^{-1}~cm^{-2}}$, implying a high equilibrium temperature of $T_{eq}=2422_{-26}^{+32}~{\rm
K}$ assuming
zero albedo and perfect redistribution.  Both the surface gravity and density of KELT-1b are
substantially higher than that of its host star, and higher than we would
expect for a stellar object.  We find that the
orbit normal of KELT-1b is well-aligned with the projected rotation
axis of its host star, with a projected alignment angle of $\lambda =
2 \pm 16$ degrees.

Even among the large and diverse menagerie of known transiting
exoplanets and low-mass companions, KELT-1b is unique.  First, it is
one of only 7 unambiguous objects with mass
the range $\sim 13-80~\mjup$ that are known to transit stars.  Among these, it has the
shortest period, and orbits the brightest host star ($V=10.7$).  In
addition, there is potentially a stellar M dwarf companion to the
primary.  For all these reasons, KELT-1b is likely to be a very
interesting object for further study, and we expect it will provide a
benchmark system to test theories of the emplacement and evolution of
short period companions, as well the physics of tidal dissipation and
irradiated atmospheres of substellar objects.  We will discuss some of
these ideas briefly.

\subsection{Brown Dwarf or Supermassive Planet? KELT-1b and the Brown Dwarf Desert}\label{sec:bdd}

Is KELT-1b a brown dwarf (BD) or a is it a suppermassive planet?
By IAU convention, brown dwarfs (BDs) are defined to have masses between
the deuterium burning limit of $\sim 13~\mjup$ \citep{spiegel2011}
and the hydrogen burning limit of $\sim 80~\mjup$ (e.g.,
\citealt{burrows1997}).  Less massive objects are defined to be
planets, whereas more massive objects are stars. By this definition,
KELT-1b is a low-mass BD.  However, it is interesting to ask
whether or not KELT-1b could have plausibly formed in a protoplanetary
disk, and therefore might be more appropriate considered a ``supermassive
planet'' \citep{schneider2011}. More generally, it is interesting to consider what KELT-1b and systems like it may
tell us about the formation mechanisms of close companions with masses
in the range of $10-100\mjup$.  

One of the first results to emerge from precision Doppler searches for
exoplanets is the existence of a BD desert, an apparent
paucity brown dwarf companions to
FGK stars with periods less than a few years, 
relative to the frequency of stellar companions in the same
range of periods \citep{marcy2000}.  Subsequent studies uncovered
planetary companions to such stars in this range of periods in
abundance \citep{cumming08}, indicating that the BD desert is a local
nadir in the mass function of companions to FGK stars.  The simplest
interpretation is that this is the gap between the largest objects
that can form in a protoplanetary disk, and the smallest objects that
can directly collapse or fragment to form a self-gravitating object in the
vicinity of a more massive protostar.   Therefore, the location
of KELT-1b with respect to the minimum of the brown dwarf
mass function might plausibly provide a clue to its origin.  

\subsubsection{Comparison Sample of Transiting Exoplanets, Brown Dwarfs, and Low-mass Stellar
Companions}\label{sec:comp}

In order to place the parameters of KELT-1b in context, we construct a
sample of transiting exoplanets, BDs, and low-mass
stellar companions to main sequence stars.  We focus only on
transiting objects, which have the advantage that both the mass and
radius of the companions are precisely known\footnote{In contrast, for
companions detected only via RVs, only the minimum mass is known.  Of
course, one can make an estimate of the posterior probability
distribution of the true mass given a measured minimum mass by
adopting a prior for the distribution of inclinations (e.g.,
\citealt{lee2011}).  However, this procedure can be particularly
misleading in the case of BDs: if BDs are indeed very rare, then
objects with minimum mass in the BD desert are more likely to be
stellar companions seen at low inclination. Anecdotally, in those
cases where constraints on the inclinations can be made, companions
with minimum mass near the middle of the brown dwarf desert often do
turn out to be stars (e.g., \citealt{sahlmann2011,fleming2012}).}.  We
collect the transiting exoplanet systems from the Exoplanet Data
Explorer (exoplanets.org, \citealt{wright2011}), discarding systems
for which the planet mass is not measured.  We supplement this list
with known transiting brown dwarfs
\citep{deleuil2008,johnson2011,bouchy2011a,bouchy2011b,anderson2011}.
We do not include the system discovered by \citep{irwin2010}, because
a radius measurement for the brown dwarf was not possible.  We also
do not include 2M0535$-$05 \citep{stassun2007}, because
it is a young, double BD system. We
add several transiting low-mass stars near the hydrogen burning limit
\citep{pont2005,pont2006b,beatty2007}. We adopt the mass of XO-3b from
the discovery paper \citep{johnskrull2008}, which is $M_P=13.1 \pm
0.4~\mjup$, which straddles the deuterium burning limit
\citep{spiegel2011}.  However, later estimates revised the mass
significantly lower to $M_P=11.8\pm 0.6$ \citep{winn2008}. We will
therefore categorize XO-3b as an exoplanet.

The disadvantage of using samples culled from transit surveys is that
the sample size is much smaller, and transit surveys have large and
generally unquantified selection biases (e.g.,
\citealt{gaudi2005,fressin2009}), particularly ground-based transit
surveys.  We emphasize that such biases are almost certainly present
in the sample we construct.  We have therefore made no effort to
be complete.  The comparisons and suggestions we make
based on this sample should not be considered definitive, but rather suggestive.

Figure \ref{fig:mpvs} places KELT-1b among the demographics of known
transiting companions to main sequence stars, focusing on massive
exoplanets, BDs, and low-mass stars with short periods of
$\la 30$ days.  KELT-1b has the tenth shortest period of any
transiting exoplanet or BD known.  It has the sixth shortest period
among giant ($M_P\ga 0.1~\mjup$) planets, with only WASP-19b,
WASP-43b, WASP-18b, WASP-12b, OGLE-TR-56b, and HAT-P-23b having
shorter periods.  KELT-1b is more massive by a factor $\sim 3$ than
the most massive of these, WASP-18b \citep{hellier2009}.  KELT-1b has
a significantly shorter period than any of the previously known
transiting brown dwarfs, by a factor of $\ga 3$.  KELT-1b therefore
appears to be located in a heretofore relatively unpopulated region of
the $M_P-P$ parameter space for transiting companions.

Although the KELT-1 system is relatively unique, it is worth asking if
there are any other known systems that bear some resemblance to it.
The $M_P\simeq 18 \mjup$, $P\simeq 1.3$ day RV-discovered companion to
the M dwarf HD 41004B \citep{zucker2003} has similar minimum mass and
orbit as KELT-1b, however the host star is obviously quite
different. Considering the host star properties as well, perhaps the
closest analogs are WASP-18b \citep{hellier2009}, WASP-33b
\citep{cameron2010}, and KOI-13b \citep{mazeh2012,mislis2012}.  All
three of these systems consist of relatively massive ($M_p\ga
3~\mjup$) planets in short ($\la 2$ day) orbits around hot $(\teff \ga
6500$~K$)$ stars.

The mass of KELT-1b ($\sim 27~\mjup$) is close to the most arid part of the BD desert,
estimated to be at a mass of $31_{-18}^{+25}~\mjup$ according to \citet{grether2006}.
Thus, 
under the assumption that the BD desert reflects the difficulty of forming
objects with this mass close to the parent star under {\it any}
formation scenario, KELT-1b may provide an interesting case to test these
various models.  For disk scenarios, gravitational
instability can likely form such massive objects, but likely only at
much larger distances \citep{rafikov2005,dodson2009,kratter2010}.  The
maximum mass possible from core accretion is poorly understood, but
may be as large as $\sim 40~\mjup$
\citep{mordasini2009}.  The possibility of significant
migration of KELT-1b from its birth location to its present position
must also be considered, particularly given the existence of a
possible stellar companion to KELT-1 (\S\ref{sec:keckao}).  This
possibility complicates the interpretation of the formation of KELT-1b
significantly.  For example, it has been suggested that brown dwarf
companions are more common at larger separations \citep{metchev2009}; thus
KELT-1b may have formed by collapse or fragmentation at a large separation, and subsequently
migrated to its current position via the Kozai-Lidov mechanism \citep{kozai1962,lidov1962}.

\begin{figure}
\epsscale{1.2}
\plotone{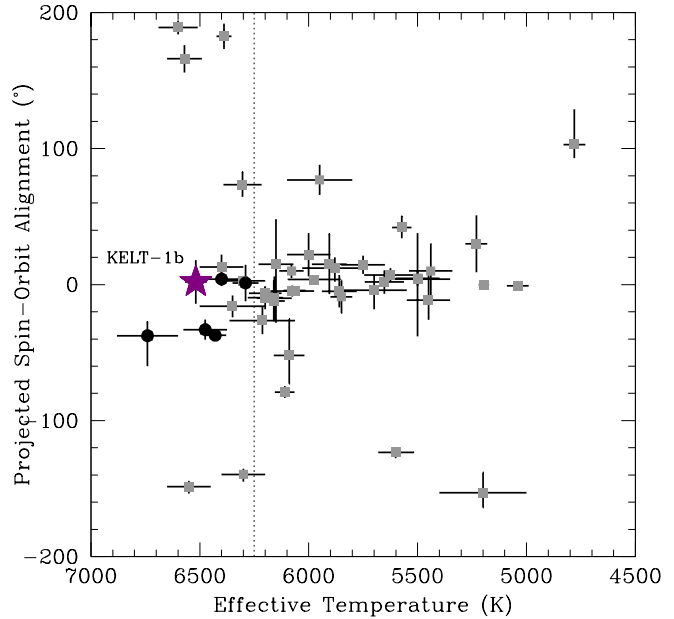}
\caption{
The projected spin-orbit alignment angle $\lambda$ for transiting planets
as measured by the RM effect, versus the effective temperature of the host
star, following \citet{winnrm2010}. The grey points show exoplanets
with mass $M_P< 5~\mjup$, whereas the black points show those with $M_P> 5 \mjup$.
KELT-1b, shown with a star, is the first transiting brown dwarf with a RM measurement.
Its orbit normal is consistent with being aligned with the projected host star spin axis. 
The dotted vertical line shows the suggested dividing line between hot and cool stars
by \citet{winnrm2010}.  
\label{fig:rmdist}}
\end{figure}

One clue to the origin of KELT-1b and the BD desert may be found by studying the
frequency of close BD companions to stars as a function of the stellar
mass or temperature.  Figure \ref{fig:mpvs} shows the mass of known
transiting short period companions as a function of the effective
temperature of the host stars.  As pointed out by \citet{bouchy2011b},
companions with $M_P\ga 5~\mjup$ appear to be preferentially found
around hot stars with $\teff \ga 6000~{\rm K}$, and KELT-1b follows
this trend.  Although these hot stars are somewhat more massive, the
most dramatic difference between stars hotter and cooler than 6000~K
is the depth of their convection zones.  This led \citet{bouchy2011b}
to suggest that tides may play an important role in shaping the
frequency and distribution of massive exoplanet and brown dwarf companions to old
stars.  Some evidence for this has been reported by \citet{winnrm2010}, who argue
that hot ($\teff \ge 6250$K) stars with close companions preferentially have high obliquities, suggesting
that if the emplacement mechanisms are similar for all stars, tidal forces must
later serve to preferentially bring cool host stars into alignment.   Figure \ref{fig:rmdist} shows the distribution of 
spin-orbit alignments for transiting planets versus the host star effective temperature.
KELT-1b falls in the group of hot stars with {\it small} obliquities.  Interestingly
the other massive $\ga 5~\mjup$ planets are also located in this group.

We discuss the possible formation and evolutionary history of
KELT-1b, and the likely role of tides in this history, in more detail
below.  We remain agnostic about the classification of KELT-1b as a brown dwarf
or supermassive planet.

\subsection{Tides, Synchronization, and Kozai Emplacement}\label{sec:tides}

\begin{figure}
\epsscale{2.2}
\plotone{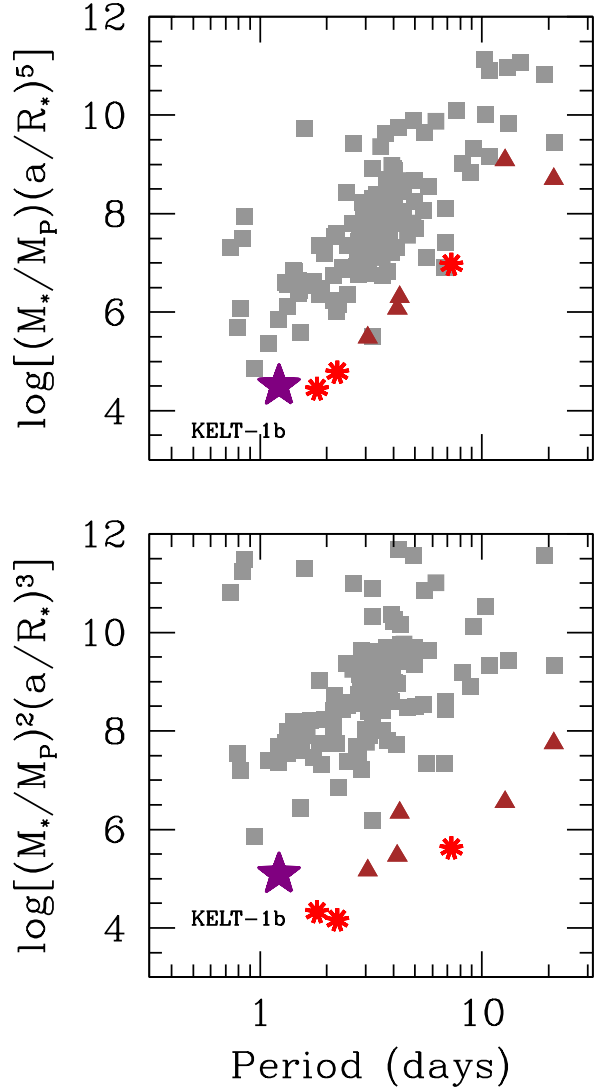}
\caption{
Dimensionless combinations of physical parameters that quantify the relative
time scale for orbital tidal decay (top panel) and stellar spin-orbit synchronization (bottom panel) for different 
binary systems, as a function of
the orbital period of the system.  See \S\ref{sec:tides} for an explanation and assumptions.
 Brown dwarfs are shown as
triangles, exoplanets as squares, and low-mass stars as asterisks.
KELT-1b is shown as the large star.  Among known transiting exoplanets
and brown dwarfs, it has the shortest characteristic time scale for orbital
decay and synchronization.
\label{fig:tidsyn}}
\end{figure}

Given the relatively large mass and short orbital period of KELT-1b,
it seems probable that tides have strongly influenced the past
evolution of the system, and may continue to be affecting its
evolution.  The literature on the influence of tides on exoplanet
systems is vast (see, e.g.,
\citealt{rasio1996,ogilvie2004,jackson2008,leconte2010,matsumura2010,hansen2010},
for but a few examples), and there appears to be little consensus on
the correct treatment of even the most basic physics of tidal
dissipation, with the primary uncertainties related to where, how, and
on what time scale the tidal energy is dissipated.  

While we are interested in evaluating the importance of tides on the evolution of 
the orbit of KELT-1b and the spin of KELT-1, delving into the rich but difficult 
subject of tides is beyond the scope of this paper.  We therefore take a somewhat
heuristic approach.  Specifically, we construct a few dimensionless quantities
that likely incorporate the primary physical properties of binary systems that
determine the scale of tidal evolution, but do not depend on the uncertain
physics of energy dissipation.  Specifically, we define,
\bea
{\cal T}_a & \equiv & \frac{M_*}{M_P} \left(\frac{a}{R_*}\right)^5,\qquad {\rm and}\\
{\cal T}_{\omega_*} & \equiv & \left(\frac{M_*}{M_P}\right)^2 \left(\frac{a}{R_*}\right)^3.
\label{eqn:tfacs}
\eea
For some classes of theories of tidal dissipation and under some
assumptions, ${\cal T}_a$ is proportional to the e-folding timescale
for decay of the orbit, and ${\cal T}_{\omega_*}$ is proportional to
the timescale for synchronization of the spin of the star with the
companion orbital period.  It is worthwhile to note
that for transiting planet systems 
the combinations of parameters $M_P/M_*$ and $a/R_*$ are generally 
much better determined than the individual
parameters. In particular, the ratio of the mass of the planet to that of the star
is closely related
to the RV semi-amplitude $K$, whereas $a/R_*$ is closely related to the
ratio of the duration of the transit to the period \citep{winn2010}. 
Figure \ref{fig:tidsyn} shows ${\cal T}_a$ and ${\cal T}_{\omega_*}$
as a function of orbital period for the sample of
transiting exoplanets, brown dwarfs, and low-mass stars discussed
previously.  KELT-1b has shorter timescales than nearly the entire sample
of systems, with the exception of a few of the low-mass stars.  We
therefore expect tidal effects to be quite important in this system.

As a specific example, under the constant time lag model
\citep{hut1981,matsumura2010}, and assuming dissipation in
the star, zero eccentricity, zero stellar obliquity, and a slowly
rotating star, the characteristic time scale for orbital decay due to
tides is $\tau_{decay} \equiv a/|{\dot a}| = (12\pi)^{-1} Q_*' {\cal
T}_a P$, where $Q_*'$ is related to the dimensionless tidal quality
factor.  For KELT-1b, ${\cal T}_a \sim 3 \times
10^{4}$, and so $\tau_{decay}\sim 0.3~{\rm
Gyr}$ for $Q_*'=10^8$, clearly much shorter than the age of the system.
Similarly, the time scale for spinning up the star by the companion is
$\tau_{synch} \equiv \omega_*/|\dot \omega_*| \propto Q_*'{\cal
T}_{\omega_*} P$ \citep{matsumura2010}, and so is also expected to be
short compared to the age of the system. 

Given the expected short synchronization time scale and the fact that
the expected time scale for tidal decay is shorter than the age of the
system, it is interesting to ask whether or not the system has
achieved full synchronization, thus ensuring the stability of KELT-1b.
The measured projected rotation velocity of the star is $\vsini=56\pm
2~\kms$, which given the inferred stellar radius corresponds to a rotation
period of $P_* = 2\pi R_*\sin I_*/\vsini = [1.322\pm 0.053]\sin
I_*$~days, which differs from the orbital period of KELT-1b by $\sim
2\sigma$ for $I_*=90^\circ$.  This is suggestive that the system is indeed synchronized.
The small discrepancy could either be due to a slightly underestimated
uncertainty on $\vsini$, or the host could be moderately inclined by
$I_* \sim [67\pm 7]^\circ$.  However, one might expect the obliquity of the star
to be realigned on roughly the same time scale as the synchronization
of its spin \citep{matsumura2010}. The stellar inclination can also be constrained
by the precise shape of the transit light curve: lower inclinations 
imply higher rotation velocities, and thus increased oblateness and gravity 
brightening \citep{barnes2009}.  Ultimately, the inclination is limited to $I_* \ga 
10^\circ$ in order to avoid break up.  

We can also ask, given the known system parameters, if the system is theoretically
expected to be able to achieve 
a stable synchronous state.  A system is ``Darwin stable'' \citep{darwin1879,hut1980}
if its total angular momentum,
\be
L_{tot} = L_{orb} + L_{\omega,*} + L_{\omega,P}
\label{eqn:ltot}
\ee
is more than the critical angular momentum of
\be
L_{crit} \equiv 4\left[\frac{G^2}{27}\frac{M_*^3M_P^3}{M_*+M_P}(C_*+C_P)\right]^{1/4},
\label{lcrit}
\ee
where $L_{orb}$ is the orbital angular momentum, $L_{\omega_*}$ is the spin angular momentum
of the star, $L_{\omega,P}$ is the spin angular momentum of the planet, and 
$C_*=\alpha_* M_* R_*^2$ and $C_P=\alpha_P M_P R_P^2$ are the moments of inertia
of the star and planet, respectively \citep{matsumura2010}.
Since $C_P/C_* \sim (M_P/M_*)(R_P/R_*)^2 \sim 10^{-3}$, the contribution from
the planet spin to the total angular momentum is negligible.  
We find $L_{tot}/L_{crit}=1.029 \pm 0.014$, marginally above the critical value
for stability.  
In addition, we find $(L_{\omega,*}+L_{\omega,P})/L_{orb} = 0.154 \pm 0.006$, which is smaller
than the maximum value of $1/3$ required for a stable equilibrium \citep{hut1980}.  Curiously, 
if we assume the star is already tidally synchronized, we instead infer $(L_{\omega,*}+L_{\omega,P})/L_{orb}=0.167 \pm 0.004$,
i.e., remarkably close to exactly one-half the critical value of 1/3.

Two additional pieces of information potentially provide clues to the
evolutionary history of this system: the detection of a possible
tertiary (\S\ref{sec:keckao}; Fig.\ref{fig:aoimage}), and the
measurement of the RM effect (Fig.~\ref{fig:RM}), demonstrating that
KELT-1 has small projected obliquity.  If the nearby companion to
KELT-1 is indeed bound, it could provide a way of emplacing KELT-1b in
a small orbit via the Kozai-Lidov mechanism
\citep{kozai1962,lidov1962}.  If KELT-1b were originally formed much
further from its host star, and on an orbit that was significantly
misaligned with that of the putative tertiary, then its orbit might
subsequently be driven to high eccentricity via secular perturbations
from the tertiary \citep{holman1997,lithwick2011,katz2011}.  If it
reached sufficiently high eccentricity such that tidal effects became
important at periastron, the orbit would be subsequently circularized
at a relatively short period \citep{fabrycky2007,wu2007,socrates2012}.
Nominally, one might expect the orbit of KELT-1b to be then left with
a relatively large obliquity \citep{naoz2011}.  The measured projected obliquity is
$\la 16$ degrees, implying that either the current true obliquity is
small, or the star is significantly inclined (i.e., $I_* \sim 0$).
However, if the star is significantly inclined, then the system cannot
be synchronized.  Perhaps a more likely alternative is that, after
emplacement by the tertiary and circularization of the orbit, the
system continued to evolve under tidal forces, with KELT-1b migrating
inward to its current orbit while damping the obliquity of KELT-1 and
synchronizing its spin period.  Clearly, detailed simulations are
needed to establish whether or not this scenario has any basis in
physical reality.

\subsection{Comparison to Theoretical Models of Brown Dwarfs}\label{sec:models}

\begin{figure}
\epsscale{1.2}
\plotone{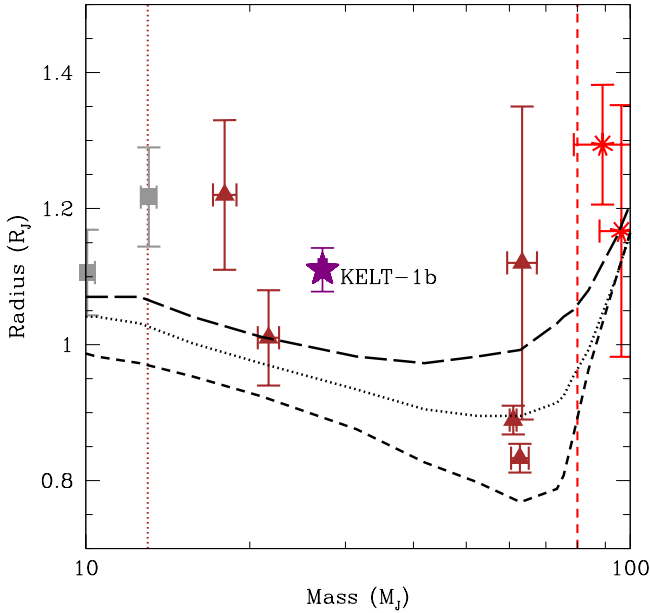}
\caption{Radius versus mass for the known transiting companions to
main-sequence stars with companion masses in the range $10-100~M_J$
that have measured radii.  An estimate of the deuterium burning limit
\citep{spiegel2011} is shown as the vertical dotted line, and the
hydrogen burning limit is shown as the vertical dashed line.  Brown
dwarfs are shown as triangles, exoplanets as squares, and low-mass
stars as asterisks.  KELT-1b is shown as the large star.  Predicted
radii as a function of mass for isolated objects from the isochrones of
\citet{baraffe2003} are shown for an age of 5 Gyr (dashed), 1 Gyr
(dotted), and 0.5 Gyr (long dashed); the true age of the KELT-1 system
is almost certainly between $1-5$ Gyr.  Although stellar insolation is
likely to increase the radii at fixed mass, \citet{bouchy2011a}
predict that the effect is small.  KELT-1b therefore has an
anomalously large radius.
\label{fig:mr}}
\end{figure}

Transiting brown dwarfs provide one of the only ways to test and
calibrate models of BD structure and evolution, which are used to
interpret observations of the hundreds of free floating brown dwarfs
for which no direct measurement of mass and radius is possible.  Given
that only 5 transiting brown dwarfs with radius measurements
were previously known, KELT-1b
potentially provides another important test of these models.  Figure
\ref{fig:mr} shows the mass-radius relation for the known transiting
companions to main-sequence stars with companion masses in the range
$10-100~M_J$.  Being close to the minimum in the brown
dwarf desert, the mass of KELT-1b begins to fill in the dearth
of know systems between
$\sim 20-60~\mjup$.  Furthermore, the formal uncertainty in its
radius is only $\sim 2.5\%$, thereby allowing for a stringent test of
models.  In contrast, the two transiting BDs with similar masses,
CoRoT-3b \citep{deleuil2008} and KOI-423b \citep{bouchy2011b}, have
much larger radius uncertainties, presumably due to the relative faintness of
the host stars.

Evolutionary models for isolated BDs generally predict that young
($\sim 0.5$ Gyr) objects in the mass range $10-100~\mjup$ should have
radii of $\sim \rjup$ (see the models of \citealt{baraffe2003} in
Fig.~\ref{fig:mr}).  As these objects cool, however, their radii
decrease, particularly for masses between 50 and 80 $\mjup$.  After
$\sim 1$~Gyr, all isolated objects with mass between 20-80~$\mjup$ are
predicted to have radii $<\rjup$.  The radius we measure for KELT-1b is $R_P =
1.110_{-0.022}^{+0.032}~\rjup$, which, at a mass of
$M_P=27.23_{-0.48}^{+0.50}~\mjup$, is $\sim 7~\sigma$ and $\sim
10~\sigma$ larger than the radius predicted by \citet{baraffe2003}
for ages of 1 Gyr and 5 Gyr, respectively.  KELT-1b is strongly
irradiated, which in principle can delay its cooling and
contraction.  However, \citet{bouchy2011a} predict that the effect of
insolation is small for brown dwarfs in this mass range, although
their models were for a much more modest insolation corresponding to
an equilibrium temperature of 1800~K (versus $\sim 2400$K for
KELT-1b).  Therefore, given the estimated $1.5-2$ Gyr age of the
system, KELT-1b is likely to be significantly inflated relative to
predictions.

\begin{figure}
\epsscale{1.2}
\plotone{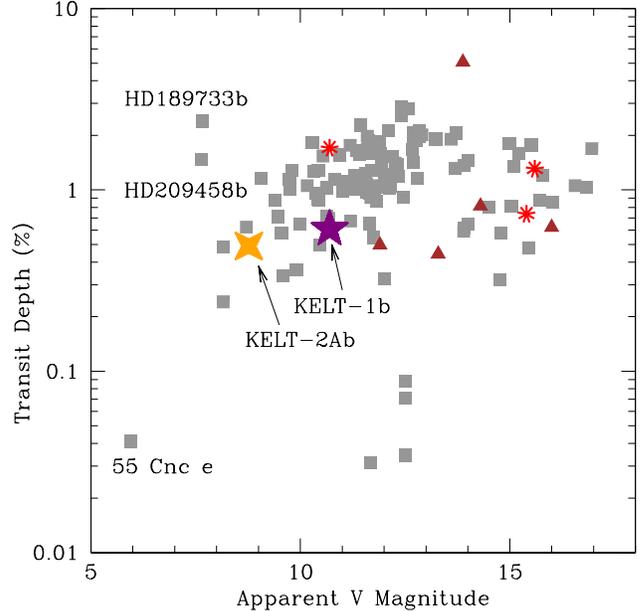}
\caption{
Transit depth assuming no limb darkening, i.e., $(R_P/R_*)^2$, as a
function of the apparent $V$ magnitude of the host star for a sample
of transiting systems. Brown dwarfs are shown as triangles, exoplanets
as squares, and low-mass stars as asterisks.  KELT-1b is shown as the
large star.  All else being equal, objects in the top left provide the
best targets for follow-up.  KELT-1b has a similar transit depth as
the other known transiting brown dwarfs, but is significantly
brighter.  Also labeled are some other benchmark systems.  KELT-2b
($M_P \sim 1.5~\mjup$) is shown as a large cross (Beatty et al., in
preparation).
\label{fig:rprsv}}
\end{figure}

Using the benchmark double transiting BD 2M0535$-$05,
\citet{gmc2009} explore models in which brown dwarfs have
large spots, which reduce the flux from their surface, thereby
decreasing their effective temperatures and increasing their radii
relative to those without spots (see also \citet{bouchy2011a}). 
 They find that these can lead to
significantly inflated radii, but only for large spot filling factors
of $\sim 50\%$, and for relatively young ($\sim 0.5$ Gyr) systems.
However, a detailed spectroscopic analysis of that system by \citet{mohanty2010}
and \citet{mohanty2012},
shows that surface spots cannot be present with such a large filling factor, 
and thus favor global structural effects such as strong
internal magnetic fields (e.g., \citealt{mullan2010}).
Many other mechanisms have been invoked to explain the inflated radii of
some giant exoplanets (see \citealt{fortney2010} for a review),
however it is not clear which, if any, of the many mechanisms that
have proposed may also be applied to inflated brown dwarfs.

We would be remiss if we did not question whether we were erroneously
inferring a large radius for the planet. In the past, such situations
have arisen when there is a discrepancy between the constraint on the
stellar density from the light curve and the constraint on the stellar
surface gravity of the star from spectroscopy (e.g.,
\citealt{johnskrull2008,winn2008}).  In our case, we find no such
tension.  The parameters of the star inferred from the spectroscopic
data alone are in nearly perfect agreement with the 
results from the global analysis of the light curve, RV data,
and spectroscopic constraints.  We note that the effect of
allowing a non-zero eccentricity also has a negligible effect on the
inferred planetary radius.  Finally, we reiterate
that the faint
companion detected in AO imaging (\S\ref{sec:keckao}), which is unresolved
in our follow-up photometry, has a 
negligible effect on our global fit and inferred parameters.
Therefore, we believe our
estimate of $R_P$ is likely robust.

We conclude by noting that there is a need for predictions of the
radii of brown dwarfs for a range of ages and stellar insolations, and
it would be worthwhile to explore whether or not the inflation
mechanisms that have been invented to explain anomalously large giant
planets might work for much more massive and dense objects as well.

\subsection{Prospects for Follow Up}

Figure \ref{fig:rprsv} compares the transit depth and apparent visual
magnitude of the KELT-1 system ($\delta \sim 0.6\%$, $V=10.7$) to the
sample of transiting systems collected in \S\ref{sec:comp} with
available $V$ magnitudes.  KELT-1 is not particularly bright compared
to the bulk of the known transiting exoplanet hosts.  However, it is
significantly brighter than the hosts of all known transiting brown
dwarfs; the next brightest is WASP-30 \citep{anderson2011}, which is
$\sim 1.2$ magnitudes fainter.  On the other hand, the depth of the
KELT-1b transit is similar to that of the other known brown dwarfs.

The prospects for follow-up of KELT-1b are exciting, not only because
of the brightness of the host, but also because of the extreme nature
of the system parameters, in particular the relatively short orbital
period, relatively large stellar radius, and relatively large amount
of stellar irradiation received by the planet.  Following
\citet{mazeh2010} and \citet{faigler2011}, we can estimate the
amplitudes of ellipsoidal variations $A_{ellip}$, Doppler beaming
$A_{beam}$ (see also \citealt{loeb2003}), reflected light eclipses and
phase variations $A_{ref}$, and thermal light eclipses and phase
variations $A_{therm}$,
\bea
A_{beam}  &=& \alpha_{beam}4\left(\frac{K}{c}\right) \sim 5.7\alpha_{beam}\times 10^{-5}\\
A_{ref} &=& \alpha_{ref} \left(\frac{R_P}{a}\right)^2\sim 4.6\alpha_{ref}\times 10^{-4}\\
A_{ellip} &=& \alpha_{ellip}\frac{M_P}{M_*} \left(\frac{R_*}{a}\right)^3 \sim 4.1\alpha_{ellip}\times 10^{-4} \\
A_{therm} &=& \alpha_{therm} \left(\frac{R_P}{R_*}\right)^2 \left(\frac{R_*}{a}\right)^{1/2} \sim 3.2 \alpha_{therm}\times 10^{-3},
\eea
where the expression for $A_{therm}$ assumes observations in the
Rayleigh-Jeans tail of both objects, and the expression for
$A_{ellip}$ assumes an edge-on orbit.  The dimensionless constants
$\alpha$ are defined in \citet{mazeh2010}, but to make contact with
the secondary eclipse analysis in \S\ref{sec:second} we note that
$\alpha_{ref}=A_g$ and $\alpha_{therm}=[f'(1-A_B)]^{1/4}$.  All of
these constants are expected to be of order unity, except for
$\alpha_{ref}$, which may be quite low for strongly irradiated
planets, depending on wavelength \citep{burrows2008}.  Based on
previous results, all of these effects with the possible exception of
Doppler beaming are likely to be detectable with precision photometry
(see, e.g., \citealt{cowan2012}).  For ellipsoidal variations in particular,
we expect $\alpha_{ellip} \sim 2$ and thus a relatively large amplitude of $A_{ellip} \sim 10^{-3}$.
Furthermore, the detection of all these signals is facilitated
by the short orbital period for KELT-1b.

The prospects for transmission spectroscopy are probably poorer, given
the relatively small planet/star radius ratio ($\sim 0.078$) and more
importantly the large surface gravity for KELT-1b.  For the
optimistic case of $T_{eq}\simeq 2400$K assuming zero albedo and perfect
redistribution, the scale height is only $H\sim kT/(\mu m_H g_P) \sim
16~{\rm km}$, and thus will only lead to changes in the transit depth
of order $\sim 2H/R_P \sim 0.04\%$.

\section{Summary}\label{sec:summary}

We have presented the discovery of KELT-1b, the first transiting
low-mass companion from the wide-field Kilodegree Extremely Little
Telescope-North (KELT-North) transit survey.  The host star KELT-1 is
a mildly evolved, solar-metallicity, rapidly-rotating, mid-F star with an age of $\sim
1.5-2$ Gyr located at a distance of $\sim 260$~pc.  The transiting companion is a
low-mass brown dwarf or supermassive planet with mass $\sim 27~\mjup$,
on a very short period, circular orbit of $P \sim 1.2$~days.

In many ways, the KELT-1 system is quite unusual and extreme: KELT-1b
receives a large amount of stellar insolation, is inflated relative to
theoretical predictions, and raises strong tides on its host.  The
obliquity of KELT-1 is consistent with zero, and there is evidence
that the spin of KELT-1 has been synchronized with the orbital period
of KELT-1.  Finally, there is a likely M-dwarf stellar companion to
the KELT-1 system with a projected separation of $\sim 150$~AU. As the
first definitively inflated transiting brown dwarf, KELT-1b
demonstrates the need for models of brown dwarfs subject to a range of
stellar insolations.

A plausible formation scenario for this system posits that KELT-1b 
formed on a much wider orbit, and was driven to a smaller
semimajor axis by the tertiary via the Kozai-Lidov mechanism.  The
system then continued to evolve under strong tidal forces, with 
KELT-1b migrating inward to its current orbit, while damping the
obliquity of KELT-1 and synchronizing its spin period.
 
The future evolution of the KELT-1 system may be spectacular.  As
KELT-1 continues to evolve and its radius increases, so will the tides
raised on it by KELT-1b.  Assuming KELT-1 is and remains tidally
locked, as it cools it will develop a deep convective envelope, but be
forced to rotate at an ever increasing rate.  In $\sim 2$ Gyr,
KELT-1 will have roughly the temperature of sun, but with a radius of
$\sim 2~\rsun$ and a rotational velocity of $\sim 100~\kms$.  At this
point, KELT-1 will likely become an active RS CVn star
\citep{walter1981}.  Eventually, as KELT-1 reaches the base of the
giant branch, it will swallow KELT-1b whole, likely resulting in a
bright UV/X-ray and optical transient \citep{metzger2012}.

\acknowledgments

We would like to particularly thank Bruce Gary for acquiring and
reducing the data from HAO, Saurav Dhital for estimating the
chance probability of a close companion, and Bence Beky and the HATNet 
team for giving up one night on the FLWO 1.2m on short notice. We thank Subo Dong, Jonathan
Fortney, Fred Rasio and Aristotle Socrates for useful discussions.  Early work on KELT-North
was supported by NASA Grant NNG04GO70G.  Work by B.S.G., J.D.E., and
T.G.B.\ was partially supported by NSF CAREER Grant AST-1056524.
E.L.N.J.\ gratefully acknowledges the support of the National Science
Foundation's PREST program, which helped to establish the Peter van de
Kamp Observatory through grant AST-0721386.  K.A.C.\ was supported by
a NASA Kentucky Space Grant Consortium Graduate Fellowship.  J.A.P.\
and K.G.S.\ acknowledge support from the Vanderbilt Office of the
Provost through the Vanderbilt Initiative in Data-intensive
Astrophysics. K.G.S.\ and L.H.\ acknowledge the support of the
National Science Foundation through PAARE grant AST-0849736 and AAG
grant AST-1009810. The TRES and KeplerCam observations were obtained
with partial support from the Kepler Mission through NASA Cooperative
Agreement NNX11AB99A with the Smithsonian Astrophysical Observatory,
D.W.L.\ PI.

This work has made use of NASA's Astrophysics Data System, the
Exoplanet Orbit Database at exoplanets.org, and the Extrasolar Planet Encyclopedia
at exoplanet.eu \citep{schneider2011}.

\begin{deluxetable}{llccc}
\tabletypesize{\scriptsize}
\tablecaption{Summary of Observations\label{tab:obs}}
\tablewidth{0pt}
\tablehead{
\colhead{Observatory} & \colhead{UT Dates} & \colhead{\# Obs} & \colhead{Wavelength} & \colhead{Section; Table}}
\startdata
\sidehead{Photometry, Primary Transit:}
PvdKO & 2011-12-03 $--$ 2011-12-04 & 151 & $i$ & \S\ref{sec:pvdko}; T\ref{tab:transit0}\\
ULMO  & 2011-12-03 $--$ 2011-12-04 & 110 & $r$ & \S\ref{sec:ulmo}; T\ref{tab:transit1} \\  
HAO   & 2011-12-10                 & 266 & $i$ & \S\ref{sec:hao}; T\ref{tab:transit2}\\
FLWO  & 2011-12-16                 & 362 & $z$ & \S\ref{sec:flwop}; T\ref{tab:transit3}\\
ULMO  & 2011-12-31 $--$ 2012-01-01 & 162 & $r$ & \S\ref{sec:ulmo}; T\ref{tab:transit4}\\
FLWO  & 2012-01-07                 & 459 & $i$ & \S\ref{sec:flwop}; T\ref{tab:transit5}\\
\sidehead{Photometry, Secondary Eclipse:}
ULMO  & 2011-12-02 & 115 & $i$ & \S\ref{sec:ulmo}\\  
FTN/LCOGT & 2011-12-30  & 72 & PS-$Z^a$ & \S\ref{sec:lcogt} \\  
ULMO  & 2012-01-04 & 121 & $i$ & \S\ref{sec:ulmo}\\  
\sidehead{Spectroscopy, Orbit:}
FLWO/TRES & 2011-11-09 $--$  2012-02-05 & 23 & 3900-8900\AA & \S\ref{sec:flwos}; T\ref{tab:rvorbit} \\
\sidehead{Spectroscopy, Rossiter-McLaughlin:}
FLWO/TRES &UT 2012-01-07  & 42 & 3900-8900\AA & \S\ref{sec:flwos}; T\ref{tab:rvrm} \\
\sidehead{Adaptive Optics Imaging:}
Keck/NIRC2 & UT 2012-01-07 &  & $H$, $K'$ & \S\ref{sec:keckao} \\
\enddata
\tablecomments{
ULMO=University of Louisville Moore Observatory;
PvdKO=Peter van de Kamp Observatory;
HAO=Hereford Arizona Observatory;
FLWO=Fred Lawrence Whipple Observatory;
FTN=Faulkes Telescope North; 
LCOGT=Las Cumbres Observatory Global Telescope Network;
\\
$^a$Pan-STARRS-$Z$
}
\end{deluxetable}

\begin{deluxetable}{lcclc}
\tabletypesize{\scriptsize}
\tablecaption{KELT-1 Stellar Properties \label{tab:hostprops}}
\tablewidth{0pt}
\tablehead{\colhead{Parameter} & \colhead{Description (Units)} & \colhead{Value} & \colhead{Source} & \colhead{Ref.}}
\startdata
Names & & TYC 2785-2130-1 & \\
      & & GSC 02785-02130 & \\
      & & 2MASS J00012691+3923017 & \\ 
$\alpha_{\rm{J2000}}$ & &  00:01:26.92 & Tycho & 1 \\
$\delta_{\rm{J2000}}$ & &  +39:23:01.7 & Tycho & 1 \\
$B_T$ & & $11.363\pm 0.065$ & Tycho-2 & 2 \\
$V_T$ & & $10.701\pm 0.057$ & Tycho-2 & 2\\
$J$ & & 9.682$\pm$0.022 & 2MASS & 3\\
$H$ & & 9.534$\pm$0.030 & 2MASS & 3\\
$K$ & & 9.437$\pm$0.019 & 2MASS & 3\\
WISE1 & & 12.077$\pm$0.022 & WISE & 4\\
WISE2 & & 12.732$\pm$0.020 & WISE & 4\\
WISE3 & & 14.655$\pm$0.030 & WISE & 4\\
$\mu_{\alpha}$ & Proper Motion in RA (mas~yr$^{-1}$) \dotfill & $-10.1 \pm 0.7$ & NOMAD &  5\\
$\mu_{\delta}$ & Proper Motion in Dec. (mas~yr$^{-1}$) \dotfill & $-9.4 \pm 0.7$ & NOMAD & 5\\
$\gamma_{\rm abs}$ & Absolute Systemic RV ($\kms$)\dotfill & $-14.2\pm 0.2$ & This Paper & \\
\dotfill & Spectral Type \dotfill & F5$\pm$1 & This Paper & \\
$d$ & Distance (pc)\dotfill & $262 \pm 14$ & This Paper & \\
\dotfill & Age (Gyr)\dotfill &  $1.75 \pm 0.25$ & This Paper & \\
$A_V$ & Visual Extinction\dotfill & $0.18 \pm 0.10$ & This Paper & \\
($U^a,V,W$) & Galactic Space Velocities (${\rm km~s}^{-1}$) \dotfill  & (19.9$\pm$1.1,-9.6$\pm$0.5, -2.6$\pm$0.9) & This Paper & \\
$M_{cz}$ & Mass of Convective Zone ($M_\odot$)  \dotfill  & $[2.8\pm 0.4] \times 10^{-5}$ & This Paper & \\
$R_{cz}$ & Base of the Convective Zone ($R_\odot$) \dotfill  & $1.30\pm 0.03$ & This Paper & \\
$C_{*}$ & Total Moment of Inertia (cgs) \dotfill  & $[1.15\pm 0.04] \times 10^{54}$ & This Paper & \\
$C_{cz}$ & Convection Zone Moment of Inertia  (cgs)\dotfill & $[3.2 \pm 0.6]\times 10^{51}$ & This Paper & \\
\enddata
\tablecomments{
1=\citet{hog1998},
2=\citet{hog2000},
3=\citet{skrutskie2006,cutri2003},
4=\citet{wright2010,cutri2012}.
5=\citet{zacharias2004}.
\\
$^{a}$ We adopt a right-handed coordinate system such that positive $U$ is toward the Galactic center.
}
\end{deluxetable}

\begin{deluxetable}{lccc}
\tabletypesize{\scriptsize}
\tablecaption{Median values and 68\% confidence intervals for the Physical and Orbital Parameters of the KELT-1 System}                                                
\tablewidth{0pt}
\tablehead{\colhead{~~~Variable} & \colhead{Description (Units)} & \colhead{Value ($e\ne 0$)}& \colhead{Value ($e\equiv 0$)}}
\startdata
\sidehead{Stellar Parameters:}
                                      ~~~$M_{*}$\dotfill &Mass (\msun)\dotfill & $1.324\pm0.026$                  &$1.322_{-0.025}^{+0.026}$\\	       
                                    ~~~$R_{*}$\dotfill &Radius (\rsun)\dotfill & $1.462_{-0.024}^{+0.037}$        &$1.452_{-0.019}^{+0.033}$\\
                                ~~~$L_{*}$\dotfill &Luminosity (\lsun)\dotfill & $3.48_{-0.18}^{+0.22}$           &$3.43_{-0.16}^{+0.20}$\\	  
                                    ~~~$\rho_*$\dotfill &Density (cgs)\dotfill & $0.597_{-0.039}^{+0.026}$        &$0.610_{-0.037}^{+0.018}$\\
                         ~~~$\logg$\dotfill &Surface gravity (cgs)\dotfill & $4.229_{-0.019}^{+0.012}$        &$4.2351_{-0.017}^{+0.0087}$\\	
                         ~~~$\teff$\dotfill &Effective temperature (K)\dotfill & $6518\pm50$                      &$6517\pm49$\\
                                        ~~~$\feh$\dotfill &Metallicity\dotfill & $0.008\pm0.073$                  &$0.009\pm0.073$\\ 
                    ~~~$v\sin{I_*}$\dotfill &Rotational velocity (m~s$^{-1}$)\dotfill & $56000\pm2000$                   &$56000\pm2000$\\  
                  ~~~$\lambda$\dotfill &Spin-orbit alignment (degrees)\dotfill & $2\pm16$                         &$1\pm15$\\	
\sidehead{Planetary Parameters:}
                                          ~~~$e$\dotfill &Eccentricity\dotfill & $0.0099_{-0.0069}^{+0.010}$      & $--$ \\
               ~~~$\omega_*$\dotfill &Argument of periastron (degrees)\dotfill & $61_{-79}^{+71}$                 & $--$ \\
                                         ~~~$P$\dotfill &Period (days)\dotfill & $1.217514\pm0.000015$            &$1.217513\pm0.000015$\\
                                  ~~~$a$\dotfill &Semi-major axis (AU)\dotfill & $0.02466\pm0.00016$              &$0.02464\pm0.00016$\\
                                        ~~~$M_{P}$\dotfill &Mass (\mj)\dotfill & $27.23_{-0.48}^{+0.50}$          &$27.24_{-0.48}^{+0.49}$\\
                                      ~~~$R_{P}$\dotfill &Radius (\rj)\dotfill & $1.110_{-0.022}^{+0.032}$        &$1.102_{-0.018}^{+0.030}$\\	
                                  ~~~$\rho_{P}$\dotfill &Density (cgs)\dotfill & $24.7_{-1.9}^{+1.4}$             &$25.2_{-1.8}^{+1.2}$\\	  
                             ~~~$\log{g_{P}}$\dotfill &Surface gravity\dotfill & $4.738_{-0.023}^{+0.017}$        &$4.744_{-0.021}^{+0.013}$\\	  
                      ~~~$T_{eq}$\dotfill &Equilibrium temperature (K)\dotfill & $2422_{-26}^{+32}$               &$2414_{-23}^{+29}$\\	
                                  ~~~$\Theta$\dotfill &Safronov number\dotfill & $0.912_{-0.028}^{+0.023}$        &$0.919_{-0.024}^{+0.019}$\\	  
                          ~~~$\fave$\dotfill &Incident flux (\fluxcgs)\dotfill & $7.81_{-0.33}^{+0.42}$           &$7.71_{-0.29}^{+0.38}$\\      
\enddata
\label{tab:physpars}
\end{deluxetable}

\begin{deluxetable}{lcc}
\tabletypesize{\scriptsize}
\tablecaption{Median values and 68\% confidence intervals for the Lighcurve and Radial Velocity Parameters of the KELT-1 System}
\tablehead{\colhead{~~~Parameter} & \colhead{Description (Units)} & \colhead{Value}}
\tablewidth{0pt}
\startdata
\sidehead{RV Parameters:}
              ~~~$T_C$\dotfill &Time of inferior conjunction (\bjdtdb)\dotfill & $2455914.1628_{-0.0022}^{+0.0023}$\\
            
          ~~~$T_{P}$\dotfill &Time of periastron (\bjdtdb)\dotfill & $2455914.07_{-0.26}^{+0.24}$\\
                               ~~~$K$\dotfill &RV semi-amplitude (m~s$^{-1}$)\dotfill & $4239\pm52$\\
                                  ~~~$K_R$\dotfill &RM amplitude (m~s$^{-1}$)\dotfill & $342\pm13$\\
                           ~~~$M_P\sin{i}$\dotfill &Minimum mass (\mj)\dotfill & $27.20_{-0.48}^{+0.49}$\\
                                  ~~~$M_{P}/M_{*}$\dotfill &Mass ratio\dotfill & $0.01964\pm0.00028$\\
                              ~~~$u$\dotfill &RM linear limb darkening\dotfill & $0.5842_{-0.0039}^{+0.0044}$\\
  ~~~$\gamma_{0}$\dotfill &zero point for Orbital RVs (Table~\ref{tab:rvorbit}) (m~s$^{-1}$)\dotfill & $-14200\pm50$~(stat.)$\pm 200$~(sys.)\\
 ~~~$\gamma_{1}$\dotfill &zero point for RM RVs (Table~\ref{tab:rvrm}) (m~s$^{-1}$)\dotfill & $-14200_{-59}^{+56}$~(stat.)$\pm 200$~(sys.)\\
                                                ~~~$\ecosw$\dotfill & \dotfill & $0.0018_{-0.0059}^{+0.0092}$\\
                                                ~~~$\esinw$\dotfill & \dotfill & $0.0041_{-0.0062}^{+0.011}$\\
                            ~~~$f(m1,m2)$\dotfill &Mass function (\mj)\dotfill & $0.01006_{-0.00037}^{+0.00038}$\\
\sidehead{Primary Transit Parameters:}
       ~~~$R_{P}/R_{*}$\dotfill &Radius of the planet in stellar radii\dotfill & $0.07801_{-0.00058}^{+0.00060}$\\
                  ~~~$a/R_*$\dotfill &Semi-major axis in stellar radii\dotfill & $3.626_{-0.080}^{+0.052}$\\
                                 ~~~$i$\dotfill &Inclination (degrees)\dotfill & $87.8_{-1.9}^{+1.3}$\\
                                      ~~~$b$\dotfill &Impact parameter\dotfill & $0.141_{-0.082}^{+0.11}$\\
                                    ~~~$\delta$\dotfill &Transit depth\dotfill & $0.006086_{-0.000089}^{+0.000094}$\\
                           ~~~$T_{FWHM}$\dotfill &FWHM duration (days)\dotfill & $0.10642\pm0.00045$\\
                     ~~~$\tau$\dotfill &Ingress/egress duration (days)\dotfill & $0.00870_{-0.00018}^{+0.00044}$\\
                            ~~~$T_{14}$\dotfill &Total duration (days)\dotfill & $0.11519_{-0.00058}^{+0.00066}$\\
          ~~~$P_{T}$\dotfill &A priori non-grazing transit probability\dotfill & $0.2558_{-0.0052}^{+0.0074}$\\
                     ~~~$P_{T,G}$\dotfill &A priori transit probability\dotfill & $0.2991_{-0.0061}^{+0.0088}$\\
~~~$T_{C,0}$\dotfill &transit time for PvdKO UT 2011-12-03 (\bjdtdb)\dotfill & $2455899.5550\pm0.0010$\\
%            ~~~$F_{0,0}$\dotfill &zero point for n20111204.i.PvdKO.dat\dotfill & $1.00332\pm0.00020$\\
%~~~$C_{0,0}$\dotfill &detrending coefficient for n20111204.i.PvdKO.dat\dotfill & $0.0026\pm0.0021$\\
 ~~~$T_{C,1}$\dotfill &transit time for ULMO UT 2011-12-03 (\bjdtdb)\dotfill & $2455899.55407\pm0.00044$\\
%             ~~~$F_{1,0}$\dotfill &zero point for n20111204.r.ULMO.dat\dotfill & $1.001721\pm0.000097$\\
% ~~~$C_{1,0}$\dotfill &detrending coefficient for n20111204.r.ULMO.dat\dotfill & $-0.00211_{-0.00087}^{+0.00088}$\\
  ~~~$T_{C,2}$\dotfill &transit time for HAO UT 2011-12-10 (\bjdtdb)\dotfill & $2455905.63860_{-0.00083}^{+0.00084}$\\
%              ~~~$F_{2,0}$\dotfill &zero point for n20111209.i.HAO.dat\dotfill & $1.00417\pm0.00018$\\
%  ~~~$C_{2,0}$\dotfill &detrending coefficient for n20111209.i.HAO.dat\dotfill & $-0.0083_{-0.0047}^{+0.0046}$\\
 ~~~$T_{C,3}$\dotfill &transit time for FLWO UT 2011-12-16 (\bjdtdb)\dotfill & $2455911.72553\pm0.00045$\\
%             ~~~$F_{3,0}$\dotfill &zero point for n20111215.z.FLWO.dat\dotfill & $1.003657_{-0.000097}^{+0.000098}$\\
% ~~~$C_{3,0}$\dotfill &detrending coefficient for n20111215.z.FLWO.dat\dotfill & $0.00002\pm0.00028$\\
 ~~~$T_{C,4}$\dotfill &transit time for ULMO UT 2011-12-31 (\bjdtdb)\dotfill & $2455927.55574_{-0.00042}^{+0.00040}$\\
%             ~~~$F_{4,0}$\dotfill &zero point for n20111231.r.ULMO.dat\dotfill & $1.003094_{-0.000084}^{+0.000085}$\\
% ~~~$C_{4,0}$\dotfill &detrending coefficient for n20111231.r.ULMO.dat\dotfill & $-0.00069\pm0.00025$\\
 ~~~$T_{C,5}$\dotfill &transit time for FLWO UT 2012-01-07 (\bjdtdb)\dotfill & $2455933.64321_{-0.00040}^{+0.00041}$\\
%             ~~~$F_{5,0}$\dotfill &zero point for n20120106.i.FLWO.dat\dotfill & $1.00431\pm0.00011$\\
% ~~~$C_{5,0}$\dotfill &detrending coefficient for n20120106.i.FLWO.dat\dotfill & $-0.00070\pm0.00033$\\
           ~~~$u_{1Sloani}$\dotfill &linear limb-darkening coefficient\dotfill & $0.2143_{-0.0041}^{+0.0045}$\\
        ~~~$u_{2Sloani}$\dotfill &quadratic limb-darkening coefficient\dotfill & $0.3152_{-0.0027}^{+0.0035}$\\
           ~~~$u_{1Sloanr}$\dotfill &linear limb-darkening coefficient\dotfill & $0.2856_{-0.0043}^{+0.0050}$\\
        ~~~$u_{2Sloanr}$\dotfill &quadratic limb-darkening coefficient\dotfill & $0.3262_{-0.0023}^{+0.0029}$\\
           ~~~$u_{1Sloanz}$\dotfill &linear limb-darkening coefficient\dotfill & $0.1645_{-0.0040}^{+0.0043}$\\
        ~~~$u_{2Sloanz}$\dotfill &quadratic limb-darkening coefficient\dotfill & $0.3062_{-0.0033}^{+0.0027}$\\
\sidehead{Secondary Eclipse Parameters:}
                         ~~~$T_{S}$\dotfill &Time of eclipse (\bjdtdb)\dotfill & $2455913.5560_{-0.0050}^{+0.0065}$\\
                                  ~~~$b_{S}$\dotfill &Impact parameter\dotfill & $0.142_{-0.083}^{+0.11}$\\
                         ~~~$T_{S,FWHM}$\dotfill &FWHM duration (days)\dotfill & $0.1073_{-0.0014}^{+0.0023}$\\
                   ~~~$\tau_S$\dotfill &Ingress/egress duration (days)\dotfill & $0.00883_{-0.00027}^{+0.00047}$\\
                          ~~~$T_{S,14}$\dotfill &Total duration (days)\dotfill & $0.1162_{-0.0017}^{+0.0026}$\\
          ~~~$P_{S}$\dotfill &A priori non-grazing eclipse probability\dotfill & $0.2525_{-0.0024}^{+0.0053}$\\
                    ~~~$P_{S,G}$\dotfill &A priori eclipse probability\dotfill & $0.2952_{-0.0029}^{+0.0064}$
\enddata
\label{tab:fitpars}
\end{deluxetable}

\begin{deluxetable}{lccrrc}
\tabletypesize{\scriptsize}
\tablecaption{KELT-1 Transit Times\label{tab:ttimes}}
\tablewidth{0pt}
\tablehead{\colhead{Epoch} & \colhead{$T_C$} & \colhead{$\sigma_{T_C}$} & \colhead{O-C} & \colhead{(O-C)/$\sigma_{T_C}$} & \colhead{Obs.}\\
\colhead{~} & \colhead{\bjdtdb} & \colhead{Seconds} & \colhead{Seconds} & \colhead{~} & \colhead{~}}
\startdata
-8 & 2455899.55497 & 87 &  188.12 &  2.14 & PvdKO\\
-8 & 2455899.55407 & 37 &  110.36 &  2.92 & ULMO\\
-3 & 2455905.63859 & 72 & -146.99 & -2.03 & HAO\\
2  & 2455911.72553 & 38 & -195.85 & -5.06 & FLWO\\
15 & 2455927.55574 & 35 &   37.16 &  1.05 & ULMO\\
20 & 2455933.64320 & 35 &   33.84 &  0.96 & FLWO\\
\enddata
\end{deluxetable}

\begin{deluxetable}{rrrrr}
\tabletypesize{\scriptsize}
\tablecaption{KELT-1 Orbital Radial Velocities and Bisectors\label{tab:rvorbit}}
\tablewidth{0pt}
\tablehead{\colhead{\bjdtdb} & \colhead{RV} & \colhead{$\sigma_{RV}$} & \colhead{Bisector} & \colhead{$\sigma_{BS}$}\\
\colhead{~} & \colhead{(m s$^{-1}$)} & \colhead{(m s$^{-1}$)} & \colhead{(m s$^{-1}$)} & \colhead{(m s$^{-1}$)}}
\startdata
2455874.861355  &  -10308.9  &  193.3  &  103.9  &  152.2 \\
2455876.751353  &  -18146.0  &  182.9  &  121.5  &  106.7 \\
2455884.783908  &  -10930.2  &  275.9  &  189.6  &  90.1 \\
2455885.723564  &  -10517.5  &  238.3  &  56.0  &  140.3 \\
2455887.750071  &  -18279.3  &  213.7  &  75.6  &  130.1 \\
2455888.709891  &  -16518.6  &  173.4  &  -61.1  &  117.0 \\
2455900.658271  &  -11722.5  &  222.0  &  -55.7  &  84.4 \\
2455901.773451  &  -10634.7  &  366.7  &  -41.6  &  108.3 \\
2455904.764448  &  -18416.8  &  180.2  &  -29.4  &  124.5 \\
2455905.598058  &  -13114.1  &  190.2  &  127.0  &  74.8 \\
2455911.618841  &  -11901.7  &  240.3  &  165.9  &  84.3 \\
2455912.633163  &  -10250.0  &  325.6  &  -33.0  &  160.8 \\
2455930.604440  &  -13938.2  &  216.2  &  67.7  &  74.6 \\
2455931.588925  &  -18257.4  &  210.1  &  131.8  &  46.3 \\
2455932.584867  &  -17277.9  &  190.9  &  -158.3  &  75.2 \\
2455934.572100  &  -10043.3  &  248.2  &  -53.4  &  82.5 \\
2455935.671466  &  -11237.1  &  334.9  &  -293.1  &  130.6 \\
2455936.607043  &  -16406.3  &  307.3  &  -34.7  &  209.3 \\
2455937.570378  &  -18299.2  &  156.2  &  4.3  &  68.5 \\
2455940.565265  &  -10308.6  &  224.9  &  -210.3  &  102.9 \\
2455957.622238  &  -10008.5  &  267.8  &  -123.9  &  97.2 \\
2455960.657124  &  -17869.8  &  297.0  &  10.2  &  171.4 \\
2455962.599920  &  -9972.1  &  235.1  &  9.8  &  121.1 \\
\enddata
\tablecomments{Radial velocities and bisector spans with uncertainties
for spectra taken with TRES at FLWO on nights outside of transit.  
Times are for mid-exposure and are in the
\bjdtdb \ standard \citep{eastman10}.  RVs are approximately on an
absolute scale, and the uncertainties have been scaled based on a
global fit to the data.  See \S\ref{sec:flwos} and
\S\ref{sec:analysis} for details.
}
\end{deluxetable}

\begin{deluxetable}{rrrrr}
\tabletypesize{\scriptsize}
\tablecaption{KELT-1 RM Radial Velocities and Bisectors\label{tab:rvrm}}
\tablewidth{0pt}
\tablehead{\colhead{\bjdtdb} & \colhead{RV} & \colhead{$\sigma_{RV}$} & \colhead{Bisector} & \colhead{$\sigma_{BS}$}\\
\colhead{~} & \colhead{(m s$^{-1}$)} & \colhead{(m s$^{-1}$)} & \colhead{(m s$^{-1}$)} & \colhead{(m s$^{-1}$)}}
\startdata
2455933.561465  &  -12405.3  &  181.6  &  53.5  &  109.8 \\
2455933.565146  &  -12596.4  &  217.0  &  -178.0  &  141.5 \\
2455933.568838  &  -12713.2  &  203.0  &  10.2  &  145.3 \\
2455933.573282  &  -12618.2  &  191.4  &  218.3  &  130.1 \\
2455933.576973  &  -12958.8  &  160.9  &  91.7  &  122.5 \\
2455933.580654  &  -13001.0  &  215.5  &  11.1  &  151.7 \\
2455933.585237  &  -12937.2  &  169.0  &  -78.0  &  124.5 \\
2455933.588917  &  -12761.6  &  153.6  &  -211.0  &  149.3 \\
2455933.592609  &  -12974.8  &  181.8  &  -59.1  &  108.3 \\
2455933.596983  &  -12945.5  &  155.8  &  50.0  &  134.5 \\
2455933.600664  &  -12643.7  &  112.7  &  -74.2  &  81.3 \\
2455933.604355  &  -13110.1  &  191.6  &  -34.8  &  113.6 \\
2455933.608765  &  -13306.8  &  94.0  &  266.9  &  129.1 \\
2455933.612457  &  -12911.2  &  136.3  &  -4.4  &  122.2 \\
2455933.616137  &  -13378.9  &  112.5  &  90.8  &  63.0 \\
2455933.621194  &  -13577.2  &  153.6  &  148.1  &  114.1 \\
2455933.624875  &  -13770.6  &  129.3  &  53.1  &  99.4 \\
2455933.628555  &  -13809.0  &  140.7  &  19.5  &  137.3 \\
2455933.633034  &  -13933.0  &  139.4  &  22.5  &  115.4 \\
2455933.636714  &  -13897.2  &  115.7  &  177.2  &  130.4 \\
2455933.640406  &  -14256.0  &  132.6  &  45.2  &  124.7 \\
2455933.645035  &  -14345.2  &  143.2  &  0.4  &  82.1 \\
2455933.648715  &  -14372.9  &  145.8  &  -20.9  &  88.2 \\
2455933.652396  &  -14371.8  &  173.9  &  -188.0  &  86.4 \\
2455933.656869  &  -14573.1  &  148.5  &  10.8  &  87.1 \\
2455933.660555  &  -14845.4  &  137.6  &  -234.2  &  104.5 \\
2455933.664235  &  -14862.9  &  126.8  &  -117.8  &  71.0 \\
2455933.668887  &  -14976.3  &  136.4  &  -183.6  &  88.4 \\
2455933.672567  &  -15112.1  &  173.7  &  37.9  &  136.2 \\
2455933.676248  &  -15073.5  &  104.0  &  -21.6  &  112.4 \\
2455933.680761  &  -15482.9  &  140.4  &  211.8  &  76.4 \\
2455933.684453  &  -15385.5  &  181.2  &  -82.2  &  198.6 \\
2455933.688133  &  -15349.1  &  141.5  &  52.1  &  108.0 \\
2455933.692820  &  -15614.2  &  156.9  &  164.2  &  135.4 \\
2455933.696501  &  -15427.2  &  206.4  &  -237.5  &  142.5 \\
2455933.700192  &  -15335.8  &  313.0  &  -699.1  &  412.4 \\
2455933.704648  &  -15667.5  &  145.1  &  157.8  &  119.1 \\
2455933.708340  &  -15366.4  &  218.3  &  -54.8  &  133.8 \\
2455933.712020  &  -15815.1  &  219.6  &  181.9  &  114.5 \\
2455933.716464  &  -15824.6  &  220.1  &  -206.9  &  147.1 \\
2455933.720145  &  -16127.7  &  180.2  &  -245.7  &  210.6 \\
2455933.723836  &  -15832.2  &  182.2  &  -33.8  &  160.4 \\
\enddata
\tablecomments{Radial velocities and bisector spans with uncertainties
for spectra taken with TRES at FLWO on the night of the primary
transit on UT 2012-01-07.  Times are for mid-exposure and are in the
\bjdtdb \ standard \citep{eastman10}.  RVs are approximately on an
absolute scale, and the uncertainties have been scaled based on a
global fit to the data.  See \S\ref{sec:flwos} and
\S\ref{sec:analysis} for details.
}
\end{deluxetable}

\begin{deluxetable}{rrr}
\tabletypesize{\scriptsize}
\tablecaption{Relative Photometry from PvdKO on UT 2011-12-03 ($i$) \label{tab:transit0}
}
\tablewidth{0pt}
\tablehead{\colhead{\bjdtdb} & \colhead{Relative Flux} & \colhead{Uncertainty}}
\startdata
2455899.454080  &  1.00112  &  0.00177 \\
2455899.455342  &  0.99986  &  0.00177 \\
2455899.456603  &  0.99606  &  0.00206 \\
2455899.457865  &  0.99556  &  0.00218 \\
2455899.459138  &  1.00147  &  0.00192 \\
2455899.460399  &  0.99946  &  0.00163 \\
2455899.461661  &  0.99686  &  0.00162 \\
2455899.462934  &  1.00004  &  0.00190 \\
2455899.464207  &  1.00294  &  0.00221 \\
2455899.465468  &  0.99780  &  0.00175 \\
\enddata
\tablecomments{This photometry has been corrected for a linear trend
with airmass, and normalized by the fitted out-of-transit flux.  See \S\ref{sec:analysis}.
This table is published in its entirety in the electronic edition.
A portion is shown here for guidance regarding its form and content.}
\end{deluxetable}

\begin{deluxetable}{rrr}
\tabletypesize{\scriptsize}
\tablecaption{Relative Photometry from ULMO on UT 2011-12-03 ($r$) \label{tab:transit1}}
\tablewidth{0pt}
\tablehead{\colhead{\bjdtdb} & \colhead{Relative Flux} & \colhead{Uncertainty}}
\startdata
2455899.475596  &  1.00155  &  0.00091 \\
2455899.476984  &  1.00058  &  0.00091 \\
2455899.478361  &  0.99852  &  0.00089 \\
2455899.479738  &  1.00015  &  0.00091 \\
2455899.481115  &  1.00175  &  0.00089 \\
2455899.482493  &  1.00039  &  0.00089 \\
2455899.483870  &  1.00020  &  0.00089 \\
2455899.485259  &  1.00144  &  0.00089 \\
2455899.486636  &  0.99967  &  0.00089 \\
2455899.488014  &  1.00099  &  0.00091 \\
\enddata
\tablecomments{This photometry has been corrected for a linear trend
with airmass, and normalized by the fitted out-of-transit flux.  See \S\ref{sec:analysis}.
This table is published in its entirety in the electronic edition.
A portion is shown here for guidance regarding its form and content.}
\end{deluxetable}

\begin{deluxetable}{rrr}
\tabletypesize{\scriptsize}
\tablecaption{Relative Photometry from HAO on UT 2011-12-10 ($i$) \label{tab:transit2}}
\tablewidth{0pt}
\tablehead{\colhead{\bjdtdb} & \colhead{Relative Flux} & \colhead{Uncertainty}}
\startdata
2455905.540572  &  0.99833  &  0.00250 \\
2455905.541132  &  0.99667  &  0.00250 \\
2455905.541712  &  1.00321  &  0.00252 \\
2455905.542272  &  1.00209  &  0.00243 \\
2455905.542842  &  1.00116  &  0.00243 \\
2455905.543412  &  0.99931  &  0.00243 \\
2455905.543972  &  1.00078  &  0.00243 \\
2455905.544542  &  0.99829  &  0.00242 \\
2455905.545112  &  0.99994  &  0.00243 \\
2455905.545692  &  0.99800  &  0.00250 \\
\enddata
\tablecomments{This photometry has been corrected for a linear trend
with airmass, and normalized by the fitted out-of-transit flux.  See \S\ref{sec:analysis}.
This table is published in its entirety in the electronic edition.
A portion is shown here for guidance regarding its form and content.}
\end{deluxetable}

\begin{deluxetable}{rrr}
\tabletypesize{\scriptsize}
\tablecaption{Relative Photometry from FLWO on UT 2011-12-16 ($z$) \label{tab:transit3}}
\tablewidth{0pt}
\tablehead{\colhead{\bjdtdb} & \colhead{Relative Flux} & \colhead{Uncertainty}}
\startdata
2455911.612185  &  1.00109  &  0.00154 \\
2455911.612983  &  1.00230  &  0.00154 \\
2455911.613516  &  1.00009  &  0.00154 \\
2455911.614002  &  1.00177  &  0.00154 \\
2455911.614488  &  0.99899  &  0.00154 \\
2455911.615078  &  1.00118  &  0.00154 \\
2455911.615564  &  1.00159  &  0.00154 \\
2455911.616051  &  0.99953  &  0.00154 \\
2455911.616560  &  1.00155  &  0.00154 \\
2455911.617081  &  1.00066  &  0.00154 \\
\enddata
\tablecomments{This photometry has been corrected for a linear trend
with airmass, and normalized by the fitted out-of-transit flux.  See \S\ref{sec:analysis}.
This table is published in its entirety in the electronic edition.
A portion is shown here for guidance regarding its form and content.}
\end{deluxetable}

\begin{deluxetable}{rrr}
\tabletypesize{\scriptsize}
\tablecaption{Relative Photometry from ULMO on UT 2011-12-31 ($r$) \label{tab:transit4}}
\tablewidth{0pt}
\tablehead{\colhead{\bjdtdb} & \colhead{Relative Flux} & \colhead{Uncertainty}}
\startdata
2455927.468704  &  0.99844  &  0.00145 \\
2455927.470335  &  0.99820  &  0.00125 \\
2455927.471713  &  0.99925  &  0.00116 \\
2455927.473090  &  1.00005  &  0.00105 \\
2455927.474467  &  0.99994  &  0.00099 \\
2455927.475844  &  1.00105  &  0.00092 \\
2455927.477233  &  0.99940  &  0.00090 \\
2455927.478610  &  1.00084  &  0.00086 \\
2455927.479987  &  0.99960  &  0.00086 \\
2455927.481365  &  0.99849  &  0.00084 \\
\enddata
\tablecomments{This photometry has been corrected for a linear trend
with airmass, and normalized by the fitted out-of-transit flux.  See \S\ref{sec:analysis}.
This table is published in its entirety in the electronic edition.
A portion is shown here for guidance regarding its form and content.}
\end{deluxetable}

\begin{deluxetable}{rrr}
\tabletypesize{\scriptsize}
\tablecaption{Relative Photometry from FLWO on UT 2012-01-07 ($i$) \label{tab:transit5}}
\tablewidth{0pt}
\tablehead{\colhead{\bjdtdb} & \colhead{Relative Flux} & \colhead{Uncertainty}}
\startdata
2455933.568776  &  1.00164  &  0.00184 \\
2455933.569101  &  0.99893  &  0.00184 \\
2455933.569459  &  1.00032  &  0.00184 \\
2455933.569830  &  1.00046  &  0.00184 \\
2455933.570200  &  0.99947  &  0.00184 \\
2455933.570536  &  1.00032  &  0.00184 \\
2455933.570871  &  1.00131  &  0.00184 \\
2455933.571195  &  1.00074  &  0.00184 \\
2455933.571531  &  0.99764  &  0.00184 \\
2455933.571867  &  1.00003  &  0.00184 \\
\enddata
\tablecomments{This photometry has been corrected for a linear trend
with airmass, and normalized by the fitted out-of-transit flux.  See \S\ref{sec:analysis}.
This table is published in its entirety in the electronic edition.
A portion is shown here for guidance regarding its form and content.}
\end{deluxetable}

\begin{deluxetable}{rrr}
\tabletypesize{\scriptsize}
\tablecaption{Relative Photometry from ULMO on UT 2011-12-02 ($i$) \label{tab:second0}}
\tablewidth{0pt}
\tablehead{\colhead{\bjdtdb} & \colhead{Relative Flux} & \colhead{Uncertainty}}
\startdata
2455897.65499  &  1.00097  &  0.00145  \\
2455897.65721  &  0.99944  &  0.00134  \\
2455897.65859  &  0.99993  &  0.00140  \\
2455897.65997  &  0.99930  &  0.00141  \\
2455897.66134  &  1.00078  &  0.00140  \\
2455897.66273  &  1.00023  &  0.00138  \\
2455897.66411  &  1.00137  &  0.00134  \\
2455897.66549  &  0.99971  &  0.00138  \\
2455897.66686  &  0.99978  &  0.00140  \\
2455897.66824  &  0.99970  &  0.00140  \\
\enddata
\tablecomments{This photometry has been normalized and corrected for a linear trend
with time.  See \S\ref{sec:second}.
This table is published in its entirety in the electronic edition.
A portion is shown here for guidance regarding its form and content.}
\end{deluxetable}

\begin{deluxetable}{rrr}
\tabletypesize{\scriptsize}
\tablecaption{Relative Photometry from FTN/LCOGT on UT 2011-12-30 (PS-$Z$) \label{tab:second1}}
\tablewidth{0pt}
\tablehead{\colhead{\bjdtdb} & \colhead{Relative Flux} & \colhead{Uncertainty}}
\startdata
2455925.730250  &  0.99855  &  0.00078  \\
2455925.731230  &  1.00045  &  0.00078  \\
2455925.732210  &  0.99953  &  0.00078  \\
2455925.733190  &  1.00032  &  0.00078  \\
2455925.734190  &  1.00069  &  0.00078  \\
2455925.735170  &  1.00000  &  0.00078  \\
2455925.739770  &  1.00089  &  0.00078  \\
2455925.740700  &  0.99925  &  0.00078  \\
2455925.741670  &  0.99874  &  0.00078  \\
2455925.742660  &  0.99933  &  0.00078  \\
\enddata
\tablecomments{This photometry has been normalized and corrected for a linear trend
with time.  See \S\ref{sec:second}.
This table is published in its entirety in the electronic edition.
A portion is shown here for guidance regarding its form and content.}
\end{deluxetable}

\begin{deluxetable}{rrr}
\tabletypesize{\scriptsize}
\tablecaption{Relative Photometry from ULMO on UT 2012-01-04 ($i$) \label{tab:second2}}
\tablewidth{0pt}
\tablehead{\colhead{\bjdtdb} & \colhead{Relative Flux} & \colhead{Uncertainty}}
\startdata
2455930.50268  &  1.00020  &  0.00111  \\
2455930.50487  &  1.00126  &  0.00110  \\
2455930.50624  &  0.99911  &  0.00110  \\
2455930.50762  &  0.99970  &  0.00111  \\
2455930.50901  &  1.00139  &  0.00116  \\
2455930.51039  &  1.00146  &  0.00111  \\
2455930.51177  &  0.99958  &  0.00111  \\
2455930.51314  &  1.00016  &  0.00111  \\
2455930.51452  &  1.00098  &  0.00111  \\
2455930.51590  &  1.00085  &  0.00111  \\
\enddata
\tablecomments{This photometry has been normalized and corrected for a linear trend
with time.  See \S\ref{sec:second}.
This table is published in its entirety in the electronic edition.
A portion is shown here for guidance regarding its form and content.}
\end{deluxetable}

\clearpage

\begin{thebibliography}{}

\bibitem[Agol et al.(2005)]{agol05} Agol, E., Steffen, J., 
Sari, R., \& Clarkson, W.\ 2005, \mnras, 359, 567 

\bibitem[Agol \& Steffen(2006)]{agol06}
Agol, E., \& Steffen, J.\ 2006, MNRAS, 374, 941

\bibitem[Alard \& Lupton (1998)]{ISIS}
Alard, C. \& Lupton, R. H.\ 1998, \apj, 503, 325

\bibitem[Alard(2000)]{alard2000} Alard, C.\ 2000, \aaps, 144, 363 

\bibitem[Alonso et al.(2004)]{alonso2004} Alonso, R., Brown, 
T.~M., Torres, G., et al.\ 2004, \apjl, 613, L153 

\bibitem[Alsubai et al.(2011)]{alsubai2011} Alsubai, K.~A., Parley, 
N.~R., Bramich, D.~M., et al.\ 2011, \mnras, 417, 709 

\bibitem[Anderson et al.(2011)]{anderson2011} Anderson, D.~R., 
Collier Cameron, A., Hellier, C., et al.\ 2011, \apjl, 726, L19 

\bibitem[Anderson et al.(2012)]{anderson2012} Anderson, D.~R., 
Collier Cameron, A., Gillon, M., et al.\ 2012, \mnras, 422, 1988 

\bibitem[Baglin(2003)]{baglin2003} Baglin, A.\ 2003, Advances in 
Space Research, 31, 345 

\bibitem[Bakos et al.(2007)]{bakos2007} Bakos, G.~{\'A}., Noyes, 
R.~W., Kov{\'a}cs, G., et al.\ 2007, \apj, 656, 552 

\bibitem[Bakos et al.(2011)]{bakos2011} Bakos, G.~{\'A}., 
Hartman, J., Torres, G., et al.\ 2011, \apj, 742, 116 

\bibitem[Bakos et al.(2012)]{bakos2012} Bakos, G.~{\'A}., 
Hartman, J.~D., Torres, G., et al.\ 2012, arXiv:1201.0659 

\bibitem[Baraffe et al.(2003)]{baraffe2003} 
Baraffe, I., Chabrier, G., Barman, T.~S., Allard, F., \& Hauschildt, P.~H.\ 2003, \aap, 402, 701 

\bibitem[Barnes(2009)]{barnes2009} Barnes, J.~W.\ 2009, \apj, 705, 
683 

\bibitem[Barnes \& Fortney(2004)]{barnes04} Barnes, J.~W., \& 
Fortney, J.~J.\ 2004, \apj, 616, 1193 

\bibitem[Batalha et al.(2012)]{batalha2012} Batalha, N.~M., Rowe, 
J.~F., Bryson, S.~T., et al.\ 2012, arXiv:1202.5852 

\bibitem[Beatty et al.(2007)]{beatty2007} Beatty, T.~G., 
Fern{\'a}ndez, J.~M., Latham, D.~W., et al.\ 2007, \apj, 663, 573 

\bibitem[Bensby et 
al.(2003)]{bensby2003} Bensby, T., Feltzing, S., \& Lundstr{\"o}m, I.\ 2003, \aap, 410, 527 

\bibitem[Borucki et al.(2010)]{borucki2010} Borucki, W.~J., Koch, 
D., Basri, G., et al.\ 2010, Science, 327, 977 

\bibitem[Bertin \& Arnouts (1996)]{SExtractor}
Bertin, E. \& Arnouts, S.\ 1996,  A\&AS, 117, 393B

\bibitem[Bouchy et 
al.(2011a)]{bouchy2011a} Bouchy, F., Deleuil, M., Guillot, T., et al.\ 2011a, \aap, 525, A68 

\bibitem[Bouchy et 
al.(2011b)]{bouchy2011b} Bouchy, F., Bonomo, A.~S., Santerne, A., et al.\ 2011b, \aap, 533, A83 

\bibitem[Brown et al.(2001)]{brown01} Brown, T.~M., 
Charbonneau, D., Gilliland, R.~L., Noyes, R.~W., \& Burrows, A.\ 2001, 
\apj, 552, 699 

\bibitem[Buchhave et al.(2010)]{buchave2010} Buchhave, L.~A., 
Bakos, G.~{\'A}., Hartman, J.~D., et al.\ 2010, \apj, 720, 1118 

\bibitem[Burke et al.(2006)]{burke2006} Burke, C.~J., Gaudi, 
B.~S., DePoy, D.~L., \& Pogge, R.~W.\ 2006, \aj, 132, 210 

\bibitem[Burrows et al.(1997)]{burrows1997} Burrows, A., Marley, 
M., Hubbard, W.~B., et al.\ 1997, \apj, 491, 856 

\bibitem[Burrows et al.(2008)]{burrows2008} Burrows, A., Ibgui, L., 
\& Hubeny, I.\ 2008, \apj, 682, 1277 


\bibitem[Butters et 
al.(2010)]{butters2010} Butters, O.~W., West, R.~G., Anderson, D.~R., et al.\ 2010, \aap, 520, L10 

\bibitem[Carter 
\& Winn(2009)]{carter2009} Carter, J.~A., \& Winn, J.~N.\ 2009, \apj, 704, 51 

\bibitem[Carter 
\& Winn(2010)]{carter2010} Carter, J.~A., \& Winn, J.~N.\ 2010, \apj, 709, 1219 

\bibitem[Carter et al.(2011)]{carter2011} Carter, J.~A., Winn, 
J.~N., Holman, M.~J., et al.\ 2011, \apj, 730, 82 

\bibitem[Charbonneau et al.(2002)]{charbonneau02} Charbonneau, D., 
Brown, T.~M., Noyes, R.~W., \& Gilliland, R.~L.\ 2002, \apj, 568, 377 

\bibitem[Charbonneau et al.(2005)]{charbonneau2005} Charbonneau, D., 
Allen, L.~E., Megeath, S.~T., et al.\ 2005, \apj, 626, 523 

\bibitem[{{Claret} \& {Bloemen}(2011)}]{claret11}
{Claret}, A., \& {Bloemen}, S. 2011, \aap, 529, A75+

\bibitem[Collier Cameron et al.(2007a)]{cameron2007a} Collier 
Cameron, A., Bouchy, F., H{\'e}brard, G., et al.\ 2007a, \mnras, 375, 951 

\bibitem[Collier Cameron et al.(2007b)]{cameron2007b} 
Collier Cameron, A., Wilson, D.~M., West, R.~G., et al.\ 2007b, \mnras, 380, 1230 

\bibitem[Collier Cameron et al.(2010)]{cameron2010} %WASP33
Collier Cameron, A., Guenther, E., Smalley, B., et al.\ 2010, \mnras, 407, 507 

\bibitem[Cody 
\& Sasselov(2002)]{cody2002} Cody, A.~M., \& Sasselov, D.~D.\ 2002, \apj, 569, 451 

\bibitem[Cowan et al.(2012)]{cowan2012} Cowan, N.~B., Machalek, 
P., Croll, B., et al.\ 2012, \apj, 747, 82 


\bibitem[Cumming et al.(2008)]{cumming08} Cumming, A., Butler, 
R.~P., Marcy, G.~W., et al.\ 2008, \pasp, 120, 531 

\bibitem[Co{\c s}kuno{\v g}lu et al.(2011)]{coskunoglu2011} Co{\c 
s}kuno{\v g}lu, B., Ak, S., Bilir, S., et al.\ 2011, \mnras, 412, 1237 


\bibitem[Cutri et al.(2003)]{cutri2003} Cutri, R.~M., Skrutskie, 
M.~F., van Dyk, S., et al.\ 2003, VizieR Online Data Catalog, 2246, 0 

\bibitem[Cutri 
\& et al.(2012)]{cutri2012} Cutri, R.~M., \& et al.\ 2012, VizieR Online Data Catalog, 2311, 0 

\bibitem[Darwin(1897)]{darwin1879}
Darwin, G.H.\ 1879, Proc.\ R.\ Soc.\ Long., 29, 168

\bibitem[Deeg et al.(2010)]{deeg2010} Deeg, H.~J., Moutou, C., 
Erikson, A., et al.\ 2010, \nat, 464, 384 

\bibitem[Deleuil et al.(2008)]{deleuil2008} 
Deleuil, M., Deeg, H.~J., Alonso, R., et al.\ 2008, \aap, 491, 889 

\bibitem[Devor(2005)]{devor2005} Devor, J.\ 2005, \apj, 628, 411 

\bibitem[Demarque et al.(2004)]{demarque2004} %Y^2 isochrones
Demarque, P., Woo,  J.-H., Kim, Y.-C., \& Yi, S.~K.\ 2004, \apjs, 155, 667 

\bibitem[Deming et al.(2005)]{deming05}
Deming, D., Seager, S., Richardson, L.~J., \& Harrington, J.\ 2005, Nature, 434, 740

\bibitem[Dhital et al.(2010)]{dhital2010} Dhital, S., West, A.~A., 
Stassun, K.~G., \& Bochanski, J.~J.\ 2010, \aj, 139, 2566 


\bibitem[Dodson-Robinson et al.(2009)]{dodson2009} 
Dodson-Robinson, S.~E., Veras, D., Ford, E.~B., 
\& Beichman, C.~A.\ 2009, \apj, 707, 79 


\bibitem[Doyle et al.(2011)]{doyle2011} Doyle, L.~R., Carter, 
J.~A., Fabrycky, D.~C., et al.\ 2011, Science, 333, 1602  

\bibitem[{{Eastman} {et~al.}(2010{\natexlab{b}}){Eastman},{Siverd}, \&{Gaudi}}]{eastman10}
{Eastman}, J., {Siverd}, R., \& {Gaudi}, B.~S. 2010{\natexlab{b}},
\pasp, 122, 935

\bibitem[Eastman, Gaudi \& Agol(2012)]{eastman12}
{Eastman}, J., {Gaudi}, B.~S., {Agol}, E. 2012, in preparation

\bibitem[Fabrycky 
\& Tremaine(2007)]{fabrycky2007} Fabrycky, D., \& Tremaine, S.\ 2007, \apj, 669, 1298 


\bibitem[Faigler 
\& Mazeh(2011)]{faigler2011} Faigler, S., \& Mazeh, T.\ 2011, \mnras, 415, 3921 

\bibitem[Fleming et al.(2012)]{fleming2012}
Fleming, S., et al.\ 2012, ApJ, submitted

\bibitem[Ford \& Gaudi(2006)]{ford06}
Ford, E.B., \& Gaudi, B.S.\ 2006, ApJ, 652, L137

\bibitem[Ford \& Holman(2007)]{ford07} 
Ford, E.~B., \& Holman, M.~J.\ 2007, \apjl, 664, L51 

\bibitem[Fortney et al.(2007)]{fortney2007} Fortney, J.~J., Marley, 
M.~S., \& Barnes, J.~W.\ 2007, \apj, 659, 1661 

\bibitem[Fortney 
\& Nettelmann(2010)]{fortney2010} Fortney, J.~J., \& Nettelmann, N.\ 2010, \ssr, 152, 423 

\bibitem[Fressin et 
al.(2009)]{fressin2009} Fressin, F., Guillot, T., \& Nesta, L.\ 2009, \aap, 504, 605 

\bibitem[Fressin et al.(2012)]{fressin2012} Fressin, F., Torres, 
G., Rowe, J.~F., et al.\ 2012, \nat, 482, 195 

\bibitem[F\H{u}r\'{e}sz(2008)]{furesz2008}
F\H{u}r\'{e}sz, G. 2008, PhD thesis, Univ. Szeged

\bibitem[Gaudi et al.(2005)]{gaudi2005} Gaudi, B.~S., Seager, S., 
\& Mallen-Ornelas, G.\ 2005, \apj, 623, 472 


\bibitem[Gaudi \& Winn(2006)]{gaudi06}
Gaudi, B.S., \& Winn, J.N.\ 2006, ApJ, 655, 550 

\bibitem[Ghez et al.(2008)]{ghez2008} Ghez, A.~M., Salim, S., 
Weinberg, N.~N., et al.\ 2008, \apj, 689, 1044 

\bibitem[G{\'o}mez Maqueo Chew et al.(2009)]{gmc2009} G{\'o}mez 
Maqueo Chew, Y., Stassun, K.~G., Pr{\v s}a, A., 
\& Mathieu, R.~D.\ 2009, \apj, 699, 1196 

\bibitem[Gould \& Morgan(2003)]{gould2003} 
Gould, A., \& Morgan, C.~W.\ 2003, \apj, 585, 1056 

\bibitem[Guenther et al.(1992)]{guenther1992} Guenther, D.~B., 
Demarque, P., Kim, Y.-C., \& Pinsonneault, M.~H.\ 1992, \apj, 387, 372 

\bibitem[Guillot(2005)]{guillot05} 
Guillot, T.\ 2005, Annual 
Review of Earth and Planetary Sciences, 33, 493

\bibitem[Gould et al.(2006)]{hot_jupiters}
Gould, A., Dorsher, S., Gaudi, B. S., \& Udalski, A.\ 2006, Acta Astron., 56, 1

\bibitem[Gould \& Morgan (2003)]{RPM_selection}
Gould, A. \& Morgan, C.\ 2003, \aj, 585, 1056

\bibitem[Grether 
\& Lineweaver(2006)]{grether2006} Grether, D., \& Lineweaver, C.~H.\ 2006, \apj, 640, 1051 


\bibitem[Hansen 
\& Barman(2007)]{handsen2007} Hansen, B.~M.~S., \& Barman, T.\ 2007, \apj, 671, 861 

\bibitem[Hansen(2010)]{hansen2010} Hansen, B.~M.~S.\ 2010, \apj, 
723, 285 

\bibitem[Hartman et al.(2004)]{hartman2004} Hartman, J.~D., Bakos, 
G., Stanek, K.~Z., \& Noyes, R.~W.\ 2004, \aj, 128, 1761 

\bibitem[Hartman et al.(2008)]{hartman2008} Hartman, J.~D., Gaudi, B.~S., Holman, M.~J., et al.\ 2008, \apj, 675, 1254 

\bibitem[Hartman et al.(2009)]{hartman2009} Hartman, J.~D., Gaudi, 
B.~S., Holman, M.~J., et al.\ 2009, \apj, 695, 336 


\bibitem[Hartman et al.(2011)]{hartman2011} Hartman, J.~D., Bakos, 
G.~{\'A}., Sato, B., et al.\ 2011, \apj, 726, 52 

\bibitem[Hauschildt et al.(1999)]{hauschildt1999} Hauschildt, P.~H., 
Allard, F., \& Baron, E.\ 1999, \apj, 512, 377 

\bibitem[Hellier et al.(2009)]{hellier2009} Hellier, C., Anderson, 
D.~R., Collier Cameron, A., et al.\ 2009, \nat, 460, 1098 


\bibitem[H{\o}g et 
al.(1998)]{hog1998} H{\o}g, E., Kuzmin, A., Bastian, U., et al.\ 1998, \aap, 335, L65 

\bibitem[H{\o}g et 
al.(2000)]{hog2000} H{\o}g, E., Fabricius, C., Makarov, V.~V., et al.\ 2000, \aap, 355, L27 

\bibitem[Holman et al.(1997)]{holman1997} Holman, M., Touma, J., 
\& Tremaine, S.\ 1997, \nat, 386, 254 


\bibitem[Holman \& Murray(2005)]{holman05} Holman, M.~J., \& 
Murray, N.~W.\ 2005, Science, 307, 1288 

\bibitem[Holman et al.(2010)]{holman2010} Holman, M.~J., Fabrycky, 
D.~C., Ragozzine, D., et al.\ 2010, Science, 330, 51 

\bibitem[Hut(1980)]{hut1980} Hut, P.\ 1980, \aap, 92, 167 

\bibitem[Hut(1981)]{hut1981} Hut, P.\ 1981, \aap, 99, 126 


\bibitem[Irwin et al.(2010)]{irwin2010} Irwin, J., Buchhave, L., 
Berta, Z.~K., et al.\ 2010, \apj, 718, 1353 

\bibitem[Jackson et al.(2008)]{jackson2008} Jackson, B., Greenberg, 
R., \& Barnes, R.\ 2008, \apj, 678, 1396 


\bibitem[Johnson 
\& Soderblom(1987)]{johnson1987} Johnson, D.~R.~H., \& Soderblom, D.~R.\ 1987, \aj, 93, 864 


\bibitem[Johnson et al.(2007)]{johnson2007} Johnson, J.~A., 
Fischer, D.~A., Marcy, G.~W., et al.\ 2007, \apj, 665, 785 

\bibitem[Johnson et al.(2011)]{johnson2011} Johnson, J.~A., Apps, 
K., Gazak, J.~Z., et al.\ 2011, \apj, 730, 79 

\bibitem[Johns-Krull et al.(2008)]{johnskrull2008} Johns-Krull, C.~M., 
McCullough, P.~R., Burke, C.~J., et al.\ 2008, \apj, 677, 657 


\bibitem[Kane et al.(2009)]{kane2009} Kane, S.~R., Mahadevan, 
S., von Braun, K., Laughlin, G., \& Ciardi, D.~R.\ 2009, \pasp, 121, 1386 
\bibitem[Katz et al.(2011)]{katz2011} Katz, B., Dong, S., 
\& Malhotra, R.\ 2011, Physical Review Letters, 107, 181101 

\bibitem[Kenyon 
\& Hartmann(1995)]{kenyon1995} Kenyon, S.~J., \& Hartmann, L.\ 1995, \apjs, 101, 117 

\bibitem[Kipping(2009)]{kipping2009} Kipping, D.~M.\ 2009, \mnras, 
392, 181 

\bibitem[Kirkpatrick et al.(2012)]{kirkpatrick2012} Kirkpatrick, J.~D., 
Gelino, C.~R., Cushing, M.~C., et al.\ 2012, arXiv:1205.2122 


\bibitem[Knutson et al.(2008)]{knutson2008} Knutson, H.~A., 
Charbonneau, D., Allen, L.~E., Burrows, A., 
\& Megeath, S.~T.\ 2008, \apj, 673, 526 

\bibitem[Kov{\'a}cs et al.(2002)]{kovacs2002} Kov{\'a}cs, G., Zucker, S., \& Mazeh, T.\ 2002, \aap, 391, 369 

\bibitem[Kov{\'a}cs et al.(2005)]{kovacs2005} Kov{\'a}cs, G., Bakos, G., \& Noyes, R.~W.\ 2005, \mnras, 356, 557 

\bibitem[Kov{\'a}cs et al.(2012)]{kovacs2012} Kov{\'a}cs, G., 
Kov{\'a}cs, T., Hartman, J.~D., et al.\ 2012, arXiv:1205.5060 


\bibitem[Kozai(1962)]{kozai1962} Kozai, Y.\ 1962, \aj, 67, 591 

\bibitem[Kratter et al.(2010)]{kratter2010} Kratter, K.~M., 
Murray-Clay, R.~A., \& Youdin, A.~N.\ 2010, \apj, 710, 1375 

\bibitem[Kraus 
\& Hillenbrand(2007)]{kraus2007} Kraus, A.~L., \& Hillenbrand, L.~A.\ 2007, \aj, 134, 2340 

\bibitem[Kurucz(1979)]{kurucz1979} Kurucz, R.~L.\ 1979, \apjs, 40, 
1 

\bibitem[Lang et al.(2010)]{lang2010} Lang, D., Hogg, D.~W., 
Mierle, K., Blanton, M., \& Roweis, S.\ 2010, \aj, 139, 1782 


\bibitem[Latham et al.(2011)]{latham2011} Latham, D.~W., Rowe, 
J.~F., Quinn, S.~N., et al.\ 2011, \apjl, 732, L24 


\bibitem[Lee et al.(2011)]{lee2011} Lee, B.~L., Ge, J., 
Fleming, S.~W., et al.\ 2011, \apj, 728, 32 


\bibitem[L{\'e}ger et 
al.(2009)]{leger2009} L{\'e}ger, A., Rouan, D., Schneider, J., et al.\ 2009, \aap, 506, 287 

\bibitem[Leggett et al.(2002)]{leggett2002} Leggett, S.~K., 
Golimowski, D.~A., Fan, X., et al.\ 2002, \apj, 564, 452 

\bibitem[Leconte et 
al.(2010)]{leconte2010} Leconte, J., Chabrier, G., Baraffe, I., \& Levrard, B.\ 2010, \aap, 516, A64 

\bibitem[Lithwick 
\& Naoz(2011)]{lithwick2011} Lithwick, Y., \& Naoz, S.\ 2011, \apj, 742, 94 

\bibitem[Loeb 
\& Gaudi(2003)]{loeb2003} Loeb, A., \& Gaudi, B.~S.\ 2003, \apjl, 588, L117 

\bibitem[Lidov(1962)]{lidov1962} Lidov, M.~L.\ 1962, \planss, 9, 719 

\bibitem[Lissauer et al.(2011)]{lissauer2011} Lissauer, J.~J., 
Fabrycky, D.~C., Ford, E.~B., et al.\ 2011, \nat, 470, 53 

\bibitem[Lomb(1976)]{lomb76}Lomb, N.R., 1976, \apss, 39, 447

\bibitem[Mandel \& Agol(2002)]{mandel2002} 
Mandel, K., \& Agol, E.\ 2002, \apjl, 580, L171 

\bibitem[Mandushev et al.(2005)]{mandushev2005} Mandushev, G., 
Torres, G., Latham, D.~W., et al.\ 2005, \apj, 621, 1061 

\bibitem[Marcy 
\& Butler(2000)]{marcy2000} Marcy, G.~W., \& Butler, R.~P.\ 2000, \pasp, 112, 137 

\bibitem[Matsumura et al.(2010)]{matsumura2010} Matsumura, S., Peale, 
S.~J., \& Rasio, F.~A.\ 2010, \apj, 725, 1995 

\bibitem[Mazeh 
\& Faigler(2010)]{mazeh2010} Mazeh, T., \& Faigler, S.\ 2010, \aap, 521, L59 

\bibitem[Mazeh et 
al.(2012)]{mazeh2012} Mazeh, T., Nachmani, G., Sokol, G., Faigler, S., \& Zucker, S.\ 2012, \aap, 541, A56 

\bibitem[Metzger et al.(2012)]{metzger2012} Metzger, B.~D., 
Giannios, D., \& Spiegel, D.~S.\ 2012, arXiv:1204.0796 

\bibitem[Mislis 
\& Hodgkin(2012)]{mislis2012} Mislis, D., \& Hodgkin, S.\ 2012, \mnras, 422, 1512 

\bibitem[McCullough et al.(2006)]{mccullough2006} McCullough, P.~R., 
Stys, J.~E., Valenti, J.~A., et al.\ 2006, \apj, 648, 1228 

\bibitem[McLaughlin(1924)]{mclaughlin1924} McLaughlin D. B. 1924, \apj, 60, 22

\bibitem[Metchev  \& Hillenbrand(2009)]{metchev2009} 
Metchev, S.~A., \& Hillenbrand, L.~A.\ 2009, \apjs, 181, 62 

\bibitem[Miller \& Fortney(2011)]{miller2011} 
Miller, N., \& Fortney, J.~J.\ 2011, \apjl, 736, L29 

\bibitem[Mohanty et al.(2010)]{mohanty2010} Mohanty, S., Stassun, 
K.~G., \& Doppmann, G.~W.\ 2010, \apj, 722, 1138 

\bibitem[Mohanty \& Stassun(2012)]{mohanty2012}
Mohanty, S., \& Stassun, K.G. 2012, \apj, submitted.

\bibitem[Mordasini et al.(2009)]{mordasini2009} 
Mordasini, C., Alibert, Y., Benz, W., \& Naef, D.\ 2009, \aap, 501, 1161 

\bibitem[Muirhead et al.(2012)]{muirhead2012} Muirhead, P.~S., 
Johnson, J.~A., Apps, K., et al.\ 2012, \apj, 747, 144 

\bibitem[Mullan 
\& MacDonald(2010)]{mullan2010} Mullan, D.~J., \& MacDonald, J.\ 2010, \apj, 713, 1249 

\bibitem[Naoz et al.(2011)]{naoz2011} Naoz, S., Farr, W.~M., 
Lithwick, Y., Rasio, F.~A., \& Teyssandier, J.\ 2011, \nat, 473, 187 

\bibitem[{Nelder \& Mead(1965)}]{nelder65}
Nelder, J.~A., \& Mead, R. 1965, The Computer Journal, 7, 308

\bibitem[Ogilvie 
\& Lin(2004)]{ogilvie2004} Ogilvie, G.~I., \& Lin, D.~N.~C.\ 2004, \apj, 610, 477 



\bibitem[{{Ohta} {et~al.}(2005){Ohta}, {Taruya}, \& {Suto}}]{ohta05}
{Ohta}, Y., {Taruya}, A., \& {Suto}, Y. 2005, \apj, 622, 1118

\bibitem[O'Donovan et al.(2006)]{odonovan2006} O'Donovan, F.~T., 
Charbonneau, D., Torres, G., et al.\ 2006, \apj, 644, 1237 

\bibitem[Pepper et al. (2003)]{KELT_THEORY}
Pepper, J., Gould, A., \& Depoy, D. L.\ 2003, Acta Astronomica, 53, 213

\bibitem[Pepper et al. (2007)]{KELT_SYNOPTIC}
Pepper, J., Pogge, R.W., DePoy, D.L., Marshall, J.L., Stanek, K.Z.,
Stutz, A.M., Poindexter, S., Siverd, R., O'Brien, T.P.,
Trueblood, M., \& Trueblood, P.\ 2007, \pasp, 119, 923

\bibitem[Pont et al.(2005)]{pont2005} Pont, F., Melo, C.~H.~F., Bouchy, F., et al.\ 2005, \aap, 433, L21 

\bibitem[Pont et al.(2006a)]{pont2006} Pont, F., Zucker, S., 
\& Queloz, D.\ 2006a, \mnras, 373, 231 

\bibitem[Pont et al.(2006b)]{pont2006b} 
Pont, F., Moutou, C., Bouchy, F., et al.\ 2006b, \aap, 447, 1035 

\bibitem[Quinn et al.(2012)]{quinn2012} Quinn, S.~N., Bakos, 
G.~{\'A}., Hartman, J., et al.\ 2012, \apj, 745, 80 

\bibitem[Rafikov(2005)]{rafikov2005} Rafikov, R.~R.\ 2005, \apjl, 
621, L69 

\bibitem[Rasio et al.(1996)]{rasio1996} Rasio, F.~A., Tout, 
C.~A., Lubow, S.~H., \& Livio, M.\ 1996, \apj, 470, 1187 

\bibitem[Rogers 
\& Seager(2010)]{rogers2010} Rogers, L.~A., \& Seager, S.\ 2010, \apj, 712, 974 

\bibitem[Rossiter(1924)]{rossiter1924} Rossiter, R. A. 1924, \apj, 60, 15

\bibitem[Rowe et al.(2010)]{rowe2010} Rowe, J.~F., Borucki, 
W.~J., Koch, D., et al.\ 2010, \apjl, 713, L150 

\bibitem[Sahlmann et 
al.(2011)]{sahlmann2011} Sahlmann, J., S{\'e}gransan, D., Queloz, D., et al.\ 2011, \aap, 525, A95 

\bibitem[Scargle(1982)]{scargle82} Scargle, J.D., 1982, \apj, 263, 835

\bibitem[Schneider et 
al.(2011)]{schneider2011} Schneider, J., Dedieu, C., Le Sidaner, P., Savalle, R., \& Zolotukhin, I.\ 2011, \aap, 532, A79 

\bibitem[Schwarzenberg-Czerny(1989)]{AoV} 
Schwarzenberg-Czerny, A.\ 1989, \mnras, 241, 153 


\bibitem[Seager  \& Sasselov(2000)]{seager2000} 
Seager, S., \& Sasselov, D.~D.\ 2000, \apj, 537, 916 

\bibitem[Seager \& Hui(2002)]{seager02} Seager, S., \& Hui, L.\ 
2002, \apj, 574, 1004

\bibitem[Seager \& Mall{\'e}n-Ornelas(2003)]{seager2003} 
Seager, S., \& Mall{\'e}n-Ornelas, G.\ 2003, \apj, 585, 1038 

\bibitem[Seager(2010)]{seager2010} Seager, S.\ 2010, Exoplanet Atmospheres: Physical Processes.~ By Sara Seager.~ Princeton University Press, 2010.~ISBN: 978-1-4008-3530-0, 

\bibitem[Seager \& Deming(2010)]{seagerd2010} 
Seager, S., \& Deming, D.\ 2010, \araa, 48, 631
 
\bibitem[Skrutskie et al.(2006)]{skrutskie2006} Skrutskie, M.~F., 
Cutri, R.~M., Stiening, R., et al.\ 2006, \aj, 131, 1163 

\bibitem[Socrates et al.(2012)]{socrates2012} Socrates, A., Katz, 
B., Dong, S., \& Tremaine, S.\ 2012, \apj, 750, 106 

\bibitem[Southworth(2009)]{southworth2009} Southworth, J.\ 2009, 
\mnras, 394, 272 


\bibitem[Spiegel et al.(2007)]{spiegel2007} Spiegel, D.~S., Haiman, 
Z., \& Gaudi, B.~S.\ 2007, \apj, 669, 1324 

\bibitem[Spiegel et al.(2011)]{spiegel2011} Spiegel, D.~S., 
Burrows, A., \& Milsom, J.~A.\ 2011, \apj, 727, 57 

\bibitem[Stassun et al.(2007)]{stassun2007} Stassun, K.~G., 
Mathieu, R.~D., \& Valenti, J.~A.\ 2007, \apj, 664, 1154 

\bibitem[Steffen \& Agol(2005)]{steffen05} Steffen, J.~H., \& 
Agol, E.\ 2005, \mnras, 364, L96 

\bibitem[Steffen et al.(2010)]{steffen2010} Steffen, J.~H., 
Batalha, N.~M., Borucki, W.~J., et al.\ 2010, \apj, 725, 1226 

\bibitem[Steffen et al.(2012)]{steffen2012} Steffen, J.~H., 
Ragozzine, D., Fabrycky, D.~C., et al.\ 2012, PNAS, in press (arXiv:1205.2309)


\bibitem[Stetson(1987)]{stetson1987} Stetson, P.~B.\ 1987, \pasp, 
99, 191 

\bibitem[Stetson(1990)]{stetson1990} Stetson, P.~B.\ 1990, \pasp, 
102, 932 


\bibitem[{{ter Braak}(2006)}]{braak06}
{ter Braak}, C. J.~F. 2006, Statistics and Computing, 16, 239

\bibitem[{{Torres} {et~al.}(2010){Torres}, {Andersen}, \& {Gim{\'e}nez}}]{torres10}
{Torres}, G., {Andersen}, J., \& {Gim{\'e}nez}, A. 2010, \aapr, 18, 67

\bibitem[Tusnski 
\& Valio(2011)]{tusnski2011} Tusnski, L.~R.~M., \& Valio, A.\ 2011, \apj, 743, 97 

\bibitem[van Saders 
\& Pinsonneault(2012)]{vansaders2012} van Saders, J.~L., \& Pinsonneault, M.~H.\ 2012, \apj, 746, 16 


\bibitem[Vidal-Madjar et al.(2003)]{vidal03} Vidal-Madjar, A., 
Lecavelier des Etangs, A., D{\'e}sert, J.-M., Ballester, G.~E., Ferlet, R., 
H{\'e}brard, G., \& Mayor, M.\ 2003, \nat, 422, 143 

\bibitem[Walter 
\& Bowyer(1981)]{walter1981} Walter, F.~M., \& Bowyer, S.\ 1981, \apj, 245, 671 


\bibitem[Welsh et al.(2012)]{welsh2012} Welsh, W.~F., Orosz, 
J.~A., Carter, J.~A., et al.\ 2012, \nat, 481, 475 

\bibitem[Winn et al.(2008)]{winn2008} Winn, J.~N., Holman, 
M.~J., Torres, G., et al.\ 2008, \apj, 683, 1076 

\bibitem[Winn(2009)]{winn2009} Winn, J.~N.\ 2009, IAU Symposium, 
253, 99 

\bibitem[Winn(2010)]{winn2010} 
Winn, J.~N.\ 2010, in Exoplanets, ed. S. Seager, Tucson:University of Arizona Press, 55 

\bibitem[Winn et al.(2005)]{winn2005} Winn, J.~N., Noyes, R.~W., 
Holman, M.~J., et al.\ 2005, \apj, 631, 1215
 
\bibitem[Winn et al.(2010)]{winnrm2010} Winn, J.~N., Fabrycky, D., 
Albrecht, S., \& Johnson, J.~A.\ 2010, \apjl, 718, L145 



\bibitem[Wright et al.(2010)]{wright2010} Wright, E.~L., 
Eisenhardt, P.~R.~M., Mainzer, A.~K., et al.\ 2010, \aj, 140, 1868 

\bibitem[Wright et al.(2011)]{wright2011} Wright, J.~T., Fakhouri, 
O., Marcy, G.~W., et al.\ 2011, \pasp, 123, 412 

\bibitem[Wright et al.(2012)]{wright2012} Wright, J.~T., Marcy, 
G.~W., Howard, A.~W., et al.\ 2012, ApJ, in press (arXiv:1205.2273)



\bibitem[Wu et al.(2007)]{wu2007} Wu, Y., Murray, N.~W., 
\& Ramsahai, J.~M.\ 2007, \apj, 670, 820 


\bibitem[Zacharias et al.(2004)]{zacharias2004} Zacharias, N., Monet, 
D.~G., Levine, S.~E., et al.\ 2004, Bulletin of the American Astronomical 
Society, 36, 1418 

\bibitem[Zucker et 
al.(2003)]{zucker2003} Zucker, S., Mazeh, T., Santos, N.~C., Udry, S., \& Mayor, M.\ 2003, \aap, 404, 775 




\end{thebibliography}
\end{document}